\documentclass[reqno, 11pt, figuresright]{amsart}
\usepackage{amssymb}

\usepackage[pdftex]{graphicx}
\usepackage{listings}
\usepackage{multirow}
\usepackage{placeins}
\usepackage{color}
\usepackage{lscape}
\usepackage{dsfont}
\usepackage{bbm}
\usepackage{graphicx}
\usepackage{subcaption}
\usepackage{makecell}
\usepackage{rotating}

\usepackage{graphicx}
\usepackage{multirow}
\usepackage{booktabs}
\usepackage{wrapfig, blindtext}
\usepackage{adjustbox}

\usepackage{rotating}
\usepackage{lipsum}
\usepackage{booktabs}

\usepackage{amsthm}
\usepackage[many]{tcolorbox}

\usepackage{ulem}

\usepackage{float}

\newtheorem{theorem}{Theorem}[section]
\tcolorboxenvironment{theorem}{
  colback=blue!5!white,
  boxrule=0pt,
  boxsep=1pt,
  left=2pt,right=2pt,top=2pt,bottom=2pt,
  oversize=2pt,
  sharp corners,
  before skip=\topsep,
  after skip=\topsep,
}
\newtheorem{corollary}{Corollary}[section]
\tcolorboxenvironment{corollary}{
  colback=blue!5!white,
  boxrule=0pt,
  boxsep=1pt,
  left=2pt,right=2pt,top=2pt,bottom=2pt,
  oversize=2pt,
  sharp corners,
  before skip=\topsep,
  after skip=\topsep,
}
\newtheorem{alg}{Algorithm}[section]
\tcolorboxenvironment{alg}{
  colback=blue!5!white,
  boxrule=0pt,
  boxsep=1pt,
  left=2pt,right=2pt,top=2pt,bottom=2pt,
  oversize=2pt,
  sharp corners,
  before skip=\topsep,
  after skip=\topsep,
}

\newcommand{\tabitem}{~~\llap{\textbullet}~~}

\newcommand{\BlackBoxes}{\global\overfullrule5pt}

\BlackBoxes

\newcommand{\R}{\mathbb{R}}

\usepackage{bm}

\theoremstyle{definition}

\numberwithin{equation}{section}

\def\0{\kern0pt\-\nobreak\hskip0pt\relax}

\makeatletter
\AtBeginDocument{%
 \def\@serieslogo{%
 \vbox to\headheight{%
 \parindent\z@ \fontsize{6}{7\p@}\selectfont
 \vss}}}

\def\makeoverbar#1#2#3#4#5#6#7{%
 \setbox0=\hbox{$\m@th#2\mkern#5mu{{}#3{}}\mkern#6mu$}%
 \setbox1=\null \dimen@=#4\fontdimen8#13 \dimen@=3.5\dimen@
 \advance\dimen@ by \ht0 \dimen@=-#7\dimen@ \advance\dimen@ by \wd0
 \ht1=\ht0 \dp1=\dp0 \wd1=\dimen@
 \dimen@=\fontdimen8#13 \fontdimen8#13=#4\fontdimen8#13
 \rlap{\hbox to \wd0{$\m@th\hss#2{\overline{\box1}}\mkern#5mu$}}
 \fontdimen8#13=\dimen@}

\def\mylabel#1#2{{\def\@currentlabel{#2}\label{#1}}}

\makeatother

\newcommand{\minus}{\scalebox{0.75}[1.0]{$-$}}

\allowdisplaybreaks

\def\v{Y_1}  
\def\w{Y_2}

\begin{document}


\title[A comparative study of factor models]{A comparative study of factor models for different periods of the electricity spot price market}

\author[C. \smash{Laudagé}]{Christian Laudagé${}^{\ast}$}
\address[C. Laudagé]{Rheinland-Pfälzische Technische Universität Kaiserslautern-Landau, DE-67663\\ Kaiserslautern, Germany}

\email{christian.laudage@rptu.de}

\author[F. \smash{Aichinger}]{Florian Aichinger${}^{\dagger,\ddagger}$}
\address[F. Aichinger]{Johann Radon Institute for Computational and Applied Mathematics (RICAM), Austrian Academy of Sciences, AT-4040 Linz, Austria/ Institute for Financial Mathematics and Applied Number Theory, University of Linz, AT-4040 Linz, Austria}

\email{florian.aichinger@ricam.oeaw.ac.at}

\author[S. \smash{Desmettre}]{Sascha Desmettre${}^{\ddagger}$}
\address[S. Desmettre]{Institute for Financial Mathematics and Applied Number Theory, University of Linz, AT-4040 Linz, Austria}

\email{sascha.desmettre@jku.at}

\subjclass[2020]{}

\thanks{}

\date{\today}

\begin{abstract}
Due to major shifts in European energy supply, a structural change can be observed in Austrian electricity spot price data starting from the second quarter of the year 2021 onward. 
In this work we study the performance of two different factor models for the electricity spot price in three different time periods.
To this end, we consider three samples of EEX data for the Austrian base load electricity spot price, one from the pre-crises from 2018 to 2021, the second from the time of the crisis from 2021 to 2023 and the whole data from 2018 to 2023.
For each of these samples, we investigate the fit of a classical $3$-factor model with a Gaussian base signal and one positive and one negative jump signal and compare it with a $4$-factor model to assess the effect of adding a second Gaussian base signal to the model. 
For the calibration of the models we develop a tailor-made Markov Chain Monte Carlo method based on Gibbs sampling. To evaluate the model adequacy, we provide simulations of the spot price as well as a posterior predictive check for the $3$- and the $4$-factor model.
We find that the $4$-factor model outperforms the $3$-factor model in times of non-crises. In times of crises, the second Gaussian base signal does not lead to a better fit of the model. To the best of our knowledge, this is the first study regarding stochastic electricity spot price models in this new market environment. Hence, it serves as a solid base for future research.   


\end{abstract}

\maketitle

\vspace{0.5cm}
\begin{minipage}{14cm}
{\small
\begin{description}
\item[\rm \textsc{ Key words} ]
{\small multi-factor models, Bayesian calibration, Markov Chain Monte Carlo, Ornstein-Uhlenbeck processes, electricity spot price, jump processes}

\item[\rm \textsc{ JEL classification} ]
{\small C15, C11, C13, Q40, C51, Q41}
\end{description}
}
\end{minipage}
\vspace{5mm}
\section*{Introduction}\label{intro}

\begin{figure}
\centering
    \includegraphics[width=140mm]{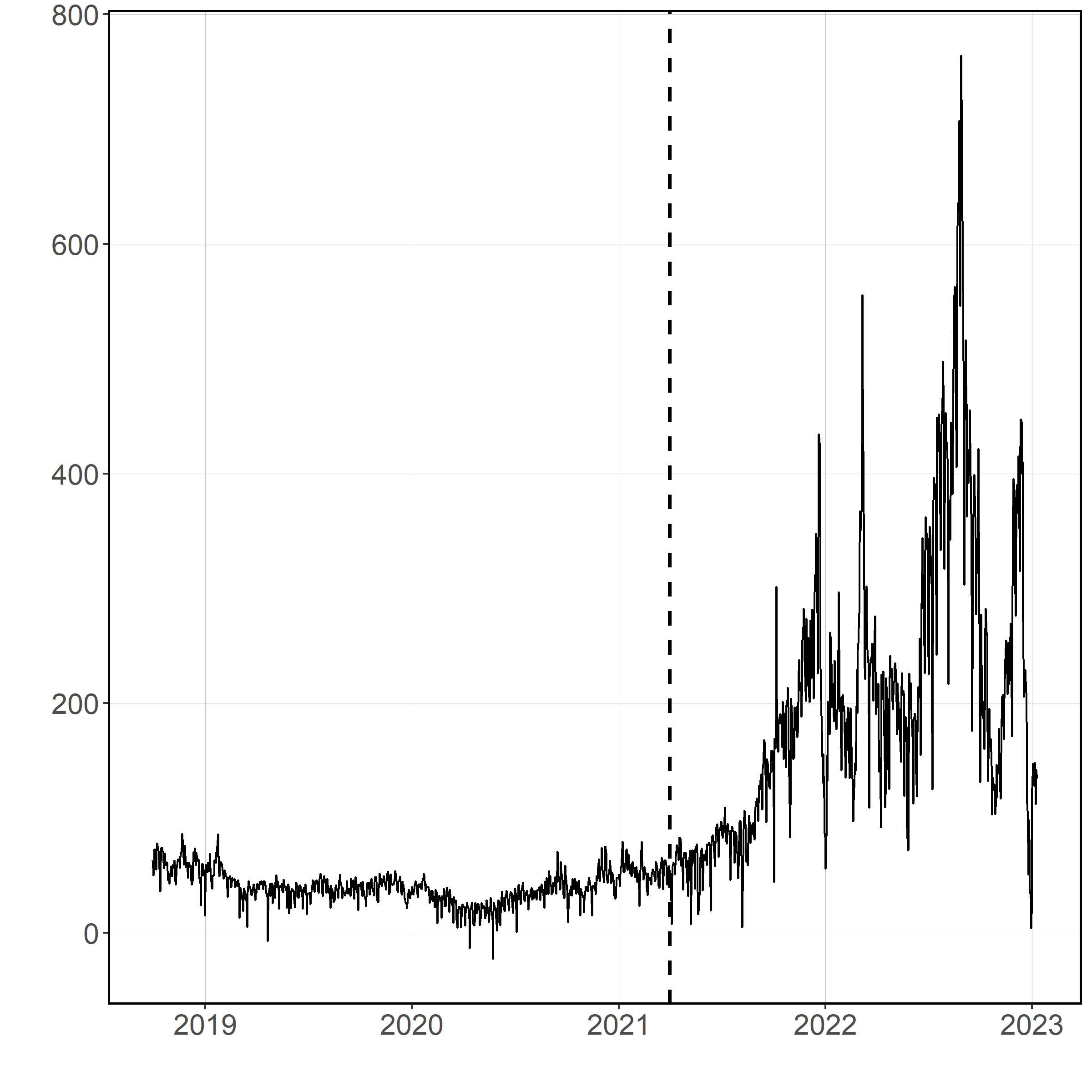}
    \caption{EEX spot price data from 30.9.2018 until 10.1.2023. We split the data at the vertical dashed line indicating the 1.4.2021, which marks a structural break in the data due to the beginning crisis.}
    \label{Data}
\end{figure}

For our study, we consider EEX data from Bloomberg for the Austrian\footnote{Since Austrian and German data is highly correlated, our results can as well be applied for modeling German electricity spot prices.} base load electricity spot price on a daily basis and including weekends for the time period from 2018 to 2023. As can be seen in Figure \ref{Data}, there is a structural break in the data around the second quarter of the year 2021, which marked the beginning of the European energy crisis. As a consequence, the energy market has become extremely volatile within the last two years, causing huge price peaks and large fluctuation of the electricity spot price. In our study, we focus on base load data (i.e. the daily average price from 0 a.m. to 12 p.m) but our modeling approach can also be extended to peak (average price on working days between 8 a.m. and 8 p.m.) and off-peak (average price of all hours that are not in the peak) data since all these data sets are strongly correlated and hence exhibit similar characteristics.
Our aim is to compare the fit of two different multi-factor electricity spot price models (cf. \cite{BKMB07}), with multiple superposed mean-reverting components and a seasonal trend for the respective market environments. To this end, we partition our data into two segments, the pre-crisis data from 2018 to 2021 and the crisis data from 2021 to 2023 (cf. Fig. \ref{Data}). 
The first spot price model we investigate is the established $3$-factor model proposed in \cite{GMP17}, where the deseasonalized spot price is modeled as a superposition of a Gaussian OU-process and a positive and a negative jump OU-process.  
The second model is a $4$-factor model, where a second Gaussian component is added to the model of \cite{GMP17}. This additional component should allow for a better distinction between small jumps and short term Gaussian fluctuations of the electricity price.
The two models are then calibrated to three different data sets, the 2018-21 data, the 2021-23 data and the whole 2018-23 data using Markov Chain Monte Carlo methods. This Bayesian approach to calibration allows a joint estimation of latent factors, taking into account possible interdependencies and also avoids the need to make strong apriori assumptions such as setting tresholds for jump sizes (cf. \cite{MBT08}). For each of these data sets, we provide model parameters along with simulations of the spot price and assessment of model adequacy through posterior predictive checking.

The paper is structured in the following way: In Section \ref{Related work} we give a non-exhaustive overview of the literature on electricity spot price models and their calibration. Table \ref{overview} provides a direct comparison of the characteristics for some of these models. Section \ref{model} introduces the $4$-factor model, which is an extension of the model of \cite{GMP17}. In Section \ref{Calibration}, we give a detailed description of the MCMC procedure, which is used for the calibration of the $4$-factor model. Section \ref{p_val} contains the calculation of the $p$-values which are crucial to assess model adequacy. Finally Section \ref{eval} provides the model parameters obtained from the MCMC algorithm together with simulations and a posterior predictive check for each model in each of the respective time periods. To conclude, we present interpretations for our results and discuss suggestions for future research.

\section{Contribution and related work}\label{Related work}

\subsection{Related work}

The first multi-factor electricity price models have been introduced by Lucia and Schwartz \cite{LS02} and Schwartz and Smith \cite{SS00}.
In these models the (log-)spot price is described as a superposition of two latent stochastic processes: The long term behavior is modeled as an arithmetic Brownian motion, the short term behavior as an OU-process. Since both components are Gaussian, the model can be calibrated using Kalman filter techniques. However, it turns out that Gaussian processes cannot appropriately describe the spikes, which frequently occur in observed spot price data. 
The one factor log-price model of Geman and Roncoroni \cite{GR06} generates the characteristic spikes by making the jump direction and intensity level-dependent: High price levels lead to high jump intensity and downward jumps are more likely, whereas if the price is low, jumps are rare and upward-directed. In the long run, the process always fluctuates around a deterministic mean value. The calibration procedure is mainly based on likelihood estimation. 
The model of Benth, Kallsen and Meyer-Brandis \cite{BKMB07} incorporates spikes and allows mean reversion to a stochastic base level by modeling the price as a sum of several OU-processes, some of which are driven by pure jump processes. A method to calibrate such a model was suggested by Tankov and Meyer-Brandis \cite{MBT08}, who investigate a superposition of two OU-processes, one driven by a Levy jump process, the other driven by a Brownian motion.  
To calibrate their model, they came up with the so-called hard thresholding technique, where first the mean reversion parameters are estimated from the autocorrelation function and then maximum likelihood methods are applied to filter out the spikes path. The same approach is used by Hinderks and Wagner \cite{HW20} for their two-factor model. The downside of calibrating the mean reversion rates separately is that some parameter interdependencies are being neglected. 
In \cite{BKN12} the models proposed in \cite{GR06} and \cite{BKMB07} as well as the one-factor mean-reversion jump-diffusion model of \cite{CF05} are calibrated to German spot price data adopting different calibration techniques.  A comparison of the resulting model properties shows that the 
factor model \cite{BKMB07} captures the fast mean-reversion of spikes and the slow
mean-reversion of the base signal very well. Another advantage of such factor models is that due to the additive linear structure, electricity forward contract prices can be calculated analytically (cf. \cite{BKMB07}). 
Gonzalez et al. \cite{GMP17} study a superposition model with one Gaussian OU-process and several jump components, each of which having its own jump size distribution, jump frequency and mean reversion rate. This additional flexibility allows to distinguish between different jump patterns caused by different underlying physical origins and simultaneously avoids the attribution of smaller jumps to the Gaussian process. 
The model is then calibrated in a Bayesian framework using Markov Chain Monte Carlo (MCMC) methods, generating samples from the posterior distributions of the model parameters.
In contrast to the MCMC methods based on time discretization of the involved processes developed in \cite{SUH07} and \cite{GN08}, Gonzalez et al. propose a MCMC algorithm for exact Bayesian inference and therefore do not have to take into account any approximation error. Table \ref{overview} on the next page provides a more detailed overview of the different spot price models and the corresponding calibration techniques mentioned above.  

\subsection{Contribution}

Based on the work of Gonzalez et al. \cite{GMP17}, we study two superposition models, 
a $3$-factor model consisting of one Gaussian and two jump OU-processes with different sign and a $4$-factor model, where a second Gaussian component is added to the 3-OU model (Section \ref{model}). To calibrate the $4$-factor model, in Section \ref{Calibration}, we develop an extension of the MCMC algorithm presented in \cite{GMP17}. 
We point out that even in the $4$-factor model, the inference performed by the MCMC algorithm is still exact at the level of distributions, since there is no discretization involved in the update of the second Gaussian component. The two models are then calibrated to EEX spot price data within different time periods and are then compared in terms of model adequacy using posterior predictive checks (Section \ref{eval}). Our study is to the best of our knowledge the first attempt to calibrate such models for the extremely volatile data in the period 2021-2023 and can serve as a starting point for future research in this direction.

\begin{sidewaystable}
\scalebox{0.9}
{
\begin{tabular}[h]{|l|c|c|c|}
    \hline
     & \textbf{Schwartz/Smith} & \textbf{Meyer-Brandis/Tankov} & \textbf{Hinderks/Wagner} \\
     \hline
     Model & $\operatorname{ln}(P_t)=X_t$ & $P_t=f_t+X_t$ & $P_t=f_t+X_t$ \\
    \hline
    deterministic part & none & \makecell{linear trend\\ trigonometric yearly seasonality\\weekend effects} & \makecell{linear trend\\ trigonometric yearly seasonality\\weekend effects} \\
    \hline
    stochastic part & \makecell{$X_t=S_t+L_t$ \\ $dL_t=\mu_l dt \sigma_L dW_t^L$\\ $dS_t=-\lambda S_tdt+\sigma_SdW_t^S$} & \makecell{$X(t)=Y_1(t)+Y_2(t)$ \\ $dY_{1}(t)=-\lambda_{1}^{-1} Y_{1}(t) dt+ dW(t)$ \\ $dY_{2}(t)=-\lambda_{2}^{-1} Y_{2}(t) dt+ dL(t)$} & \makecell{$X(t)=Y_1(t)+Y_2(t)$ \\ $dY_{1}(t)=\sigma dW(t)$ \\ $dY_{2}(t)=-\lambda Y_{2}(t) dt+ dL(t)$} \\
    \hline
    Long term modeling & Brownian motion & Gaussian OU-Process & Brownian motion \\
    \hline
    Short term modeling & Gaussian OU-Process & Jump OU-Process & Jump OU-Process\\
    \hline
    Jump specifications & no jumps & \makecell{$L$ is a Levy process\\ possibly time varying intensity } & \makecell{$L(t)=\sum\limits_{i=1}^{N(t)}B_i\cdot D_i$ \\ $B_i$ are iid Bernoulli distributed \\ $D_i$ are iid Gamma distributed}\\
    \hline
    Method of calibration & Kalman filtering & Hard thresholding & Hard thresholding \\
    \hline
    Remarks & \makecell{\tabitem Kalman Filtering does not work\\ if jump components are involved} & \makecell{\tabitem Does not work for more than\\ one Gaussian component} & \makecell{\tabitem Does not work for more than\\ one Gaussian component} \\
    \hline\\[1cm]\hline
    & \textbf{Seifert/Uhrig-Homburg} & \textbf{Gonzalez et al.}  & \textbf{4-OU model} \\
    \hline
    Model & $\operatorname{ln}(P_t)=f_t+X_t$ & $P_t=e^{f(t/260)}X_t$ & $P_t=f_t+X_t$\\
    \hline
    deterministic part & \makecell{linear trend\\ trigonometric yearly seasonality\\weekend effects} & \makecell{linear trend\\ trigonometric yearly seasonality} & \makecell{linear trend\\ trigonometric yearly seasonality}\\
    \hline
    stochastic part & \makecell{$X_t=S_t+L_t$ \\ $dL_t=\sigma_L dW_t^L$\\ $dS_t=\lambda(\mu-S_t)dt+\sigma_SdW_t^S+\xi_t dP_t$} & \makecell{$X_t= \sum\limits_{i=0}^n w_i Y_i(t),\, w_i\in\{-1,1\}$ \\ $dY_0(t)=-\lambda_0^{-1} (\mu-Y_0(t)) dt+\sigma dW_0(t)$\\ $dY_{i}(t)=-\lambda_{i}^{-1} Y_{i}(t) dt+ d\Pi_i(t),\, i=1,\dots,n$}   & \makecell{$X_t= Y_1(t) + Y_2(t) + J_{1}(t)-J_{2}(t)$ \\ $dY_i(t)=-\lambda_i^{-1} Y_i(t) dt+\sigma_{i}dW_i(t),\, i=1,2$\\ $dJ_{i}(t)=-\lambda_{i}^{-1} J_{i}(t) dt+ d\Pi_i(t),\, i=1,2$} \\
    \hline
    Long term modeling & Brownian motion  & Gaussian OU-Process & Gaussian OU-process\\
    \hline
    Short term modeling & OU-Process with jumps and BM & $n$ Jump OU-Processes & one Gaussian and two Jump OU-Processes\\
    \hline
    Jump specifications & \makecell{$P$ is a Poisson process\\ constant intensity\\ normally distributed jump sizes} & \makecell{$\Pi_i$ is a compound Poisson process\\ constant or time dependent intensity\\ exponentially distributed jump size}  & \makecell{$\Pi_i$ is a compound Poisson process\\ constant intensity\\ exponentially distributed jump size}\\
    \hline
    Method of calibration & MCMC methods & MCMC methods & MCMC methods\\
    \hline
    Remarks & \makecell{\tabitem Brownian fluctuations and jumps\\ both incorporated in process $S$\\ \tabitem same mean reversion} & \makecell{\tabitem separate mean reversion \\for each jump component} & \makecell{\tabitem separate mean reversion \\for each component}\\
    \hline
    
\end{tabular}
}
\caption{Detailed comparison of different electricity spot price models in the literature.}
\label{overview}
\end{sidewaystable}
\newpage
\section{The 3- and 4-factor model}\label{model}

In this section, we present two electricity spot price models, where the price process is modeled as a superposition of a deterministic seasonality function and multiple stochastic components. In the 3-factor model, the stochastic part consists of one Gaussian OU-process and two jump OU-processes and hence belongs to the class of models studied in \cite{GMP17}. In the general framework of \cite{GMP17}, the deseasonalized spot price is modeled as a superposition of one Gaussian OU process and up to $n$ (positive and negative) jump OU processes. However, for their calibration, only up to two positive jump components are considered, since their data does not include negative jumps. Since we also have to deal with negative jumps in our data, we take the model with one positive and one negative jump component as our base model. 
The 4-factor model is then a natural extension of the 3-factor model, where a second Gaussian OU component is added to the stochastic part. The idea to use two Gaussian components to model long and short-term behavior of the spot price goes back to the work of \cite{LS02} and has also been used in \cite{SUH07}. Incorporating more Gaussian components or jump processes into the model is straightforward from a theoretical perspective, but due to the increased complexity, convergence of the MCMC algorithm becomes significantly slower leading to problems in terms of computational tractability.

\subsection{The 3-factor model} We consider a probability space $(\Omega,\mathcal{F},\mathbb{P})$ with filtration $\mathbb{F}=(\mathcal{F}_t)_{t\geq 0}$, satisfying the usual conditions. On this probability space, we introduce an $\mathbb{F}$-adapted Brownian motions $W^{\v}$ and two compound Poisson processes $\Pi_1$ and $\Pi_2$ with constant jump intensity rates $\theta_1$, $\theta_2$ and exponentially distributed jump sizes with parameters $\beta_1$ and $\beta_2$ respectively. In the 3-factor model the electricity spot price 
\begin{equation}
P_t=f(t)+X_t
\end{equation}
is modeled as the sum of a deterministic seasonal function $f(t)$
and a stochastic part $X_t$, where the deseasonalized spot price $X_t$ is a superposition of three stochastic processes
\begin{equation}\label{3OU}
X(t)= {\v}(t)  + J_{1}(t)-J_{2}(t),
\end{equation}
with dynamics
\begin{equation}\label{GOU1}
d{\v}(t)=-\lambda_{\v}^{-1} {\v}(t) dt+\sigma_{{\v}}dW_{{\v}}(t),\quad \v(0)=0,
\end{equation}
\begin{equation}\label{Jump1}
dJ_{1}(t)=-\lambda_{J_1}^{-1} J_{1}(t) dt+ d\Pi_1(t),\quad J_1(0)=0,
\end{equation}
\begin{equation}\label{Jump2}
dJ_{2}(t)=-\lambda_{J_2}^{-1} J_{2}(t) dt+ d\Pi_2(t),\quad J_2(0)=0.\vspace{3mm}
\end{equation}
The process $\v$ admits the explicit solution
\begin{equation}\label{Gauss-OU solution}
Y_1(t)=\int_0^t \sigma e^{-\lambda_{Y_1}^{-1}(t-s)}dW_s,
\end{equation}
and is called a Gaussian OU process since the conditional distribution $Y_{t+s}$ given $Y_t$ is normally distributed with
\[
\mathbb{E}[Y_1(t+s)|Y_1(t)=y]=ye^{-\lambda_{Y_1}^{-1}s},\qquad \mathbb{V}[Y_1(t+s)|Y_1(t)=y]=\lambda_{Y_1}\sigma_{Y_1}^2\left(1-e^{-2\lambda_{Y_1}^{-1}s}\right)/2.
\]
In equation \eqref{GOU1}, the long term mean  is $0$, i.e.,~the process tends to revert to $0$. The parameter $\lambda_{Y_1}^{-1}$ is responsible for the speed of the mean reversion and $\sigma_{Y_1}$ governs the volatility of the respective process.
The processes $J_1$ and $J_2$ are driven by the compound Poisson processes $\Pi_1$ and $\Pi_2$ with interval representation $\Pi_i(t)=\sum_{j=1}^{\infty}\xi_j^i\mathbbm{1}_{\{t\geq\tau_j^i\}}$, $i=1,2$, where $\tau_j^i$ are the arrival times of a Poisson process and $\xi_j^i$ is the jump size at time $\tau_j^i$. $J_1$ and $J_2$ are called jump OU processes and they have explicit solutions
\[
J_i(t)=\sum\limits_{j:0\leq \tau_j^i\leq t}e^{-\lambda_{J_i}^{-1}(t-\tau_j^i)}\xi_j^i.
\]
The parameters $\lambda_{J_1}^{-1}$ and $\lambda_{J_2}^{-1}$ in equations \eqref{Jump1} and \eqref{Jump2} are the mean reversion speeds of the respective jump processes. The mean level is again $0$. This $3$-factor model is a special case of the class of models described in \cite{GMP17} for the case of one positive and one negative jump component.

\subsection{The 4-factor model} 
We now consider an extension of the 3-factor model, where a second Gaussian component is added to model the stochastic part.
The idea to use two Gaussian components to model long and short-term behavior of the spot price goes back to the work of \cite{LS02} and has also been used in \cite{SUH07}.
Let $W^{\w}$ be an $\mathbb{F}$-adapted Brownian motion independent of $W^{\v}$. We again model the electricity spot price 
\begin{equation}
P_t=f(t)+X_t
\end{equation}
as the sum of a deterministic seasonal function $f(t)$
and a stochastic part $X_t$, where the deseasonalized spot price $X_t$ is now a superposition of four stochastic processes
\begin{equation}\label{4OU}
X(t)= {\v}(t) + {\w}(t) + J_{1}(t)-J_{2}(t),
\end{equation}
with $\v$, $J_{1}$ and $J_2$ defined via \eqref{GOU1}, \eqref{Jump1}, \eqref{Jump2}, and $Y_2$ being a second Gaussian component with dynamics
\begin{equation}\label{GOU2}
d{\w}(t)=-\lambda_{\w}^{-1} {\w}(t) dt+\sigma_{{\w}}dW_{\w}(t),\quad \w(0)=0,
\end{equation}
In equation \eqref{GOU2}, the long term mean is again $0$, i.e.,~the process tends to revert to $0$. The parameter $\lambda_{Y_2}^{-1}$ is responsible for the speed of the mean reversion and $\sigma_{Y_2}$ governs the volatility of the respective process. Note that there is no apriori assumption on which of the two Gaussian components models the long- and which models the short-term fluctuations. This distinction is only made after the calibration by examining the posterior parameters and paths of the hidden variables.  We further point out that setting $\lambda_{Y_2}=\infty$ leads to a model, where the second Gaussian component is modeled as an arithmetic Brownian motion in the spirit of \cite{SS00}. We will also study this variant of our model in Section \ref{eval}.

\subsection{Futures prices}
In this section, we give an explicit formula for the price of a futures contract with maturity $T$ in the 4-factor model. In our model, there are six sources of uncertainty, two of them are generated by the diffusion factors of the respective Gaussian OU processes and two generated by each of the jump processes, namely jump-intensity risk and jump-size risk. We assume that the market price of risk arising from the diffusion of the Gaussian OU processes is constant, moreover we make the simplifying assumption that all the jump risk premia are captured by jump-size risk (cf. \cite{V03}). Thus, if $\theta^*$ denotes the jump-intensity under the risk neutral measure, we assume that $\theta^*=\theta$. Therefore, under the risk neutral measure $\mathbb{Q}$, the spot price process is of the form  
\begin{equation}\label{P-Q}
P_t=f(t)+\v(t)+\w(t)+J_{1}(t)-J_{2}(t),
\end{equation}
with
\begin{equation}\label{L-Q}
d\v(t)=(-\lambda_{\v}^{-1}\v(t)-\phi_{\v})dt+\sigma_{\v}d W_{\v}^{\mathbb{Q}}(t),
\end{equation}
\begin{equation}\label{S-Q}
d\w(t)=(-\lambda_{\w}^{-1}\w(t)-\phi_{\w})dt+\sigma_{\w}d W_{\w}^{\mathbb{Q}}(t),
\end{equation}
\begin{equation}\label{J1-Q}
dJ_{1}(t)=-\lambda_{J_1}^{-1} J_{1}(t) dt+ d\Pi_1(\theta_1,\beta_1^*),
\end{equation}
\begin{equation}\label{J2-Q}
dJ_{2}(t)=-\lambda_{J_2}^{-1} J_{2}(t) dt+ d\Pi_2(\theta_2,\beta_2^*),\vspace{3mm}
\end{equation}
where $\phi_L$ and $\phi_S$ are the risk premia for the long term and short term uncertainty and $\beta_1^*$, $\beta_2^*$ are the risk neutral jump size parameters. The following result provides an explicit formula for the price of a futures contract in the $4$-factor model. The proof can be found in the Appendix.
\vspace{3mm}
\begin{theorem}\label{ThmFutures}
If the electricity spot price $P$ is modelled by the $4$-factor model \eqref{4OU} with risk neutral dynamics \eqref{L-Q}-\eqref{J2-Q}, the price of a future contract on the spot price with maturity $T$ at time $t$ is given as
\begin{equation*}
    \begin{aligned}
    & F(t,T,P)=\mathbb{E}^{\mathbb{Q}}[P_T|\mathcal{F}_t]\\
    & = f(T)+\v(t) e^{-\lambda_{\v}^{-1}(T-t)}-\frac{\phi_{\v}}{\lambda_{\v}^{-1}}(1-e^{-\lambda_{\v}^{-1}(T-t)})+\w(t) e^{-\lambda_{\w}^{-1}(T-t)}-\frac{\phi_{\w}}{\lambda_{\w}^{-1}}(1-e^{-\lambda_{\w}^{-1}(T-t)})\\
    &\quad + J_{1}(t) e^{-\lambda_{J_1}^{-1}(T-t)}+\frac{\theta_1}{\beta_1^*}\lambda_{J_1}(1-e^{-\lambda_{J_1}^{-1}(T-t)})- J_{2}(t) e^{-\lambda_{J_2}^{-1}(T-t)}-\frac{\theta_2}{\beta_2^*}\lambda_{J_2}(1-e^{-\lambda_{J_2}^{-1}(T-t)}).
    \end{aligned}
\end{equation*}
\end{theorem}
\vspace{3mm}
The derivation of the futures prices in the 3OU model is now straightforward, since all terms related to the second Gaussian component $Y_2$ in the formula just vanish, while the other terms remain unchanged. The effect of a different number of Gaussian model components has also been studied in \cite[Section 3.1/3.3]{LS02}.
\vspace{3mm}
\begin{corollary}\label{CorFutures}
If the electricity spot price $P$ is modelled by the $3$-factor model \eqref{3OU} with risk neutral dynamics \eqref{L-Q},\eqref{J1-Q},\eqref{J2-Q}, the price of a future contract on the spot price with maturity $T$ at time $t$ is given as
\begin{equation*}
    \begin{aligned}
    & F(t,T,P)=\mathbb{E}^{\mathbb{Q}}[P_T|\mathcal{F}_t]= f(T)+\v(t) e^{-\lambda_{\v}^{-1}(T-t)}-\frac{\phi_{\v}}{\lambda_{\v}^{-1}}(1-e^{-\lambda_{\v}^{-1}(T-t)})\\
    &\quad + J_{1}(t) e^{-\lambda_{J_1}^{-1}(T-t)}+\frac{\theta_1}{\beta_1^*}\lambda_{J_1}(1-e^{-\lambda_{J_1}^{-1}(T-t)})- J_{2}(t) e^{-\lambda_{J_2}^{-1}(T-t)}-\frac{\theta_2}{\beta_2^*}\lambda_{J_2}(1-e^{-\lambda_{J_2}^{-1}(T-t)}).
    \end{aligned}
\end{equation*}
\end{corollary}

\vspace{5mm}
\section{Calibration}\label{Calibration}

In this section, we explain the calibration of the superposition model introduced in the previous section. After calibrating the deterministic seasonality function applying a least-squares method, we use a Bayesian approach to calibrate the parameters of the stochastic components to the deseasonalized data. To this end, we
apply MCMC methods to obtain samples from the posterior distributions of the model parameters. The algorithm we use is based on the algorithm described in \cite[Section 3.3]{GMP17}, extended by the update steps for the additional Gaussian component in the $4$-factor model. 

\subsection{Calibration of the seasonality function}

\begin{figure}[htbp]
\centering
    \begin{subfigure}{0.3\linewidth}
        \includegraphics[width=\linewidth]{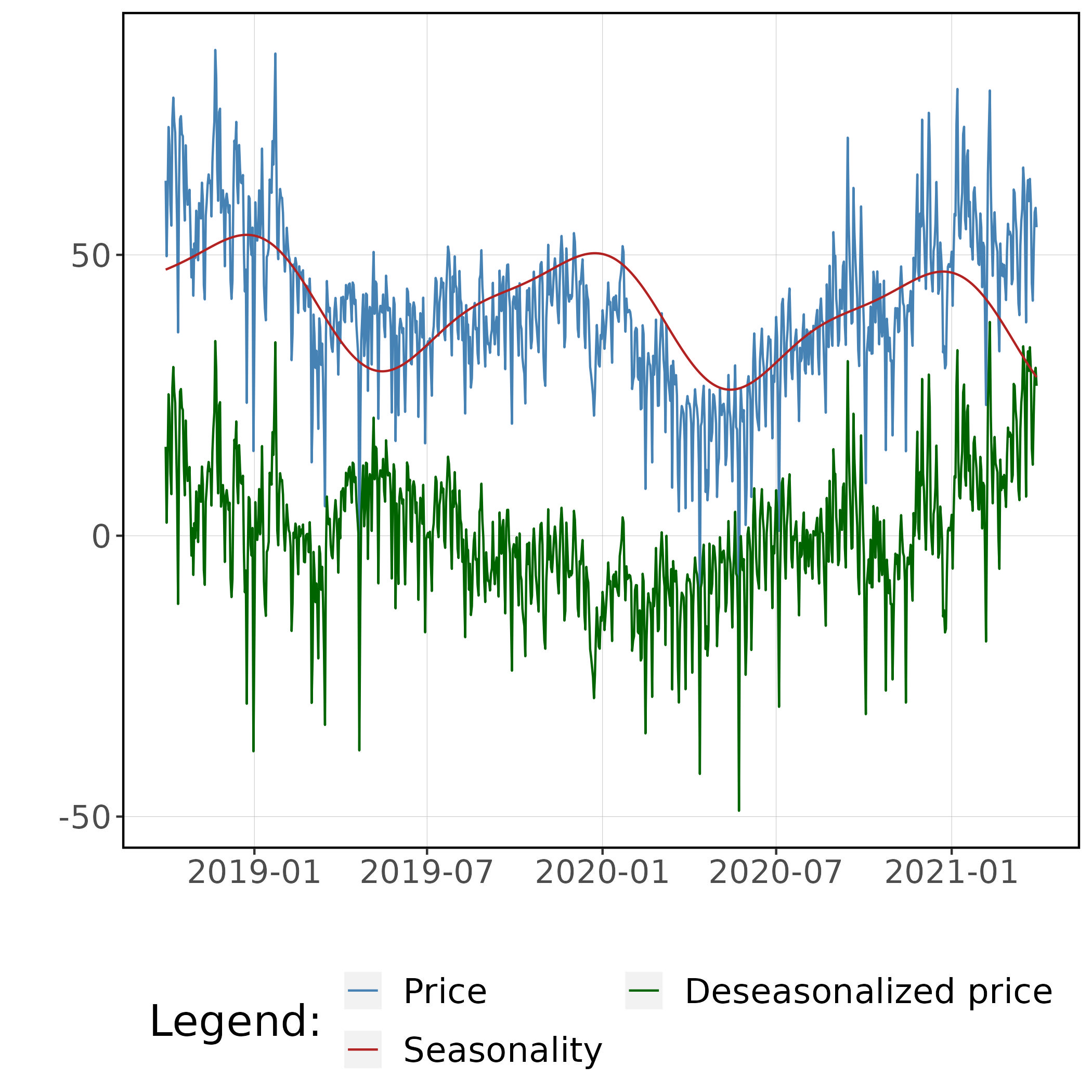}
    \caption{2018-21 Data}
    \end{subfigure}
\hfil
    \begin{subfigure}{0.3\linewidth}
        \includegraphics[width=\linewidth]{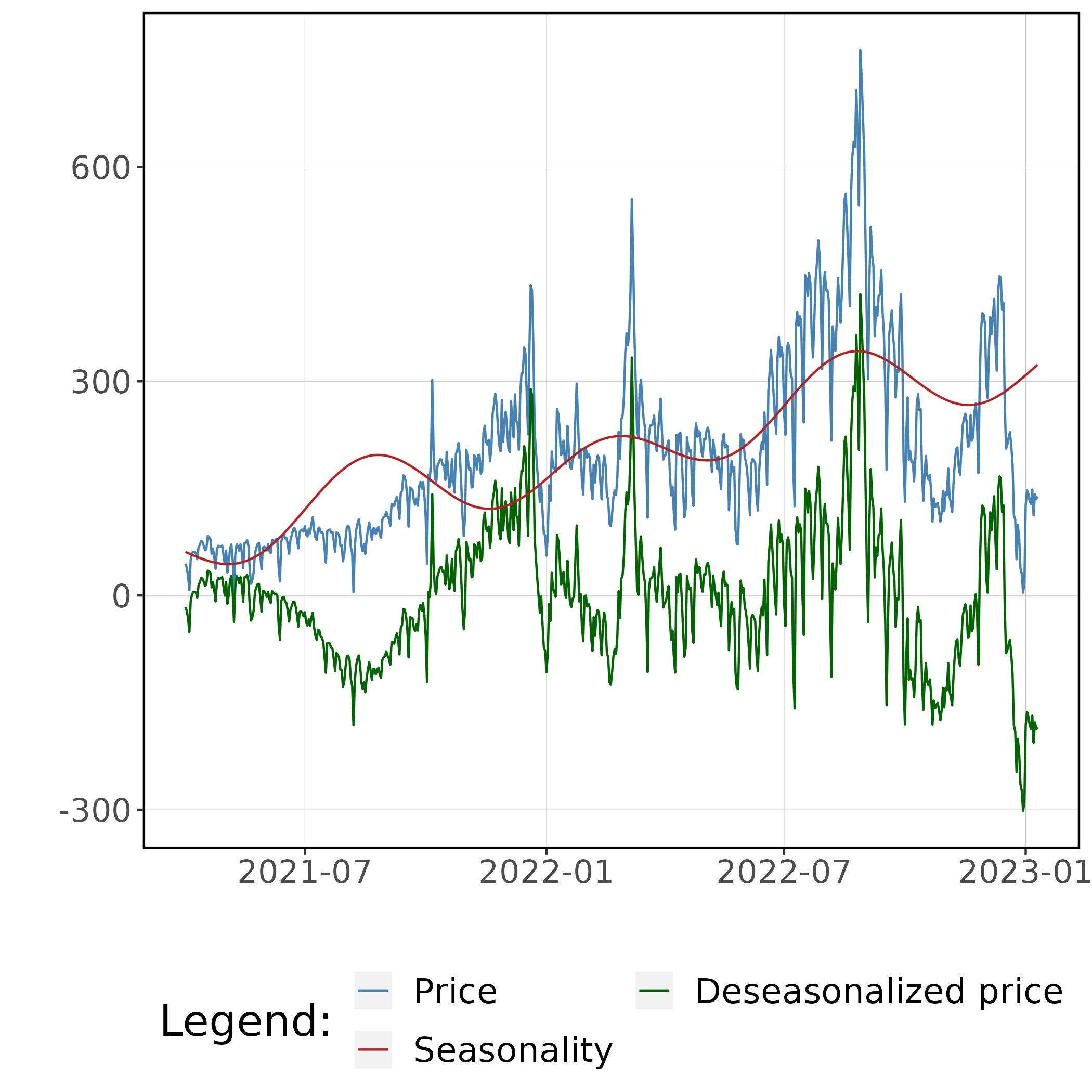}
    \caption{2021-23 Data}
    \end{subfigure}
\hfil
    \begin{subfigure}{0.3\linewidth}
        \includegraphics[width=\linewidth]{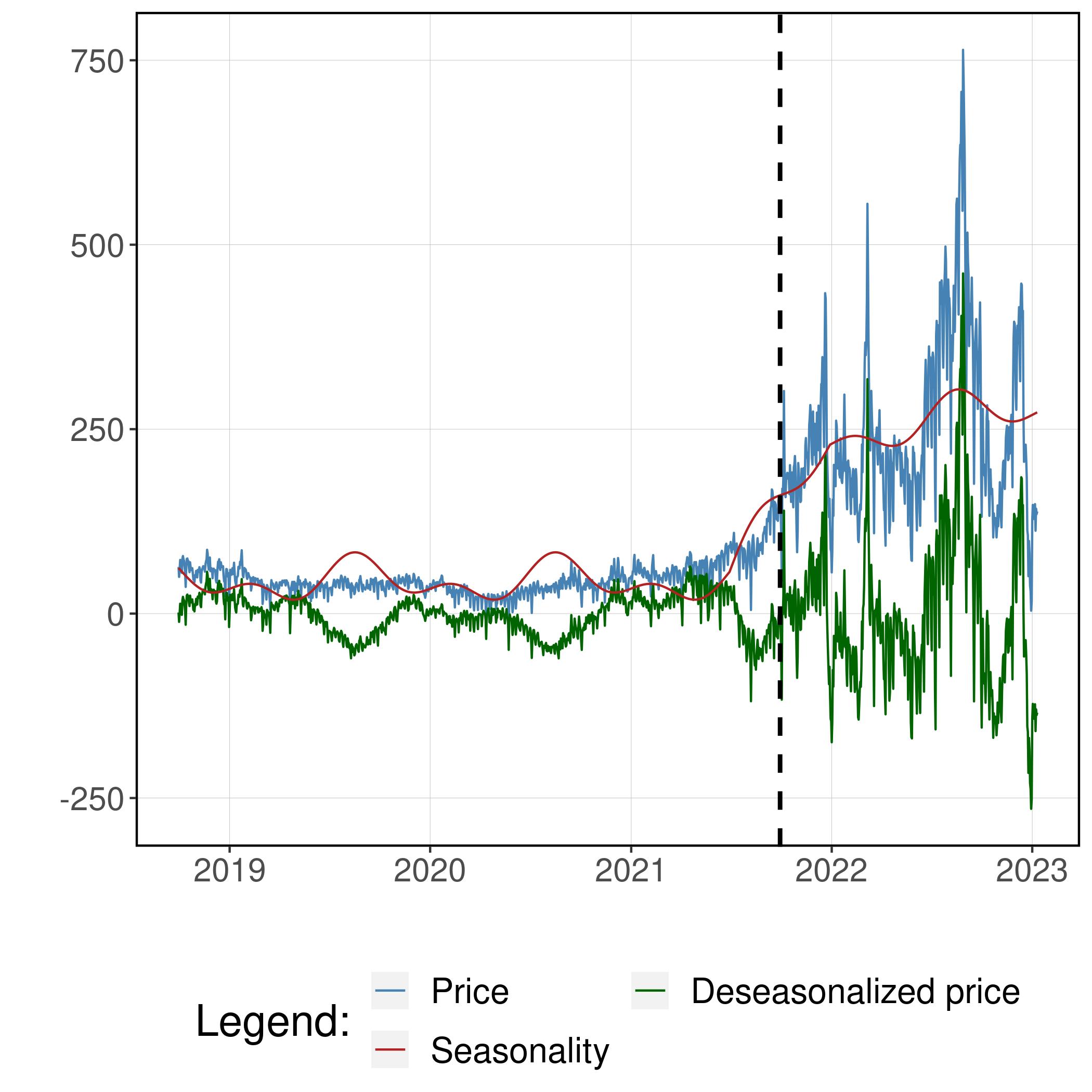}
    \caption{2018-23 Data}
    \end{subfigure}

\caption{Seasonal functions and deseasonalized data for different time periods}
\label{seasonality}
\end{figure}

We first calibrate the deterministic seasonality function $f$. We assume that $f$ is a superposition of a linear trend and trigonometric functions modeling half-yearly and quarterly seasonal variations of the spot price:
\vspace{3mm}
\begin{equation*}
\begin{aligned}
    {}
    & f(t;a_1,\dots,a_6)=a_1 + a_2t + a_3 \operatorname{sin}(2\pi t)   + a_4 \operatorname{cos}(2 \pi t) + a_5 \operatorname{sin}(4 \pi t) + a_6 \operatorname{cos}(4 \pi t).
\end{aligned}    
\end{equation*}
 The parameters $a_1,\dots, a_6$ are then obtained by applying least-squares methods.
 Finally $f$ gets subtracted from the data to get the deseasonalized data. Figure \ref{seasonality} shows the calibrated seasonal functions for the three different data samples we consider. 
 We did not include weekend effects in our seasonality function, since especially for the extremely volatile crisis data, these effects are negligible.
 In order to take into account the change of the mean level in the data after the beginning of the crisis, the linear part of the seasonal function was calibrated piecewise for three different time periods. This is important since for the next step of the calibration, we assume that the deseasonalized data has mean level zero. In the rightmost plot in Figure \ref{seasonality}, the dates on which the gradient changes are the 29.6.2021 and the 29.12.2021.

\subsection{Calibration of the stochastic part}
We calibrate the model parameters $\lambda_{\v}$,$\sigma_{\v}$,$\lambda_{\w}$,$\sigma_{\w}$,\\$\lambda_{J_1}$,$\lambda_{J_2}$,$\theta_1$,$\theta_2$,$\beta_1$,$\beta_2$ simultaneously in a Bayesian framework,
i.e.,~our aim is to obtain the joint posterior distribution \[\pi(\lambda_{\v},\sigma_{\v},\lambda_{\w},\sigma_{\w},\lambda_{J_1},\lambda_{J_2},\theta_1,\theta_2,\beta_1,\beta_2|\chi)
\]
given the deseasonalized spot price data $\chi$. To this end, we use MCMC methods, constructing a Markov chain whose stationary distribution is the posterior distribution of the model parameters. For a detailed introduction to the topic see for example \cite{BGJM11} and \cite{GCDVR13}.
For our calibration procedure, we apply Gibbs sampling, i.e.,~we iteratively generate samples from the distribution of each variable conditioned on the current values of the other variables. 
We start with samples 
\[\left(\lambda_{\v}^{(0)},\lambda_{\w}^{(0)},\sigma_{\v}^{(0)},\sigma_{\w}^{(0)},\lambda_{J_1}^{(0)},\lambda_{J_2}^{(0)},\theta_1^{(0)},\theta_2^{(0)},\beta_1^{(0)},\beta_2^{(0)}\right)
\]
from the apriori distribution of the parameters. Based on a sample \[\left(\lambda_{\v}^{(k)},\lambda_{\w}^{(k)},\sigma_{\v}^{(k)},\sigma_{\w}^{(k)},\lambda_{J_1}^{(k)},\lambda_{J_2}^{(k)},\theta_1^{(k)},\theta_2^{(k)},\beta_1^{(k)},\beta_2^{(k)}\right),
\]
we then obtain \[\left(\lambda_{\v}^{(k+1)},\lambda_{\w}^{(k+1)},\sigma_{\v}^{(k+1)},\sigma_{\w}^{(k+1)},\lambda_{J_1}^{(k+1)},\lambda_{J_2}^{(k+1)},\theta_1^{(k+1)},\theta_2^{(k+1)},\beta_1^{(k+1)},\beta_2^{(k+1)}\right)
\]
via the following algorithm:
\vspace{3mm}

\begin{alg}[Gibbs Sampling]{}\label{GS}
Starting from a given initial state of the chain, the MCMC algorithm cycles through the following steps:
    \begin{enumerate}
    \item ${\sigma_{\v}^2}^{(k+1)}\sim \pi\left({\sigma_{\v}^2}^{(k+1)}|\sigma_{\w}^{(k)},\lambda_{\v}^{(k)},\lambda_{\w}^{(k)},\lambda_{J_1}^{(k)},\lambda_{J_2}^{(k)},\theta_1^{(k)},\theta_2^{(k)},\beta_1^{(k)},\beta_2^{(k)},\mathcal{X}\right)$
    \item ${\sigma_{\w}^2}^{(k+1)}\sim \pi\left({\sigma_{\w}^2}^{(k+1)}|\sigma_{\v}^{(k+1)},\lambda_{\v}^{(k)},\lambda_{\w}^{(k)},\lambda_{J_1}^{(k)},\lambda_{J_2}^{(k)},\theta_1^{(k)},\theta_2^{(k)},\beta_1^{(k)},\beta_2^{(k)},\mathcal{X}\right)$
    \item $\lambda_{{\v}}^{(k+1)}\sim \pi\left(\lambda_{\v}^{(k+1)}|\sigma_{\v}^{(k+1)},\sigma_{\w}^{(k+1)},\lambda_{\w}^{(k)},\lambda_{J_1}^{(k)},\lambda_{J_2}^{(k)},\theta_1^{(k)},\theta_2^{(k)},\beta_1^{(k)},\beta_2^{(k)},\mathcal{X}\right)$
    \item $\lambda_{\w}^{(k+1)}\sim\pi\left(\lambda_{\w}^{(k+1)}|\sigma_{\v}^{(k+1)},\sigma_{\w}^{(k+1)},\lambda_{\v}^{(k+1)},\lambda_{J_1}^{(k)},\lambda_{J_2}^{(k)},\theta_1^{(k)},\theta_2^{(k)},\beta_1^{(k)},\beta_2^{(k)},\mathcal{X}\right)$
    \item $\lambda_{J_1}^{(k+1)}\sim \pi\left(\lambda_{J_1}^{(k+1)}|\sigma_{\v}^{(k+1)},\sigma_{\w}^{(k+1)},\lambda_{\v}^{(k+1)},\lambda_{\w}^{(k+1)},\lambda_{J_2}^{(k)},\theta_1^{(k)},\theta_2^{(k)},\beta_1^{(k)},\beta_2^{(k)},\mathcal{X}\right)$
    \item $\lambda_{J_2}^{(k+1)}\sim \pi\left(\lambda_{J_2}^{(k+1)}|\sigma_{\v}^{(k+1)},\sigma_{\w}^{(k+1)},\lambda_{\v}^{(k+1)},\lambda_{\w}^{(k+1)},\lambda_{J_1}^{(k+1)},\theta_1^{(k)},\theta_2^{(k)},\beta_1^{(k)},\beta_2^{(k)},\mathcal{X}\right)$
    \item $\theta_1^{(k+1)}\sim \pi\left(\theta_1^{(k+1)}|\sigma_{\v}^{(k+1)},\sigma_{\w}^{(k+1)},\lambda_{\v}^{(k+1)},\lambda_{\w}^{(k+1)},\lambda_{J_1}^{(k+1)},\lambda_{J_2}^{(k+1)},\theta_2^{(k)},\beta_1^{(k)},\beta_2^{(k)},\mathcal{X}\right)$
    \item $\theta_2^{(k+1)}\sim \pi\left(\theta_2^{(k+1)}|\sigma_{\v}^{(k+1)},\sigma_{\w}^{(k+1)},\lambda_{\v}^{(k+1)},\lambda_{\w}^{(k+1)},\lambda_{J_1}^{(k+1)},\lambda_{J_2}^{(k+1)},\theta_1^{(k+1)},\beta_1^{(k)},\beta_2^{(k)},\mathcal{X}\right)$
    \item $\beta_1^{(k+1)}\sim \pi\left(\beta_1^{(k+1)}|\sigma_{\v}^{(k+1)},\sigma_{\w}^{(k+1)},\lambda_{\v}^{(k+1)},\lambda_{\w}^{(k+1)},\lambda_{J_1}^{(k+1)},\lambda_{J_2}^{(k+1)},\theta_1^{(k+1)},\theta_2^{(k+1)},\beta_2^{(k)},\mathcal{X}\right)$
    \item $\beta_2^{(k+1)}\sim \pi\left(\beta_2^{(k+1)}|\sigma_{\v}^{(k+1)},\sigma_{\w}^{(k+1)},\lambda_{\v}^{(k+1)},\lambda_{\w}^{(k+1)},\lambda_{J_1}^{(k+1)},\lambda_{J_2}^{(k+1)},\theta_1^{(k+1)},\theta_2^{(k+1)},\beta_1^{(k+1)},\mathcal{X}\right)$
\end{enumerate}
\end{alg}

\vspace{3mm}
\subsection{Data augmentation}\label{Data augmentation}
 The Gibbs sampler would yield samples of the posterior distribution of the parameters, however, in order to calculate densities involved in the Gibbs algorithm, we need the likelihood $l(\chi|\lambda_{\v},\sigma_{\v},\lambda_{\w},\sigma_{\w},\lambda_{J_1},\lambda_{J_2},\theta_1,\theta_2,\beta_1,\beta_2)$ of the data given the parameters, which for our superposition model cannot be calculated analytically or numerically. Thus, to overcome this hurdle, we use so-called data augmentation methods (cf. \cite{BNS01}, \cite{RPD04}). We point out that the augmentation method we use is an extension of the procedure described in \cite{GMP17} when adding a second Gaussian component. \\
Let $\chi=\{x_0,\dots,x_N\}$ be the vector of observations of the deseasonalized spot price at times $0=t_0,\dots,t_N=T$ and denote by $\Delta t=t_i-t_{i-1}$ the time increment between two observations. In order to get an explicit expression for the likelihood $l(\chi|\lambda_{\v},\sigma_{\v},\lambda_{\w},\sigma_{\w},\lambda_{J_1},\lambda_{J_2},\theta_1,\theta_2,\beta_1,\beta_2)$, we augment the state space with observations $\mathcal{Y}_2=\{y_{2,0},\dots,y_{2,N}\}$ of the short-term process and observations $\mathcal{J}_1=\{j_{1,0},\dots, j_{1,N}\}$ and $\mathcal{J}_2=\{j_{2,0},\dots, j_{2,N}\}$ of the jump processes.
The likelihood of the data $\mathcal{X}$ given $\mathcal{Y}_2$, $\mathcal{J}_1$ and $\mathcal{J}_2$ becomes independent of $\lambda_{\w},\sigma_{\w},\beta_1, \beta_2, \theta_1, \theta_2$ and can be calculated explicitly as 

\begin{equation}\label{density}
\begin{aligned}
    {}
    & l(\mathcal{X}|\lambda_{\v},\sigma_{\v},\mathcal{Y}_2,\mathcal{J}_1,\mathcal{J}_2)\\
    &\qquad\qquad =\frac{1}{\sqrt{2\pi}\prod_{i=1}^N \sqrt{\lambda_{\v} \sigma_{\v}^2\left(1-e^{2\lambda_{\v}^{-1}\Delta t}\right)/2}}\operatorname{exp}\left\{-\frac{1}{2}\sum\limits_{i=1}^N \frac{\left(y_{1,i}-y_{1,i-1} e^{-\lambda_{\v}^{-1}\Delta t}\right)^2}{\lambda_{\v} \sigma_{\v}^2\left(1-e^{-2\lambda_{\v}^{-1}\Delta t}\right)/2} \right\},\\
    & \text{with }y_{1,i}=x_i-y_{2,i}-j_{1,i}+j_{2,i}.
\end{aligned}    
\end{equation}

Since for $N$ iid random variables $\epsilon_{i}\sim \mathcal{N}(0,1)$, $i=1,\dots, N$, the transition densities of the Gaussian OU-process $Y_2$ are given by 
\[
Y_2(j\Delta t)=Y_2((j-1)\Delta t)e^{-\lambda_{Y_2}^{-1}\Delta t}+\left(\frac{\sigma_{Y_2}^2\lambda_{Y_2}}{2}\left(1-e^{-2\lambda_{Y_2}^{-1}\Delta t}\right)\right)^{1/2}\epsilon_{2,i},\quad i=1,\dots,N.
\]
Instead of treating the process $Y_2(t)$ itself as missing data, the random vector $\mathcal{E}=\{\epsilon_{2,1},\dots,\epsilon_{2,N}\}$ is considered as a hidden variable. Similarly the sets of pairs $\Phi_1=\{(\tau_{1,j},\xi_{1,j})\}_{1\leq j\leq N_{T_1}}$ and $\Phi_2=\{(\tau_{2,j},\xi_{2,j})\}_{1\leq j\leq N_{T_2}}$ of jump times and corresponding jump sizes are treated as hidden variables instead of $J_1(t)$ and $J_2(t)$. This has the advantage, that the parameters are independent of the latent variables. The sets $\Phi_i$ can be interpreted as realisations of marked Poisson processes taking values in $[0,T]\times (0,\infty)$ and their likelihood $l(\Phi_i|\theta_i,\beta_i)$ can be calculated with respect to a dominating measure (for details see Appendix B).
Thus with 

\begin{equation}\label{observations}
\begin{aligned}
y_{2,0}=0;\quad y_{2,i}=y_{2,i-1}e^{-\lambda_{Y_2}^{-1}\Delta t}+\left(\frac{\sigma_{Y_2}^2\lambda_{Y_2}}{2}\left(1-e^{-2\lambda_{Y_2}^{-1}\Delta t}\right)\right)^{1/2}\epsilon_{i},\quad i=1,\dots,N,\\
j_{1,i}=\sum\limits_{n=1}^{N_T^1}e^{-\lambda_{J1}^{-1}(i\cdot\Delta t -\tau_{1,n})}\xi_{1,n} \mathbbm{1}_{\{\tau_{1,n}\leq i\cdot\Delta t\}},\quad i=0,\dots,N,\\
j_{2,i}=\sum\limits_{n=1}^{N_T^2}e^{-\lambda_{J1}^{-1}(i\cdot\Delta t -\tau_{2,n})}\xi_{2,n} \mathbbm{1}_{\{\tau_{2,n}\leq i\cdot\Delta t\}},\quad i=0,\dots,N,
\end{aligned}
\end{equation}
Thus the likelihood of the data given the model parameters and our hidden variables can be calculated explicitly via
\begin{equation*}
\begin{aligned}
    {}
    & l(\mathcal{X}|\lambda_{\v},\sigma_{\v},\lambda_{\w},\sigma_{\w},\lambda_{J_1},\lambda_{J_2},\Phi_1,\Phi_2,\mathcal{E})= \eqref{density}
\end{aligned}    
\end{equation*}
with the observations $\mathcal{Y}_2$,  $\mathcal{J}_1$ and $\mathcal{J}_2$ obtained via \eqref{observations}.
To see, why the explicit likelihood of the data is particularly important for the Gibbs sampling, consider the following factorization which is frequently used within the algorithm:

\begin{small}
\begin{equation}\label{factorization}
    \begin{aligned}
    {}
    & \pi(\lambda_{\v},\sigma_{\v},\lambda_{\w},\sigma_{\w},\lambda_{J_1},\lambda_{J_2},\theta_1,\theta_2,\beta_1,\beta_2,\Phi_1,\Phi_2,\mathcal{E}|\mathcal{X}) \propto l(\mathcal{X}|\lambda_{\v},\sigma_{\v},\lambda_{\w},\sigma_{\w},\lambda_{J_1},\lambda_{J_2},\Phi_1,\Phi_2,\mathcal{E})\\
    & \cdot l(\Phi_1|\theta_1,\beta_1)\cdot l(\Phi_2|\theta_2,\beta_2)\cdot \pi(\lambda_{\v},\sigma_{\v},\lambda_{\w},\sigma_{\w},\lambda_{J_1},\lambda_{J_2},\theta_1,\theta_2,\beta_1,\beta_2,\mathcal{E}).
   \end{aligned}
\end{equation}
\end{small}






\subsection{Classes of prior distributions}
We specify the classes of prior distributions for the model parameters. As in \cite{GMP17}, conjugate priors are chosen whenever it is possible for the sake of computational efficiency. The starting values for the first MCMC iteration are the starting values of \cite{GMP17} scaled to the yearly time framework. Here $\operatorname{IG}(a,b)$ denotes the inverse Gamma distribution with shape parameter $a$ and scale parameter $b$ and $\operatorname{Ga}(a,b)$ denotes the Gamma distribution with shape parameter $a$ and scale parameter $b$.

\begin{table}[h]
\begin{adjustbox}{width=\columnwidth,center}
\centering
\begin{tabular}{l|c|c|c|c}
\toprule
 {} &{Parameter} & {Prior distribution} & {Starting value (3OU)} & {Starting value (4OU)}\\
\midrule
Volatility of 1. Gaussian OU process & $\sigma_{\v}$ & $\operatorname{IG}(1.5,0.005\cdot 365)$ & $0.2\sqrt{365}$ & $0.1\sqrt{365}$  \\
Volatility of 2. Gaussian OU process & $\sigma_{\w}$ & - & - & $10$ \\
Mean reversion speed of 1. Gaussian OU process & $\lambda_{\v}$ & $\operatorname{IG}(1,1)$  & $\frac{5}{365}$ & $0.001$\\
Mean reversion speed of 2. Gaussian OU process & $\lambda_{\w}$ & $\operatorname{IG}(1,1)$  & $1$ & $1$ \\
Mean reversion speed of pos. jump process & $\lambda_{J_1}$ & $\operatorname{IG}(1,1)$ & $\frac{5}{365}$ & $\frac{1}{365}$\\
Mean reversion speed of pos. jump process & $\lambda_{J_2}$ &$\operatorname{IG}(1,1)$& $\frac{1}{365}$ & $\frac{1}{365}$\\
Positive jump intensity & $\theta_1$ & $\operatorname{Ga}(1,\frac{10}{365})$ & $0.001\cdot 365$ & $0.001\cdot 365$\\
Negative jump intensity & $\theta_2$ & $\operatorname{Ga}(1,\frac{10}{365})$ & $0.001\cdot 365$ & $0.001\cdot 365$ \\
Positive jump size & $\beta_1$ & $\operatorname{IG}(1,1)$ & $0.5$ & $0.5$  \\
Negative jump size & $\beta_2$ & $\operatorname{IG}(1,1)$ & $0.5$ & $0.5$  \\

\bottomrule
\end{tabular}
\end{adjustbox}
\caption{Prior distributions and starting values for the model parameters. For the prior distribution of $\sigma_{\w}$ we refer to Section~\ref{MCMC}.}
\end{table}

\subsection{MCMC algorithm for the 4-factor model}\label{MCMC}

We use Gibbs sampling to iteratively generate samples from the distribution of each variable conditioned on the current values of the other variables. Compared to the algorithm in \cite[Section 3.3]{GMP17}, we have additional update steps for the parameters $\sigma$ and $\lambda$ of the second Gaussian OU process and its path (the additional hidden variable $\mathcal{E}$). The update steps for the jump paths (the hidden variables $\Phi_i$) have been developed in \cite{RPD04} and \cite{FS09} in the context of calibrating volatility models and have first been used for electricity spot price models in \cite{GMP17}.
To ensure that the mixing is of the same order for small and large values of $\lambda$, a transformation of $\lambda$ to $\rho:=e^{-\lambda^{-1}}$ is applied in the inference procedure. A detailed explanation of each updated step is given in Appendix A.
\vspace{3mm}

\begin{alg}[Augmented Gibbs Sampling]{}\label{AGS}
Starting from a given initial state of the chain, the MCMC algorithm cycles through the following steps. 
\begin{enumerate}
    \item $\sigma_{\v}^2\sim \pi(\sigma_{\v}^2|\lambda_{\v},\lambda_{\w},\sigma_{\w},\lambda_{J_1},\lambda_{J_2},\Phi_1,\Phi_2,\mathcal{E},\mathcal{X})$
    \item $\sigma_{\w}^2\sim \pi(\sigma_{\w}^2|\lambda_{\v},\sigma_{\v},\lambda_{\w},\lambda_{J_1},\lambda_{J_2},\Phi_1,\Phi_2,\mathcal{E},\mathcal{X})$
    \item $\rho_{{\v}}\sim \pi(\rho_{\v}|\sigma_{\v},\lambda_{\w},\sigma_{\w},\lambda_{J_1},\lambda_{J_2},\Phi_1,\Phi_2,\mathcal{E},\mathcal{X})$
    \item $\rho_{\w}\sim\pi(\rho_{\w}|\lambda_{\v},\sigma_{\v},\sigma_{\w},\lambda_{J_1},\lambda_{J_2},\Phi_1,\Phi_2,\mathcal{E},\mathcal{X})$
    \item $\rho_{J_1}\sim \pi(\rho_{J_1}|\sigma_{\v},\lambda_{{\v}},\lambda_{\w},\sigma_{\w},\lambda_{J_2},\Phi_1,\Phi_2,\mathcal{E},\mathcal{X})$
    \item $\rho_{J_2}\sim \pi(\rho_{J_2},\sigma_{\v},\lambda_{{\v}},\lambda_{\w},\sigma_{\w},\lambda_{J_1},\Phi_1,\Phi_2,\mathcal{E},\mathcal{X})$
    \item $\theta_1\sim \pi(\theta_1|\Phi_1)$
    \item $\theta_2\sim \pi(\theta_2|\Phi_2)$
    \item $\beta_1\sim \pi(\beta_1|\Phi_1)$
    \item $\beta_2\sim \pi(\beta_2|\Phi_2)$
    \item $\mathcal{E} \sim \pi(\mathcal{E}|\lambda_{\v},\sigma_{\v},\lambda_{\w},\sigma_{\w},\lambda_{J_1},\lambda_{J_2},\Phi_1,\Phi_2,\mathcal{X})$
    \item $\Phi_1 \sim \pi(\Phi_1|\lambda_{\v},\sigma_{\v},\lambda_{\w},\sigma_{\w},\lambda_{J_1},\lambda_{J_2},\theta_1,\beta_1,\Phi_2,\mathcal{E},\mathcal{X})$
    \item $\Phi_2 \sim \pi(\Phi_2|\lambda_{\v},\sigma_{\v},\lambda_{\w},\sigma_{\w},\lambda_{J_1},\lambda_{J_2},\theta_2,\beta_2,\Phi_1,\mathcal{E},\mathcal{X})$\\
\end{enumerate}
\end{alg}

\vspace{3mm}

\section{Posterior predictive check}\label{p_val}

We assess the adequacy of the latent variables involved in our models by calculating $p$-values for the trajectories of the posterior distribution of the hidden variables, which we obtain throughout our MCMC procedure. As demonstrated in \cite{GMP17}, this allows us to perform posterior predictive checks in the sense of \cite{R84}.
Let $m$ be the total number of iterations of the MCMC algorithm and $n$ be the number of iterations within the burn-in period. For each iteration $n< k \leq m$ calculate $p$-values in the following way: 

\subsection{$p$-values for the Gaussian OU-process $Y_1$}
For $N$ i.i.d.~random variables $\epsilon_{1,j}\sim \mathcal{N}(0,1)$, $j=1,\dots, N$, the transition densities of the Gaussian OU-process $Y_1^{(k)}$ are given by 
\[
Y_1^{(k)}(t_j)=Y_1^{(k)}(t_{j-1})e^{-\left(\lambda_{Y_1}^{(k)}\right)^{-1}\Delta_j}+\left(\frac{\left(\sigma_{Y_1}^{(k)}\right)^2\lambda_{Y_1}^{(k)}}{2}\left(1-e^{-2\left(\lambda_{Y_1}^{(k)}\right)^{-1}\Delta_j}\right)\right)^{1/2}\epsilon_{1,j}^{(k)},\quad j=1,\dots,N.
\]
Take the realisations $\left(y_{1,0}^{(k)},\dots, y_{1,N}^{(k)}\right)$ of $Y_1^{(k)}$ obtained in the MCMC step and calculate noise data
\[
\epsilon_{1,j}^{(k)}=\left(y_{1,j}^{(k)}-y_{1,j-1}^{(k)}e^{-\left(\lambda_{Y_1}^{(k)}\right)^{-1}\Delta_j}\right)\left(\frac{\left(\sigma_{Y_1}^{(k)}\right)^2\lambda_{Y_1}^{(k)}}{2}\left(1-e^{-2\left(\lambda_{Y_1}^{(k)}\right)^{-1}\Delta_j}\right)\right)^{-1/2},\quad j=1,\dots,N. 
\]
Use the noise data $\left\{\epsilon_{1,j}^{(k)}\right\}_{j=1,\dots,N}$ as input for a Kolmogorov-Smirnov test for the standard normal distribution, yielding the corresponding $p$-value $p_{Y_1}^{(k)}$. 

\subsection{$p$-values for the Gaussian OU-process $Y_2$}
Subject the components $\left\{\epsilon_{2,j}^{(k)}\right\}_{j=1,\dots,N}$ of the hidden variable $\mathcal{E}^{(k)}$ obtained in the $k$-th step of the MCMC algorithm to a Kolmogorov-Smirnov test for the standard normal distribution, yielding the corresponding $p$-value $p_{Y_2}^{(k)}$.

\subsection{$p$-values for the jump processes}
For the jump data $\Phi_i^{(k)}$, $i\in\{1,2\}$ sampled from the Markov chain in the $k$-th iteration we have
\[
\Phi_i^{(k)}=\left\{\left(\tau_{i,j}^{(k)},\xi_{i,j}^{(k)}\right)\right\}_{0\leq j\leq N_{T_i}^{(k)}}.
\]
Subject the jump sizes $\left\{\xi_{i,j}^{(k)}\right\}_{j=0,\dots,N_{T_i}}$ to a Kolmogorov-Smirnov test for the exponential distribution with mean $\beta_i^{(k)}$ to get the $p$-value $p_{\xi_i}^{(k)}$. For the jump locations $\left\{\tau_{i,j}^{(k)}\right\}_{j=1,\dots,N_{T_i}}$ calculate the set of inter-arrival times $\left\{\Delta\tau_{i,j}^{(k)}\right\}_{j=1,\dots,N_{T_i}}$ with $\Delta\tau_{i,j}^{(k)}=\tau_{i,j}^{(k)}-\tau_{i,j-1}^{(k)}$. Perform a Kolmogorov-Smirnov test for the exponential distribution with mean $\left(\theta_i^{(k)}\right)^{-1}$ on the set of inter-arrival times $\left\{\Delta\tau_{i,j}^{(k)}\right\}_{j=1,\dots,N_{T_i}}$ to obtain the $p$-value $p_{\theta_i}^{(k)}$.

\subsection{Calculation of posterior predictive $p$-values}
For each iteration step $k$, $n< k\leq m$, we obtain $6$ $p$-values
\[
\left\{p_{Y_1}^{(k)},p_{Y_2}^{(k)},p_{\xi_1}^{(k)},p_{\theta_1}^{(k)},p_{\xi_2}^{(k)},p_{\theta_2}^{(k)}\right\}.
\]
The posterior predictive $p$-values $\left\{p_{Y_1},p_{Y_2},p_{\xi_1},p_{\theta_1},p_{\xi_2},p_{\theta_2}\right\}$ are now obtained by calculating the mean values 
\[
\left\{p_{Y_1},p_{Y_2},p_{\xi_1},p_{\theta_1},p_{\xi_2},p_{\theta_2}\right\}=\left\{\frac{\sum\limits_{k=n+1}^m p_{Y_1}^{(k)}}{m-n},\frac{\sum\limits_{n+1}^m p_{Y_2}^{(k)}}{m-n},\frac{\sum\limits_{n+1}^m p_{\xi_1}^{(k)}}{m-n},\frac{\sum\limits_{n+1}^m p_{\theta_1}^{(k)}}{m-n},\frac{\sum\limits_{n+1}^m p_{\xi_2}^{(k)}}{m-n},\frac{\sum\limits_{n+1}^m p_{\theta_2}^{(k)}}{m-n}\right\}.
\]
\vspace{5mm}
\section{Evaluation of the model fit}\label{eval}
In this section, we compare the fit of the $3$-factor and $4$-factor model for three different time periods namely the pre-crisis period 2018-2021, the crisis 2021-23 and the whole interval 2018-2023. The EEX data we use for our studies starts at 30.9.2018 and ends at 10.1.2023. When we separate our data to investigate the model adequacy for the respective time intervals, we consider the 1.4.2021 as the start of the crisis period (cf. Figure \ref{Data}). This choice is partly motivated in order not to include the strong upwards trend observed from the second quarter of 2021 onward into the calibration of the linear trend in the seasonality function.
In contrast to the studies carried out in \cite{GMP17}, our price data exhibits negative jumps in every time period we investigate. This is the reason why our model features a negative stochastic jump component. The generalization to a model with multiple jump components as described in \cite{GMP17} is straightforward, but was not considered in our study for the sake of computational tractability, see also the runtimes in Table~\ref{details}.

Fast convergence of the MCMC algorithm is heavily dependent on finding appropriate paths of the latent variables, and hence it is important to update the latent variables in an efficient way. In Table \ref{details}, we provide details on the number of update steps, we used within every iteration of the algorithm. For the birth-death step and the multiplicative update step, we orientate ourselves on a number of 5 loops which is the same number as used in \cite{GMP17}. For the spot price data in the time interval 2021-2023, the calibration is more challenging as expected. Here, we reduce the number of loops to 1. This allows us to put more focus on the update of the latent OU-variable. For the time interval 2018-2021 it is possible to update more increments in a block (100 increments per loop), because the data fluctuates more regularly. If we include the spot price data of the time interval 2021-2023, then we find that the convergence rate improves by updating only a single increment per loop. The length of the burn-in period is inspired by \cite{GMP17}, who chose 5 Mio.~iterations for their 3-factor model. Consequently we opted for 10 Mio.~total iterations, in order to ensure convergence of the MCMC algorithm, as in our 4-factor model there are more hidden variables involved. The number of 10 Mio.~iterations is for consistency reasons then also kept for the 3-factor model.


\begin{table}[h]
	\begin{adjustbox}{width=\columnwidth,center}
		\centering
		\begin{tabular}{l|cc|cc|cc}
			\toprule
			\multicolumn{1}{c}{} & \multicolumn{2}{c}{\textbf{2018-21}} & \multicolumn{2}{c}{\textbf{2021-23}} & \multicolumn{2}{c}{\textbf{2018-23}}\\
			\cmidrule(rl){2-3} \cmidrule(rl){4-5} \cmidrule(rl){6-7}
			\multicolumn{1}{c}{} & \multicolumn{1}{c}{\textbf{3-OU}} & \multicolumn{1}{c}{\textbf{4-OU}}& \multicolumn{1}{c}{\textbf{3-OU}} & \multicolumn{1}{c}{\textbf{4-OU}} & \multicolumn{1}{c}{\textbf{3-OU}} & \multicolumn{1}{c}{\textbf{4-OU}}\\
			\midrule
			Number of Iterations  & 10 Mio. & 10 Mio. & 10 Mio. & 10 Mio. & 10 Mio. & 10 Mio.\\
			Burn-In-Period  & 9 Mio. & 9 Mio. & 9 Mio. & 9 Mio. & 9 Mio. & 9 Mio.\\
			Number of Loops for Birth-Death step & 5 & 5 & 5 & 1 & 5 & 5\\
			Number of Loops for Multiplicative Update & 5 & 5 & 5 & 1 & 5 & 5\\
			Number of Loops for Latent OU-Variable & - & 30 & - & 200 & - & 30\\
			Number of New Increments per Loop & - & 100 & - & 1 & - & 1\\
			Number of Permutations Increments per Loop & - & 100 & - & 2 & - & 2\\
			Runtime (days)  & 3.9958 & 3.3188 & 4.0901 & 5.4426 & 9.6146 & 8.2958\\
			\bottomrule
		\end{tabular}
	\end{adjustbox}
 \caption{Information about the updates of the hidden variables.}
 \label{details}
\end{table}

\subsection{Model fit for different time periods}
In what follows, we present calibration results together with some illustrations and interpretations for each time interval we considered.
We always start by summarizing the posterior properties (mean and standard deviation) of the respective calibrated parameters.
For our analysis, we then also provide posterior sample paths of the latent variables as well as simulations based on the calibrated parameters. Model adequacy is then assessed by calculating the posterior predictive $p$-values as described in Section \ref{p_val}. In addition, we also provide further statistical analysis by comparing the autocorrelation function of the data to those obtained for our simulations. Each subsection ends with an interpretation of the obtained results and what conclusions can be drawn from them.

\subsubsection{2018-21 spot price data:} We calibrate the 3-factor model and the 4-factor model to the spot-price data in the time interval 2018-2021. We start with an overview of the posterior properties of the model parameters obtained from the MCMC procedure  described in Section~\ref{MCMC}. Later in this section, we present a more detailed analysis of our calibration results.

\begin{table}[H]
\begin{adjustbox}{width=0.8\textwidth}
\centering
\begin{tabular}{c|cccc|cc}
\toprule
\multicolumn{1}{c}{} & \multicolumn{4}{c}{\textbf{standard model}}  & \multicolumn{2}{c}{\textbf{Brownian motion}}\\
 \cmidrule(rl){2-5} \cmidrule(rl){
 6-7} 
\multicolumn{1}{c}{}  & \multicolumn{2}{c}{\textbf{3-OU}} & \multicolumn{2}{c}{\textbf{4-OU}} & \multicolumn{2}{c}{\textbf{BB+3OU}}\\

  \cmidrule(rl){2-5}  \cmidrule(rl){6-7} 
\textbf{Parameter}  & {Mean} & {SD} & {Mean} & {SD} & {Mean} & {SD}\\
\midrule
$\sigma_{\v}$  & 42.741 & 9.518 & 116.943 & 11.112 & 118.509 & 10.909 \\
$\sigma_{\w}$  & - & - & 22.051 & 6.700  & 16.836 & 3.763 \\
$\lambda_{\v}$  & 0.202 & 3.038 & 0.005 & 0.001 & 0.005 & 0.001 \\
$\lambda_{\w}$  & - & - & 1.882 & 29.056  & - & - \\
$\lambda_{J_1}$ & 0.006 & 0.001 & 0.008 & 0.003 & 0.009 & 0.004\\
$\lambda_{J_2}$ & 0.002 & 0 & 0.002 & 0 & 0.002 & 0\\
$\theta_1$ & 333.864 & 82.143 & 129.844 & 80.131 & 126.394 & 55.924\\
$\theta_2$ & 336.260 & 59.495 & 86.999 & 30.244 & 88.135 & 35.188\\
$\beta_1$  & 4.019 & 0.524 & 4.816 & 0.131 & 4.605 & 1.263\\
$\beta_2$  & 7.359 & 0.877 & 13.547 & 3.156 & 13.837 & 3.531\\
\bottomrule
\end{tabular}
\end{adjustbox}

\caption{Posterior properties of the model parameters in the 2018-21 time period. We present the mean and the standard deviation (SD) for all model parameters.}
\label{parameters_18_21}
\end{table}

\emph{3- and 4-factor model:} In the 3-factor model, the Gaussian component models the long term behavior of the spot price process (cf. Figure \ref{Fig18-21}(A)), while short term fluctuations are incorporated into the jump process (cf. Figure \ref{Fig18-21}(B)). This observation corresponds to the high jump intensity parameters in Table \ref{parameters_18_21}. 
In the 4-factor model, one Gaussian component still models the long term behavior of the spot price process, while the second is used to model short term fluctuations (cf. Figure \ref{Fig18-21}(C)). This leads to sparser paths in terms of jump arrival times of the jump process (cf. Figure \ref{Fig18-21}(D)), where jumps of larger size dominate. This observation corresponds to the lower jump intensity parameters together with higher jump size parameters in Table \ref{parameters_18_21}. 

 \begin{figure}[H]
\centering
   \begin{subfigure}{0.45\linewidth}
       \includegraphics[width=\linewidth]{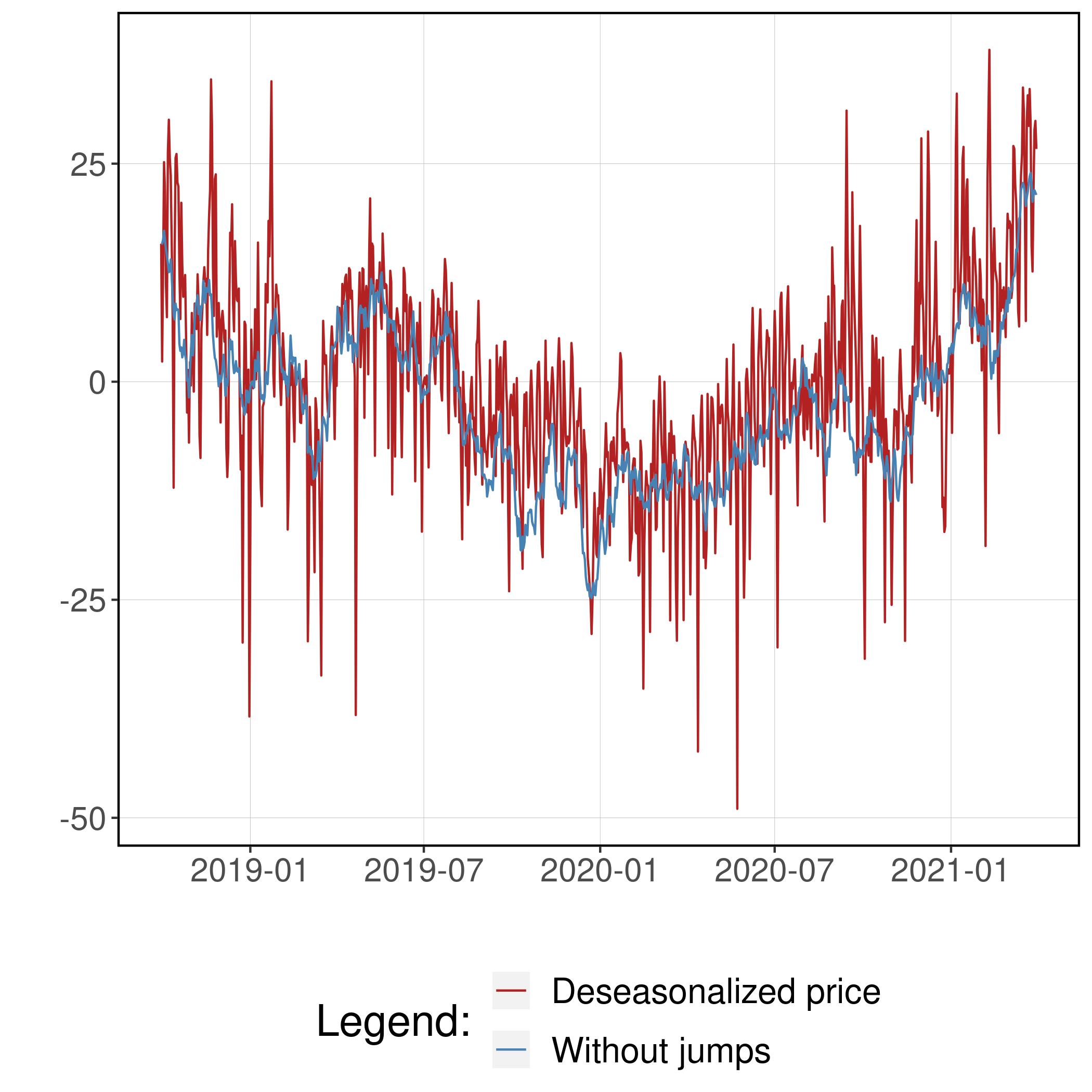}
    \caption{Data and Gaussian residuals}
    \end{subfigure}
\hfil
   \begin{subfigure}{0.45\linewidth}
        \includegraphics[width=\linewidth]{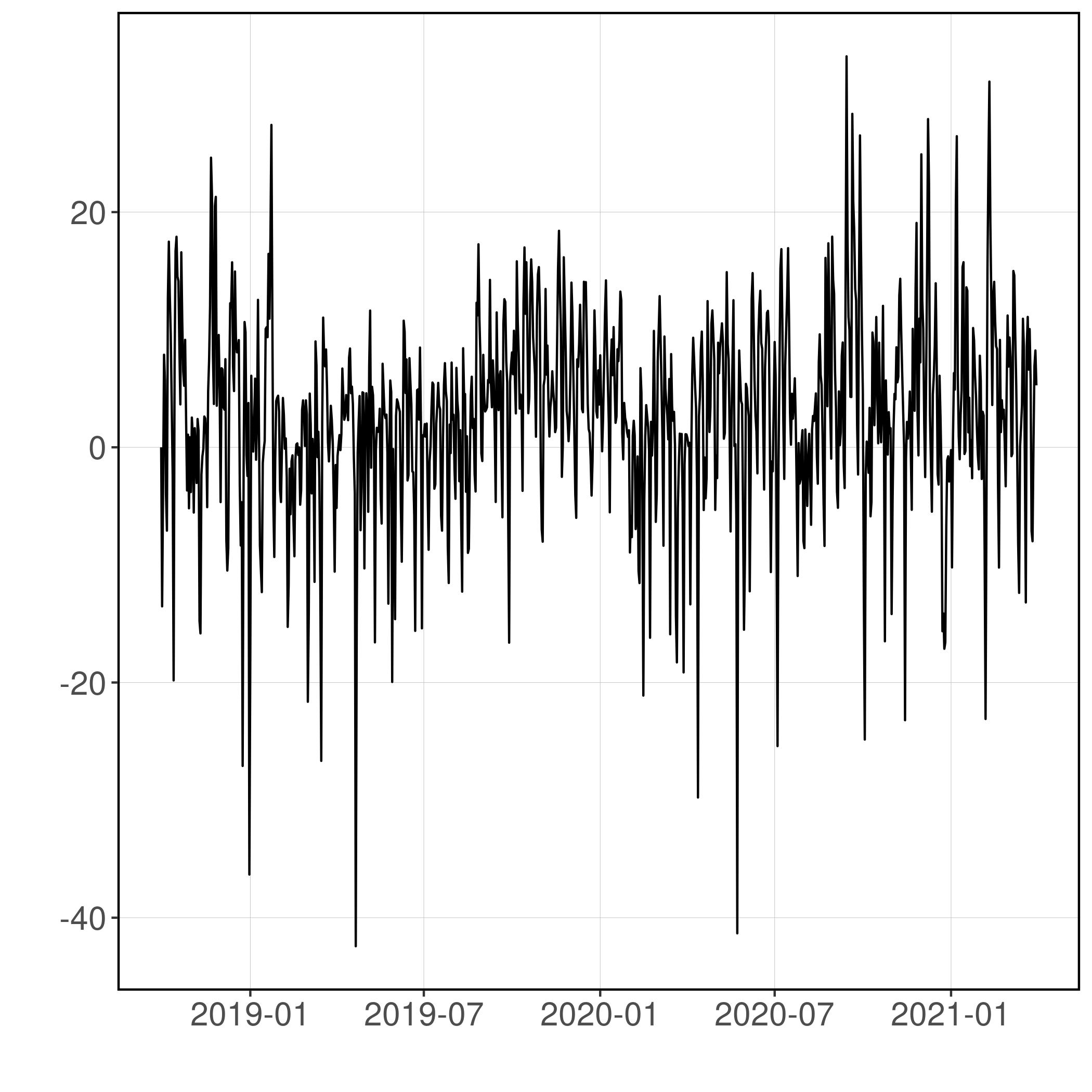}
    \caption{Sample path of the jump process}
    \end{subfigure}
\end{figure}    
\begin{figure}[H]\ContinuedFloat
   \begin{subfigure}{0.45\linewidth}
        \includegraphics[width=\linewidth]{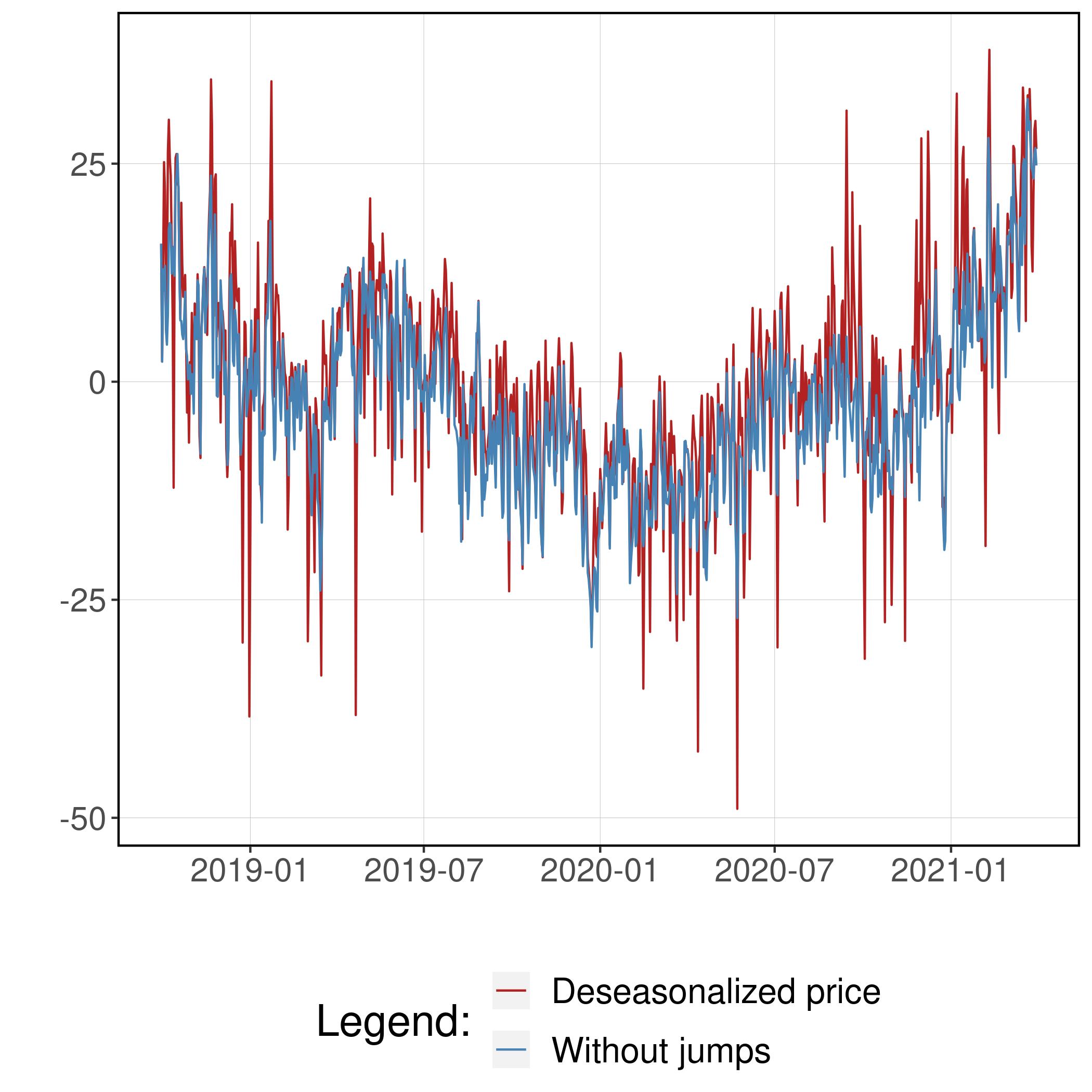}
    \caption{Data and Gaussian residuals}
    \end{subfigure}
\hfil
    \begin{subfigure}{0.45\linewidth}
        \includegraphics[width=\linewidth]{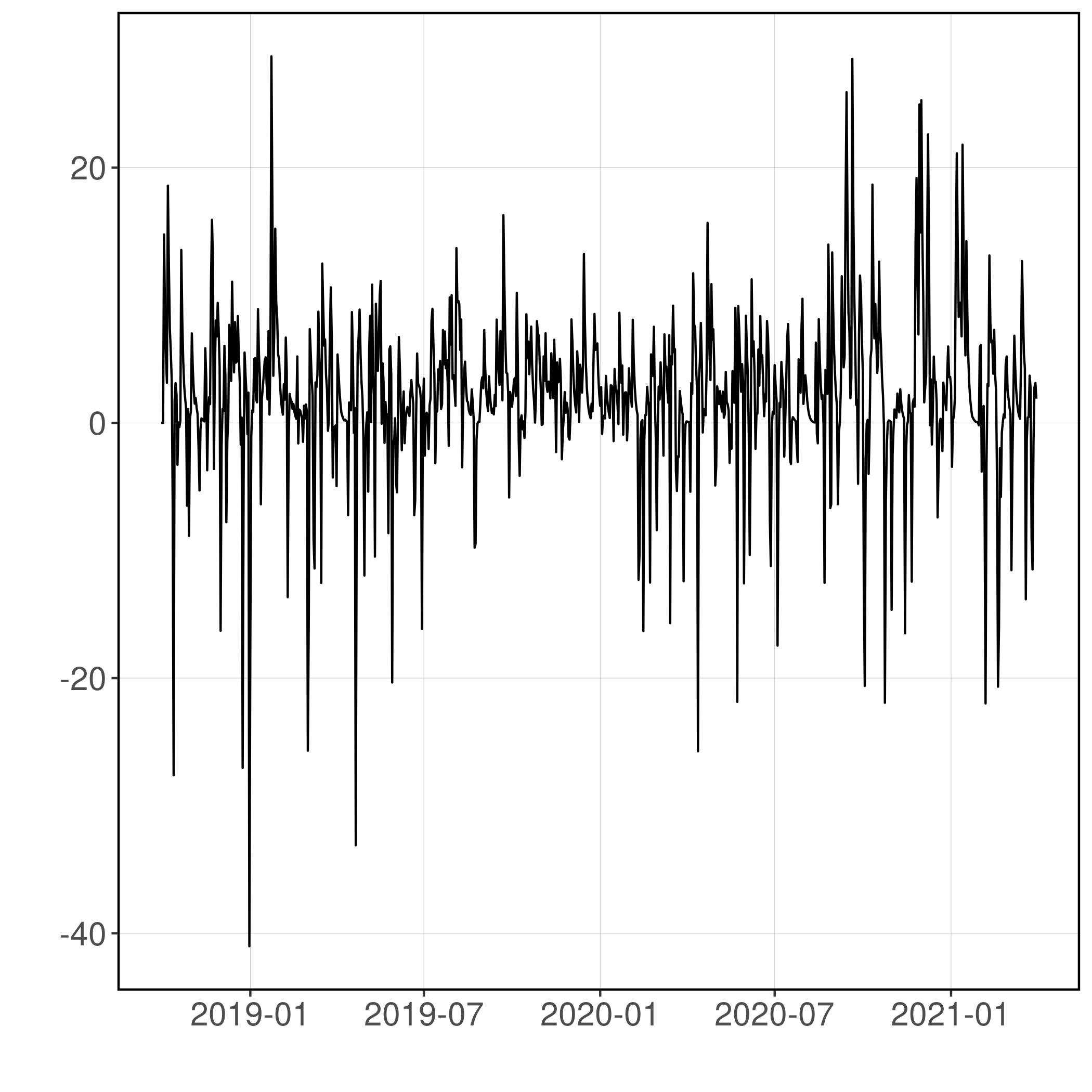}
   \caption{Sample path of the jump process}
    \end{subfigure}

\caption{Paths of the hidden variables in the 3-factor model (top) and 4-factor model (bottom) in the time period 2018-21.}
\label{Fig18-21}
    \end{figure}

The high $p$-value for the additional OU-process (cf. Figure \ref{p-values 18-21}(B)) indicates that short term fluctuations of the spot price process are indeed Gaussian. In direct comparison with Figure \ref{p-values 18-21}(A), it can also be observed, that the 4-OU model leads to increased $p$-values for the jump intensity parameters. The decay of the ACF of the data is more accurately reproduced by the simulations in the 4-OU model as can be seen in the comparison of Figure \ref{p-values 18-21}(C) and Figure \ref{p-values 18-21}(D).

  \begin{figure}[H]
\centering
    \begin{subfigure}{0.39\linewidth}
        \includegraphics[width=\linewidth]{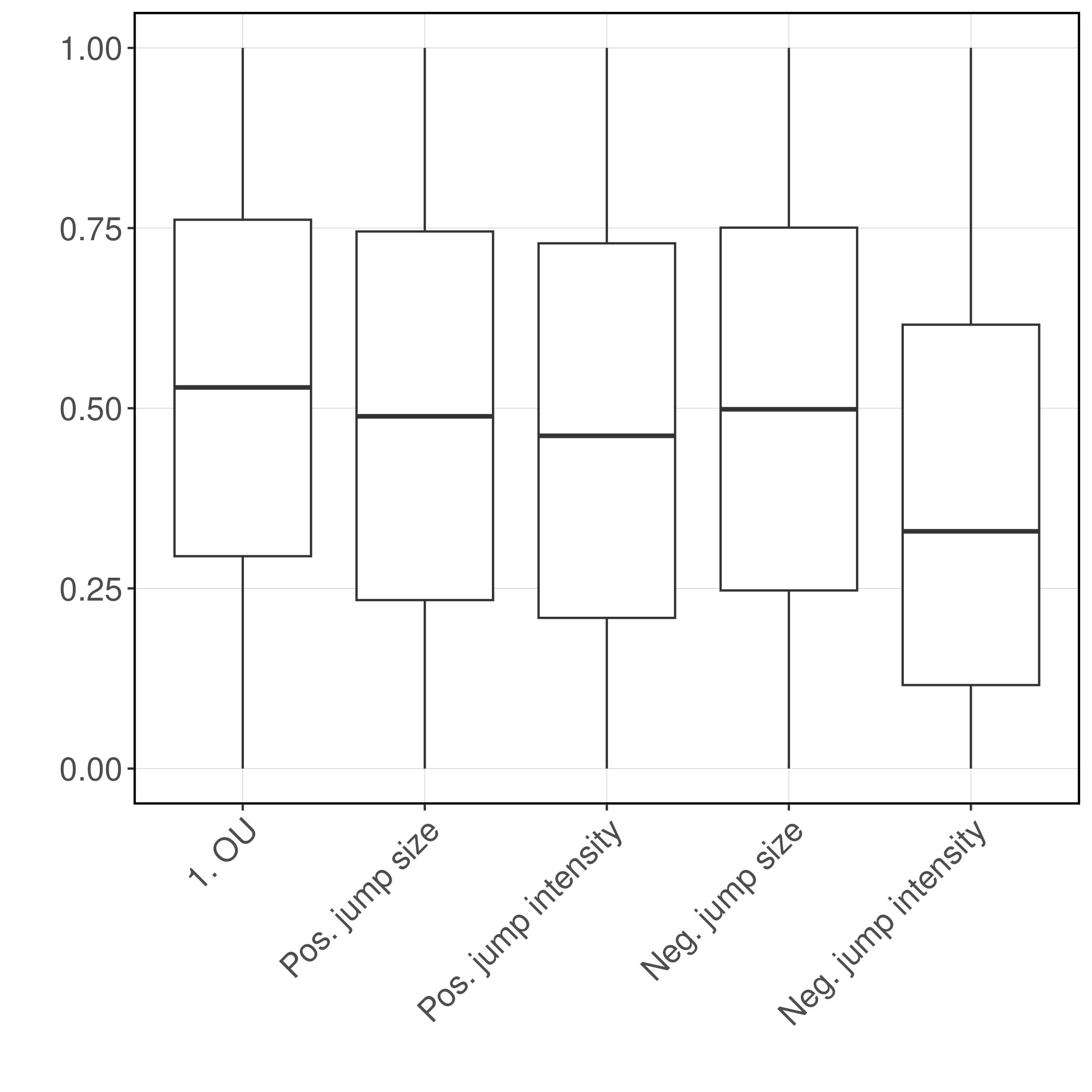}
    \caption{Boxplot for the $p$-values}
    \end{subfigure}    
\hfil
     \begin{subfigure}{0.39\linewidth}
        \includegraphics[width=\linewidth]{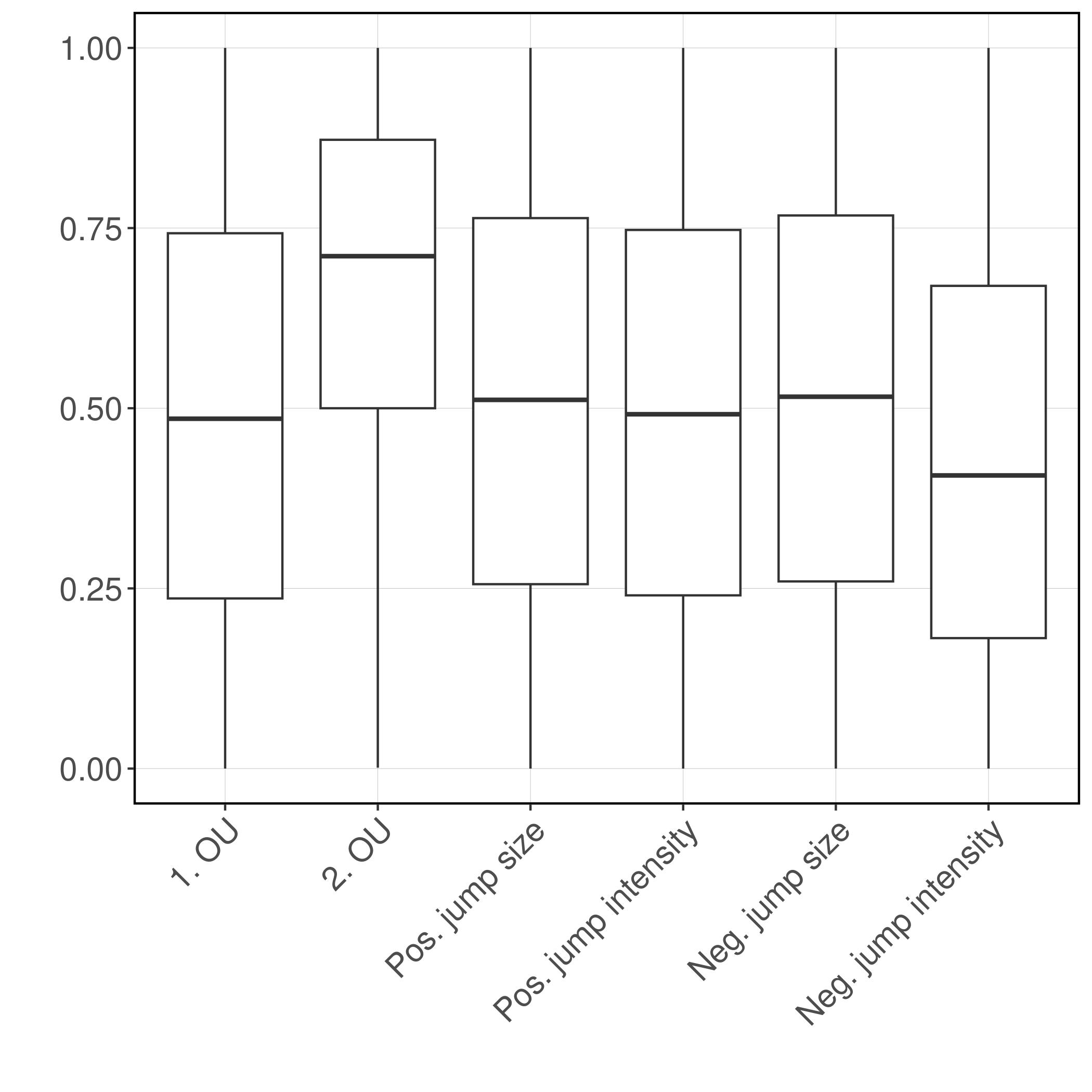}
    \caption{Boxplot for the $p$-values}
   \end{subfigure}    
 \end{figure}       
  \begin{figure}[H]\ContinuedFloat
\centering
    \begin{subfigure}{0.39\linewidth}
        \includegraphics[width=\linewidth]{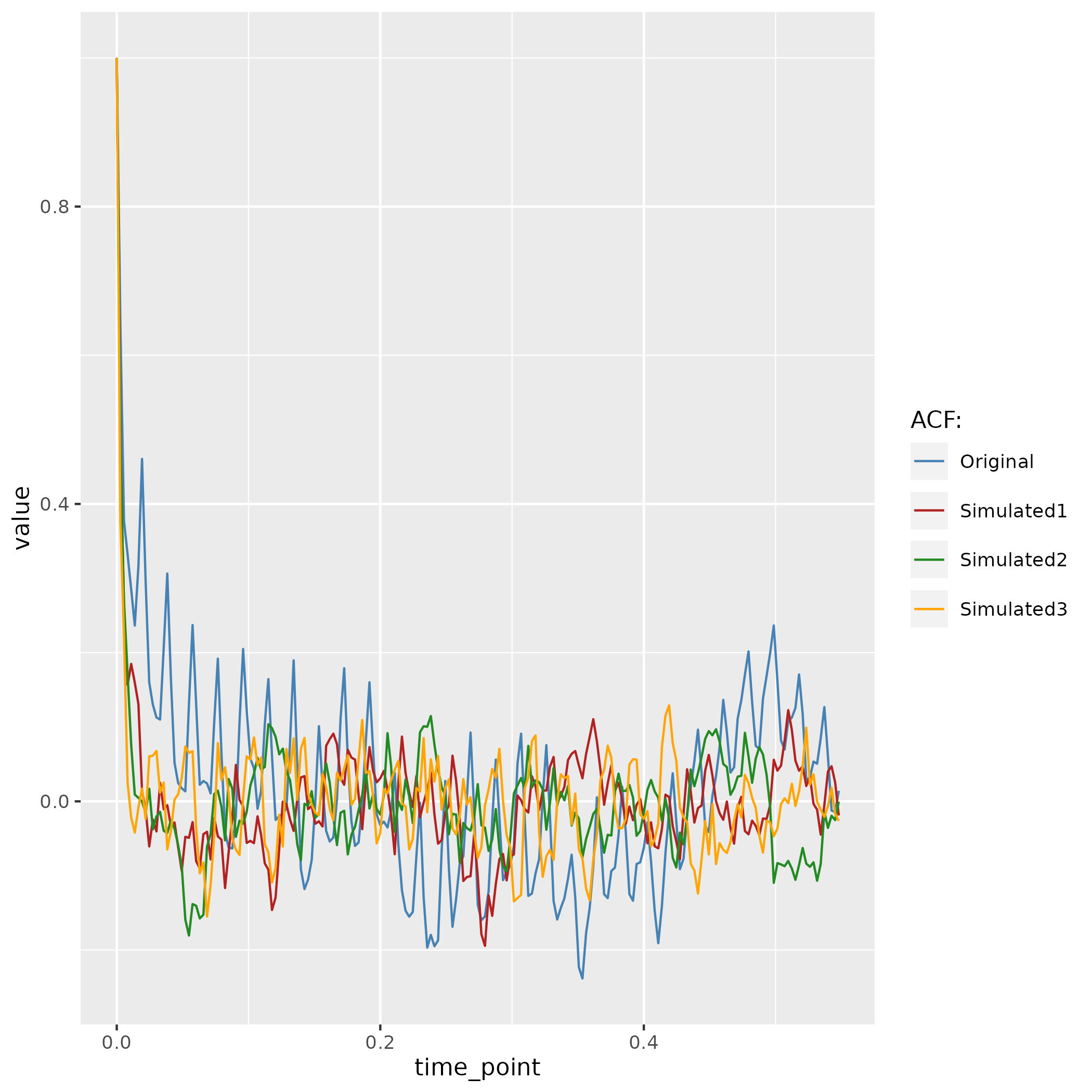}
    \caption{ACF for data and simulations }
    \end{subfigure}    
\hfil
     \begin{subfigure}{0.39\linewidth}
        \includegraphics[width=\linewidth]{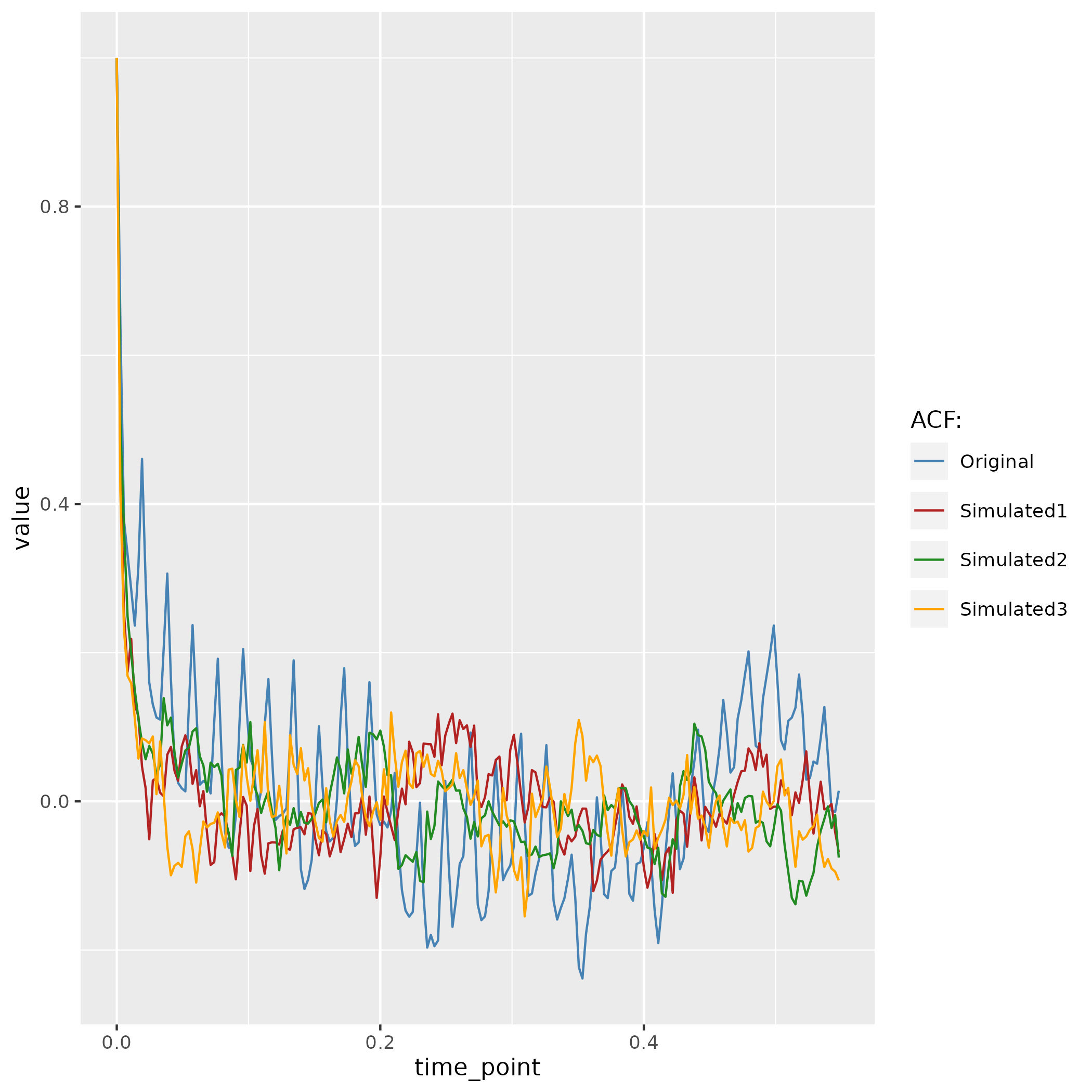}
    \caption{ACF for data and simulations }
   \end{subfigure}
   \end{figure}
     \begin{figure}[H]\ContinuedFloat
\centering
    \begin{subfigure}{0.39\linewidth}
        \includegraphics[width=\linewidth]{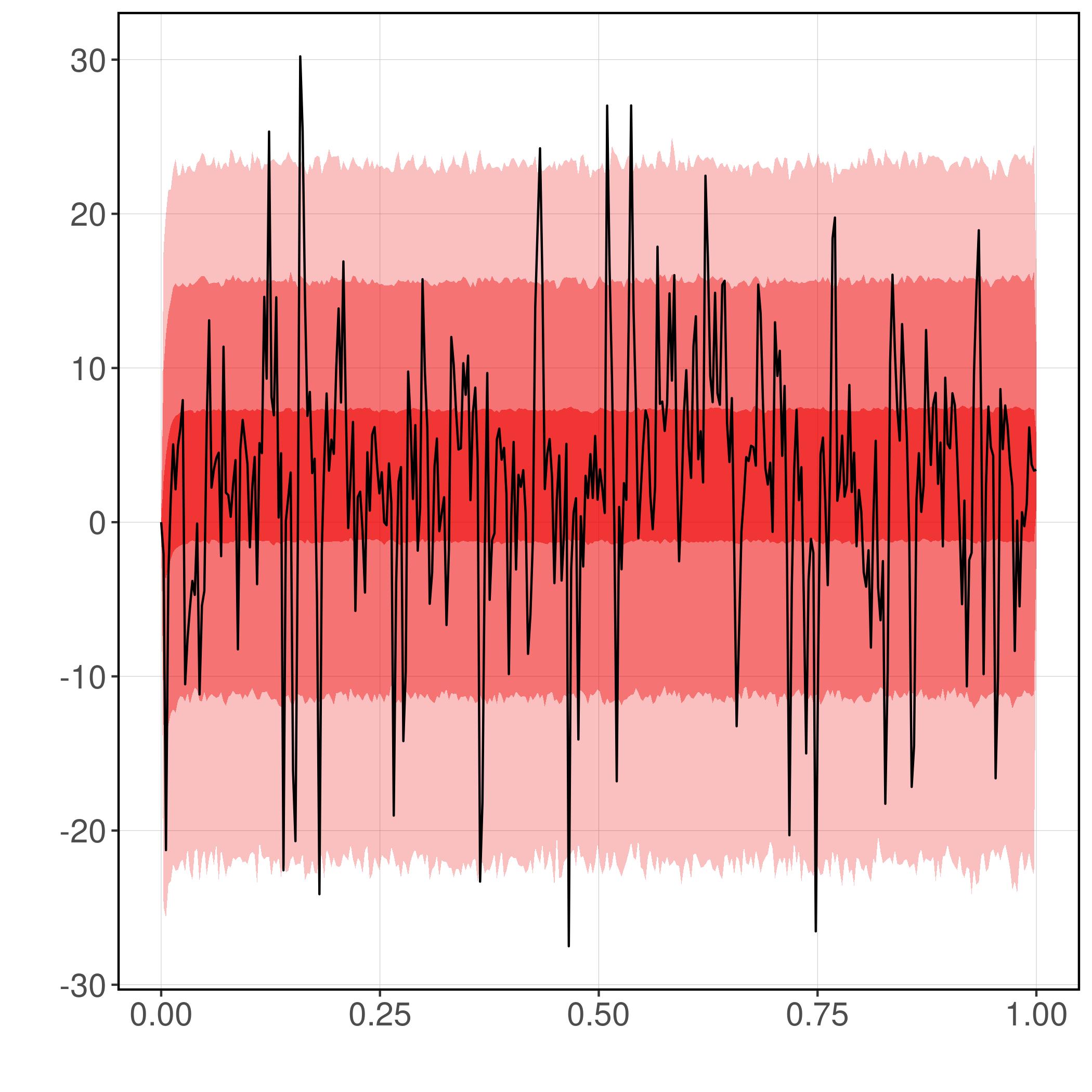}
    \caption{Simulated path and quantiles (dark to light red: 25-75\%, 5-95\%, 1-99\%)}
    \end{subfigure}    
\hfil
     \begin{subfigure}{0.39\linewidth}
        \includegraphics[width=\linewidth]{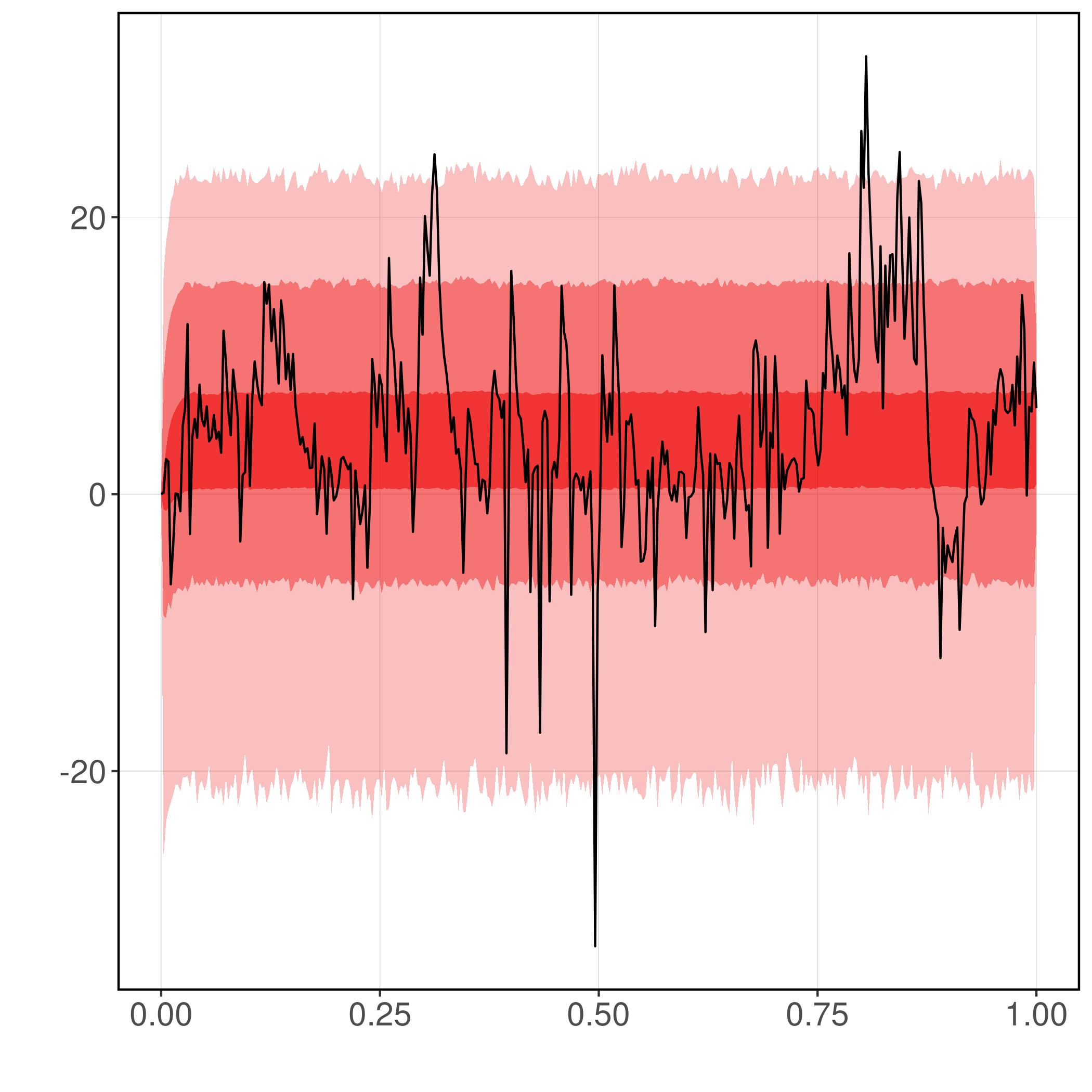}
    \caption{Simulated path and quantiles (dark to light red: 25-75\%, 5-95\%, 1-99\%)}
   \end{subfigure}       

\caption{Statistical evaluation of the $3$-factor (left) and $4$-factor (right) model in the time period 2018-21.}
\label{p-values 18-21}
    \end{figure}

\emph{Long-term Brownian motion:}
We also consider a version of the 4-factor model, in which the mean reversion parameter $\lambda_{\w}$ of the second Gaussian component \eqref{GOU2} is fixed as zero. Thus the second Gaussian component is modeled as a driftless (the drift
component is already captured in the deterministic part) arithmetic Brownian motion
\[
d\w (t)=\sigma_{\w} dW_{\w}(t)
\]
in the spirit of \cite{LS02}. This approach leads to larger long term deviations of the spot price process from the mean level, as it can be observed in the simulations (cf. Figure \ref{Fig4OU_18-21BB}(C)). While the $p$-values for this model are still good (cf. Figure \ref{Fig4OU_18-21BB}(A)), the decay in the ACF of the historical data is not accurately reproduced by the simulations (cf. Figure \ref{Fig4OU_18-21BB}(B)).

\begin{figure}[H]
    \begin{subfigure}{0.45\linewidth}
        \includegraphics[width=\linewidth]{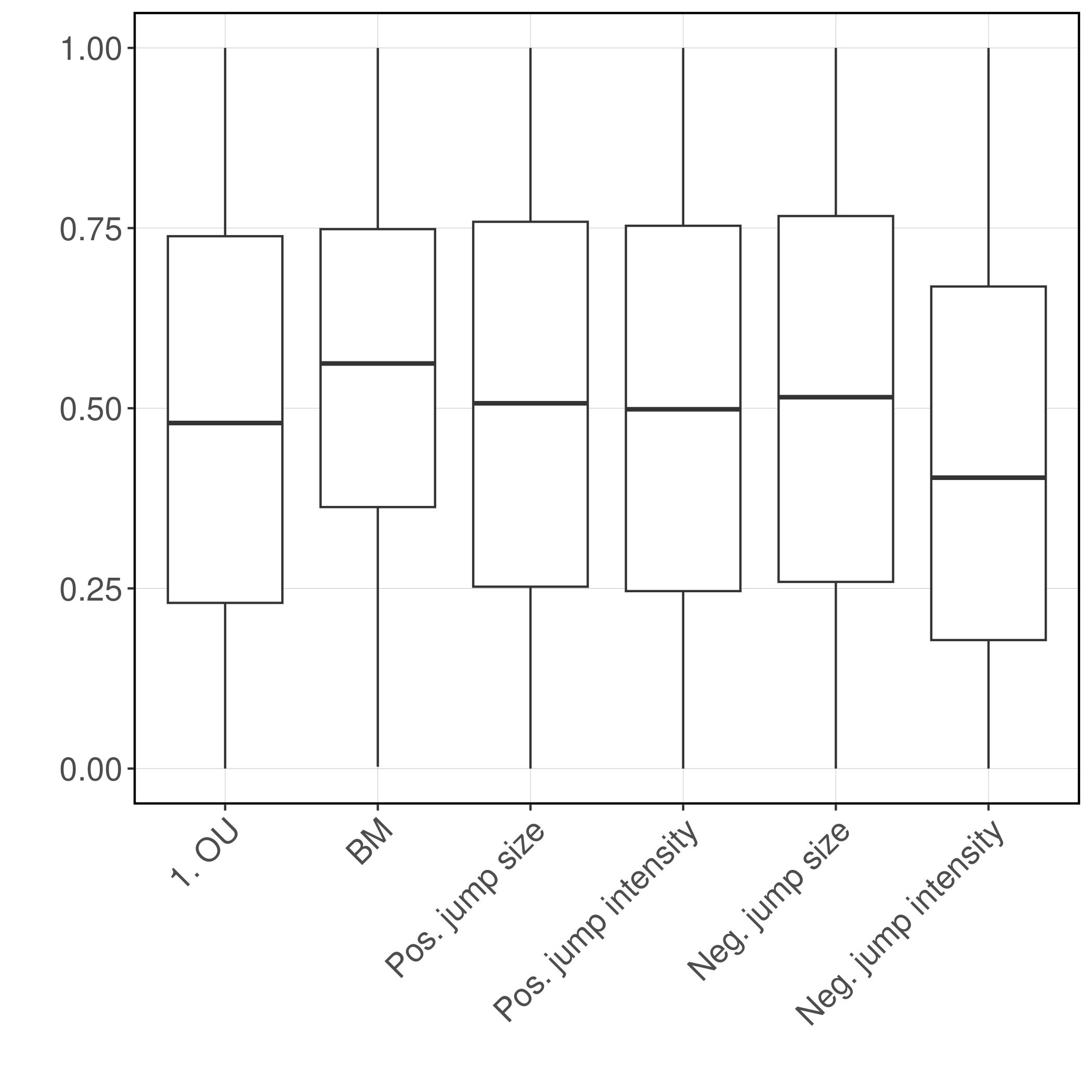}
    \caption{Boxplot of the p-values}
    \end{subfigure}
\hfil
\begin{subfigure}{0.45\linewidth}
        \includegraphics[width=\linewidth]{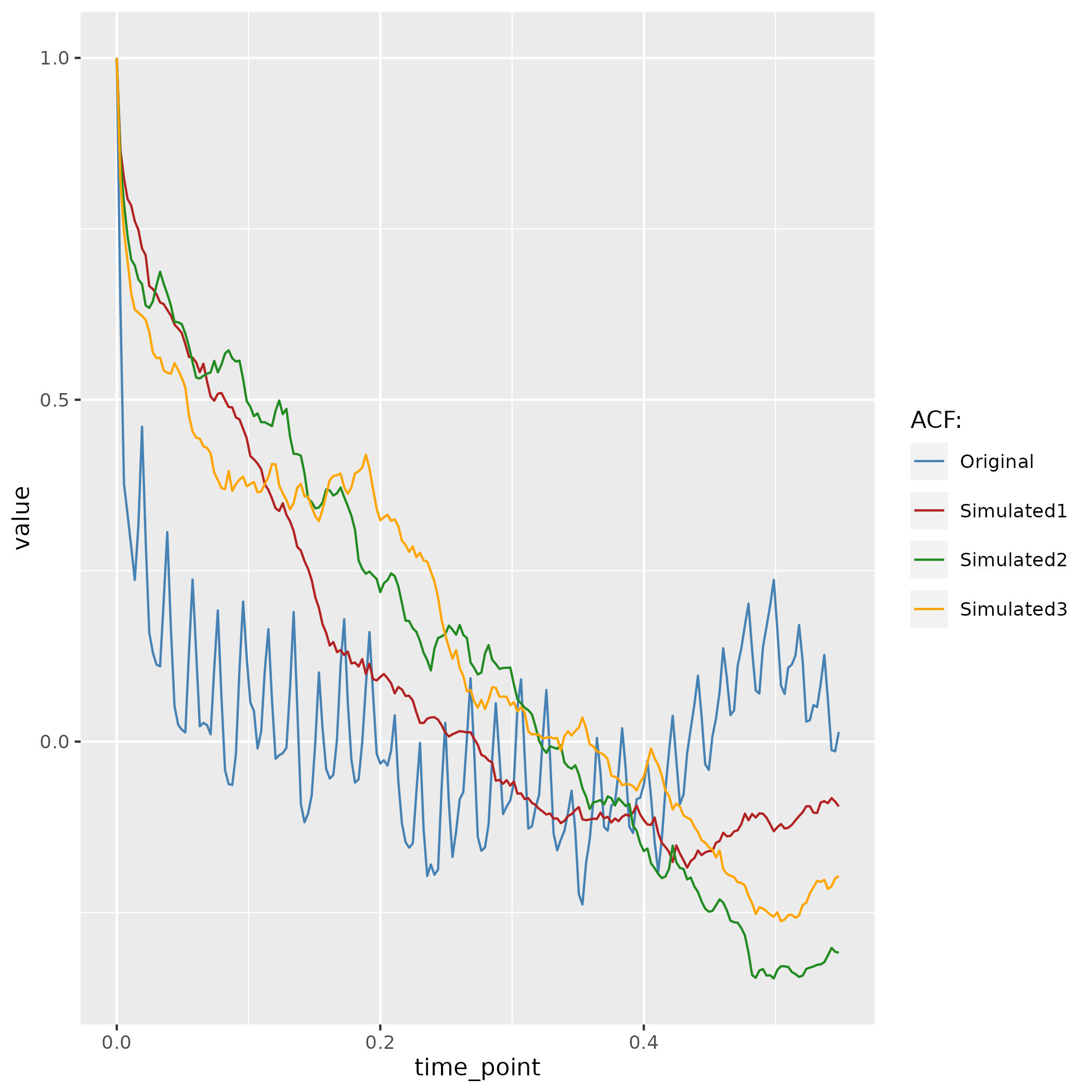}
    \caption{ACF for data and simulations}
    \end{subfigure}
    
\end{figure}
\begin{figure}[H]\ContinuedFloat
  \begin{subfigure}{0.45\linewidth}
        \includegraphics[width=\linewidth]{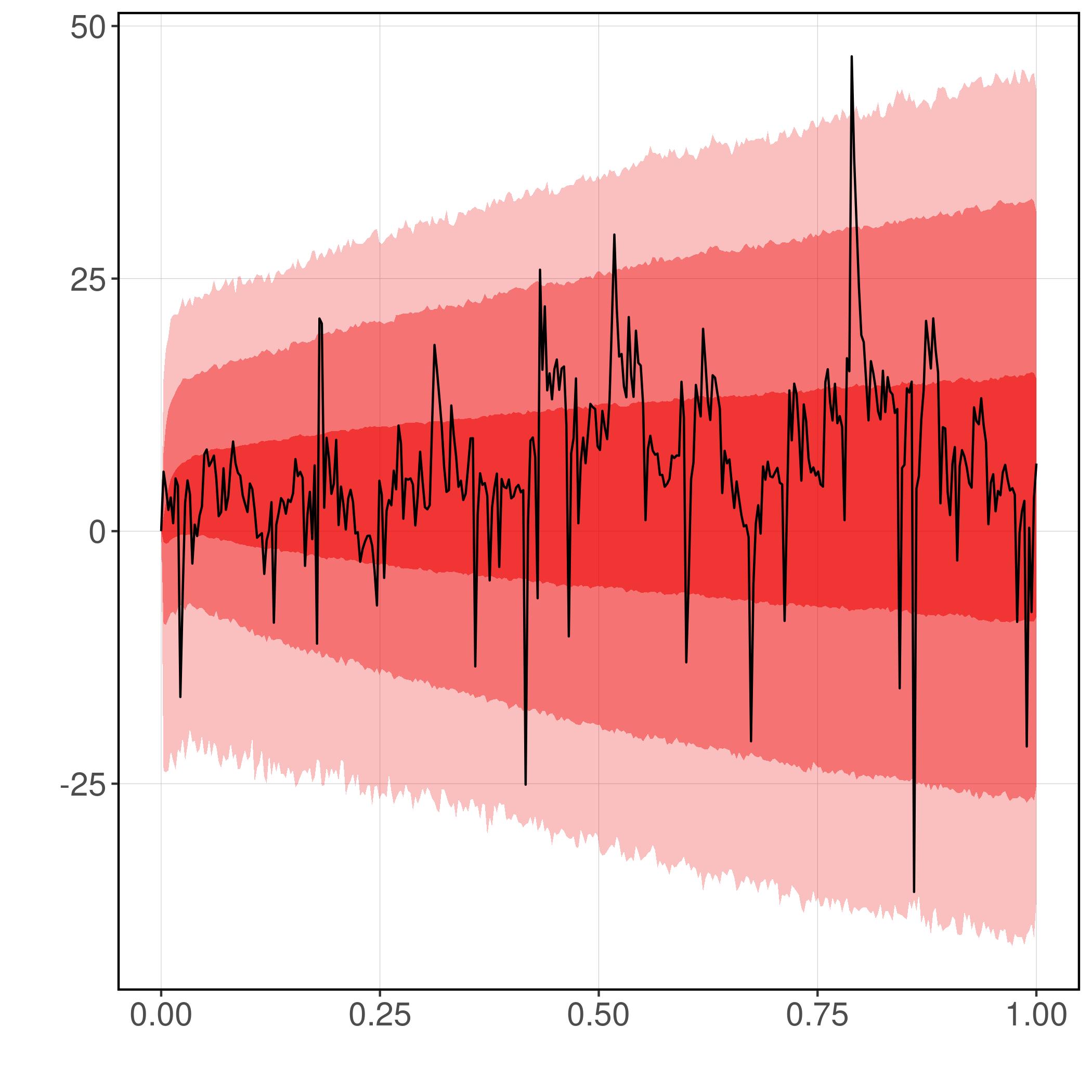}
    \caption{Simulated path and quantiles }
    \end{subfigure}  
\hfil
    \begin{subfigure}{0.45\linewidth}
        \includegraphics[width=\linewidth]{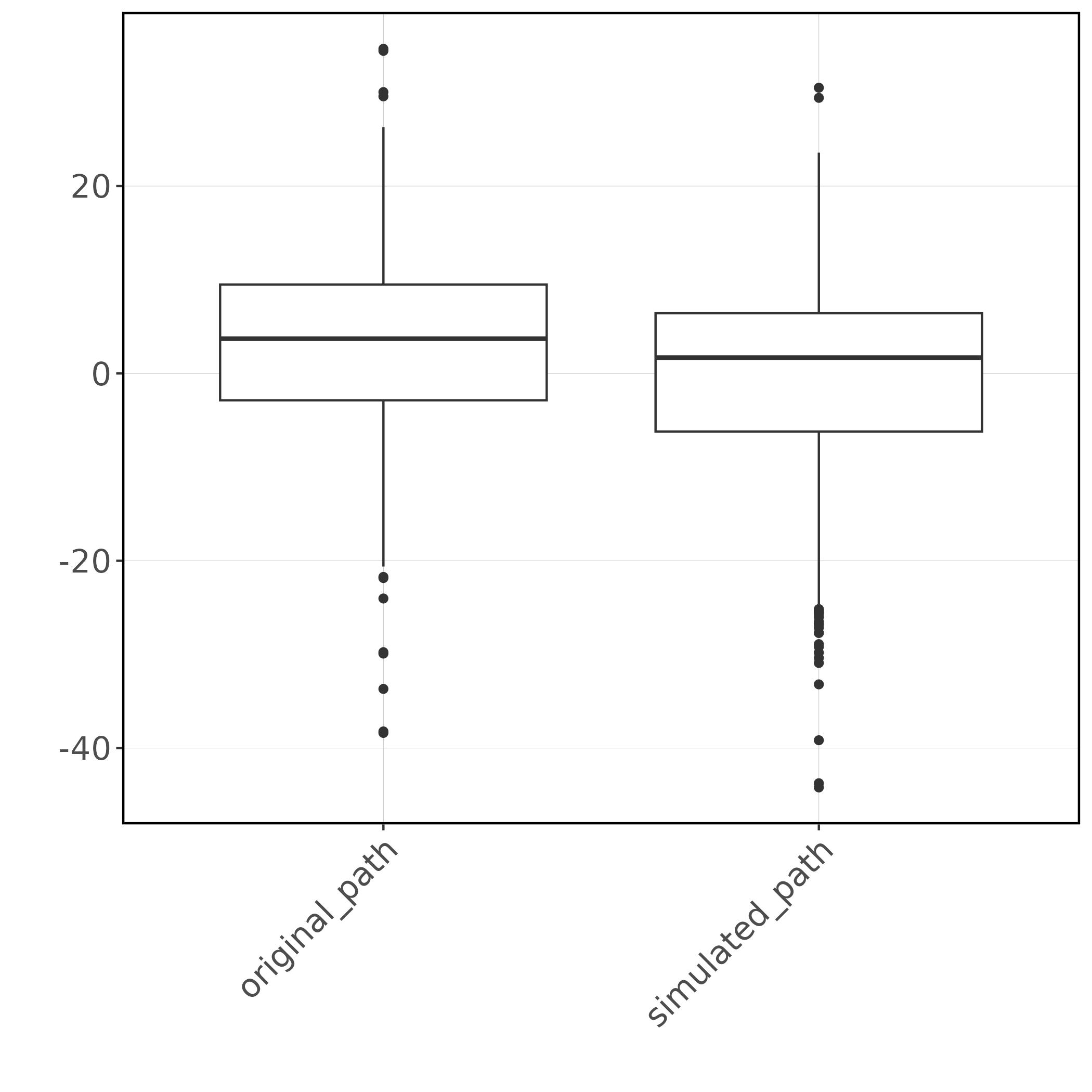}
    \caption{Boxplots for data and simulations}
    \end{subfigure}

\caption{Statistical evaluation of the 4-factor model with long-term Brownian motion in the period 2018-21.}
\label{Fig4OU_18-21BB}
    \end{figure}

\emph{Conclusion for the period 2018-21:} In the pre-crisis period the statistical indicators as $p$-values and autocorrelation analysis indicate a good overall fit for both the 3- and the 4-factor model with a tendency to the 4-factor model. The main difference between the two approaches is, that in the 3-factor model, the Gaussian component is only responsible for modeling the long term behavior of the price process while smaller fluctuations are modeled by the jump processes with high intensity rates. In the 4-factor model, these short term fluctuations are now modeled by the second Gaussian component, whereas the intensity of the jump rates is considerably lower, leading to sparser paths of the jump processes. The very high $p$-value for the additional Gaussian component in the 4-factor model, as well as the increased $p$-values for the jump intensity rates indicate that it makes sense to separately model the short term fluctuations of the spot price with a Gaussian process.\vspace{5mm}

\subsubsection{2021-23 spot price data} We calibrate the 3-factor model and the 4-factor model to the spot-price data in the time interval 2021-2023. We start with an overview of the posterior properties of the model parameters obtained from the MCMC procedure  described in Section~\ref{MCMC}. Later in this section, we present a more detailed analysis of our calibration results.

\begin{table}[H]
\begin{adjustbox}{width=0.8\textwidth}
\centering
\begin{tabular}{c|cccc|cc}
\toprule
\multicolumn{1}{c}{} & \multicolumn{4}{c}{\textbf{standard model}}  & \multicolumn{2}{c}{\textbf{Brownian motion}}\\
 \cmidrule(rl){2-5} \cmidrule(rl){
 6-7} 
\multicolumn{1}{c}{}  & \multicolumn{2}{c}{\textbf{3-OU}} & \multicolumn{2}{c}{\textbf{4-OU}} & \multicolumn{2}{c}{\textbf{BB+3OU}}\\

  \cmidrule(rl){2-5}  \cmidrule(rl){6-7} 
\textbf{Parameter}  & {Mean} & {SD} & {Mean} & {SD} & {Mean} & {SD}\\
\midrule
$\sigma_{\v}$  & 357.870 & 47.435 & 1.431 & 0.303 & 13.279 & 1.135\\
$\sigma_{\w}$  & - & - & 305.894 & 0.216 & 280.524 & 0.610\\
$\lambda_{\v}$  & 0.010 & 0.054 & 0.037 & 0.017 & 0.005 & 0.001\\
$\lambda_{\w}$  & - & - & 0.970 & 0.004 & - & -\\
$\lambda_{J_1}$ & 0.035 & 0.006 & 0.020 & 0 & 0.020 & 0\\
$\lambda_{J_2}$ & 0.074 & 0.014 & 0.004 & 0 & 0.006 & 0\\
$\theta_1$ & 178.611 & 41.323 & 207.293 & 10.790 & 253.268 & 9.984\\
$\theta_2$ & 132.444 & 31.951 & 131.031 & 8.565 & 100.426 & 6.906\\
$\beta_1$  & 37.123 & 4.855 & 29.093 & 1.519 & 29.119 & 1.418\\
$\beta_2$  & 24.943 & 3.756 & 38.703 & 2.634 & 37.768 & 2.981\\
\bottomrule
\end{tabular}
\end{adjustbox}

\caption{Posterior properties of the model parameters in the 2021-23 time period. We present the mean and the standard deviation (SD) for all model parameters.}
\label{parameters_21_23}
\end{table}

\emph{3- and 4-factor model:} In the crisis period 2021-23, the Gaussian component of the 3-factor model is only responsible for the short term fluctuations (cf. Figure \ref{Fig3OU_21-23}(A)), while the large long term deviations are modeled as a superposition of positive and negative jumps (cf. Figure \ref{Fig3OU_21-23}(B)). These observations for the sample paths of the underlying processes are in line with the estimated model parameters we obtain in Table \ref{parameters_21_23}, where we have high values for both positive and negative jump sizes.  
This leads to the conclusion that the rapid up- and downward movements in the desaisonalized data in this time period are not Gaussian any more but are rather modeled by a succession of exponentially distributed jumps of large size. The adequacy of these calibration results are confirmed by the relatively high $p$-values for both jump sizes and intensity (cf. Figure \ref{Fig3OU_21-23}(C)). The simulation in Figure \ref{Fig3OU_21-23}(E) reproduces quite well the characteristic behavior of the historic data (cf. \ref{Fig3OU_21-23}(A)) and the confidence intervals correspond to the range of the historic data. The model adequacy is finally also confirmed by a comparison of the autocorrelation fuction (ACF) of the data with the ACF of our simulations (cf. \ref{Fig3OU_21-23}(D)).

  \begin{figure}[H]
\centering
      \begin{subfigure}{0.39\linewidth}
        \includegraphics[width=\linewidth]{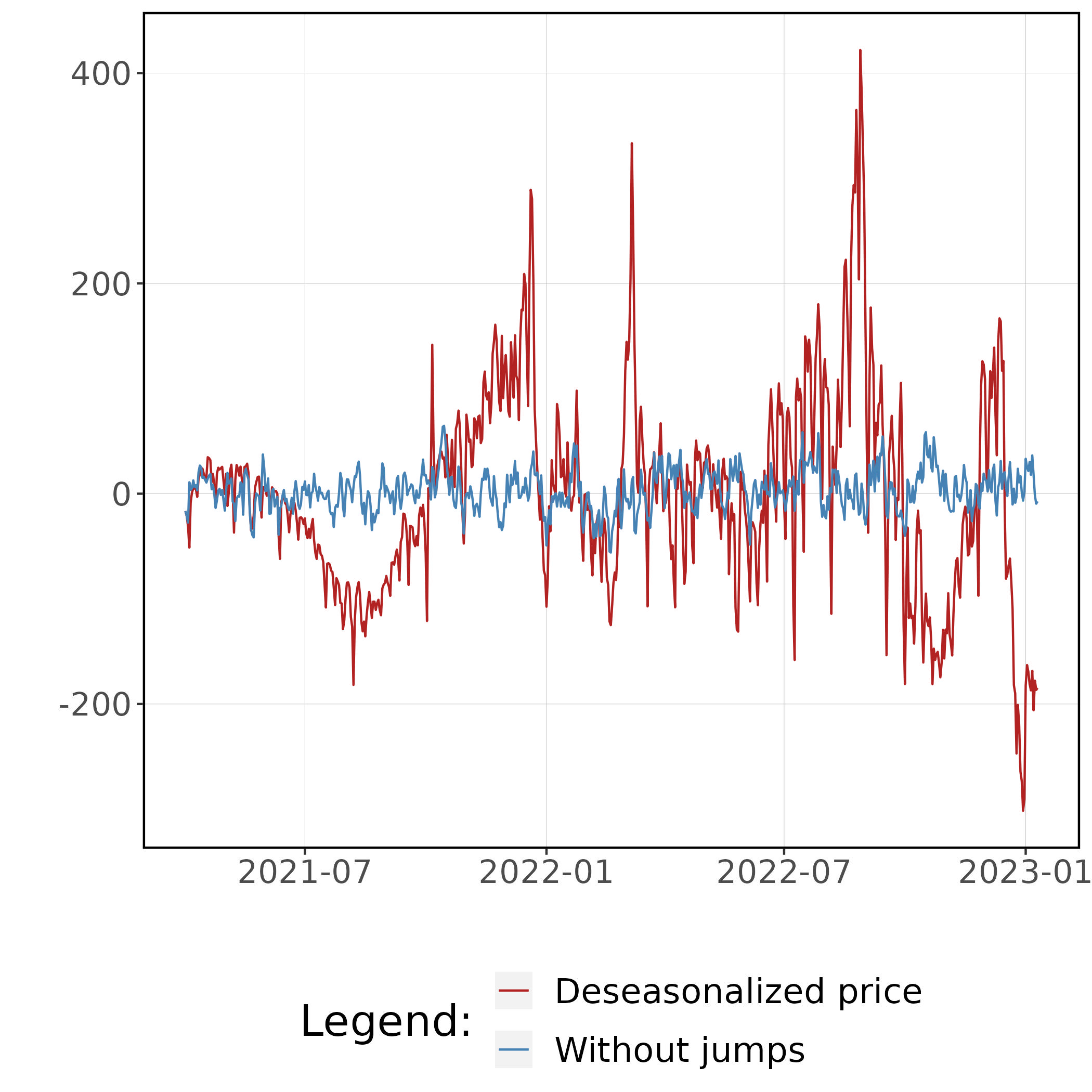}
    \caption{Data and Gaussian residuals after jump removal}
    \end{subfigure}
\hfil
    \begin{subfigure}{0.39\linewidth}
        \includegraphics[width=\linewidth]{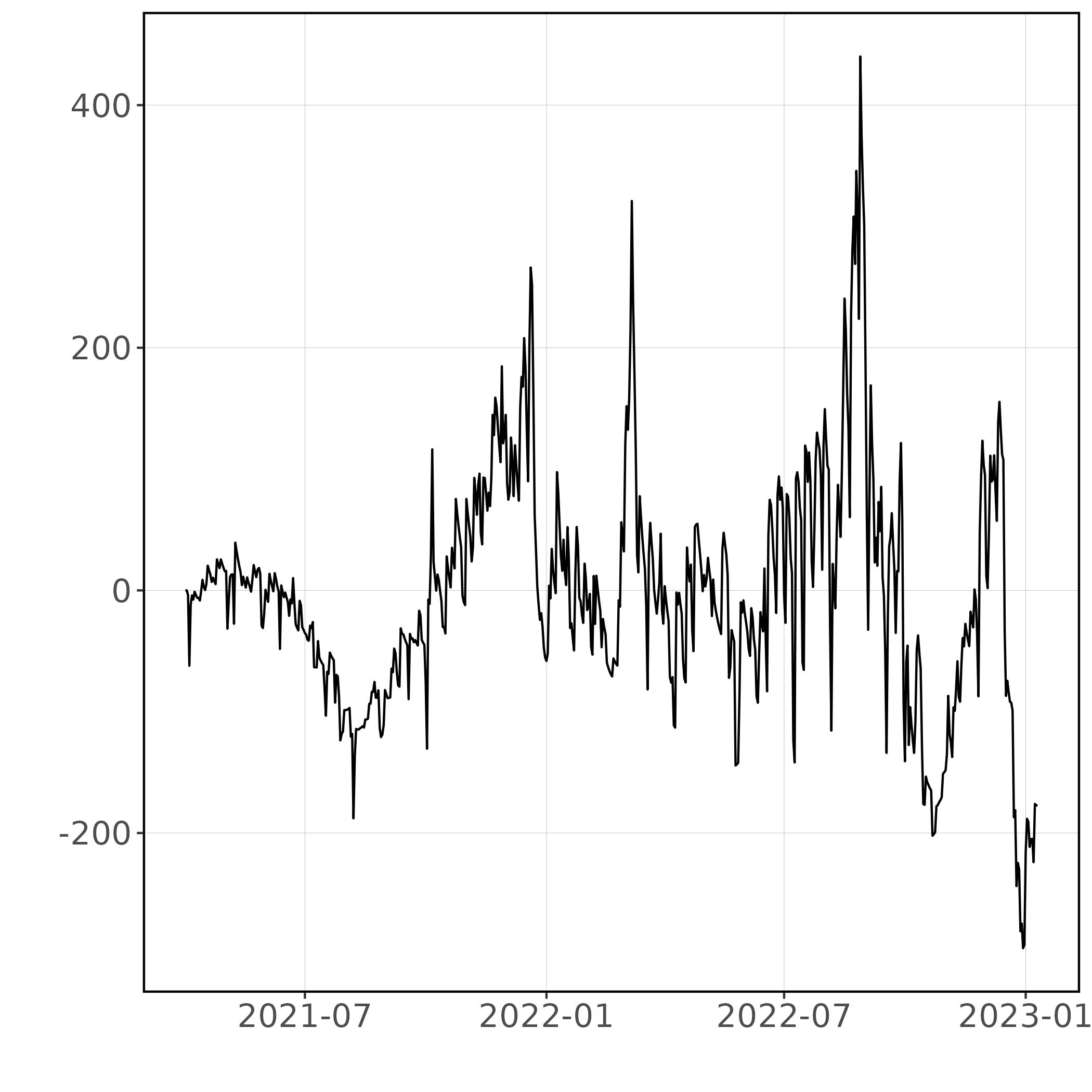}
    \caption{Sample path of the jump process}
    \end{subfigure}
\end{figure}    
\begin{figure}[H]\ContinuedFloat
  \begin{subfigure}{0.39\linewidth}
        \includegraphics[width=\linewidth]{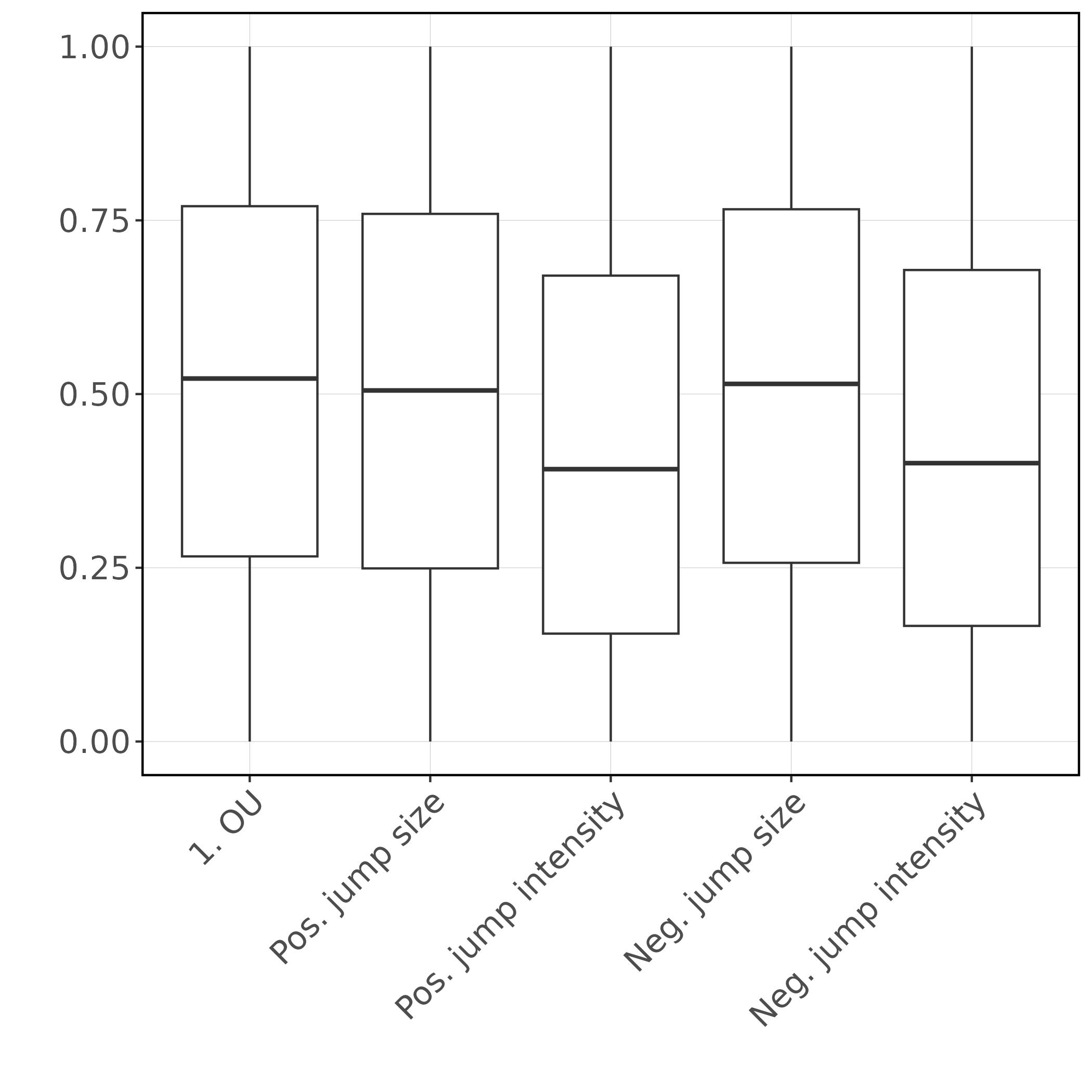}
    \caption{Boxplot of the p-values}
    \end{subfigure}
\hfil
 \begin{subfigure}{0.39\linewidth}
        \includegraphics[width=\linewidth]{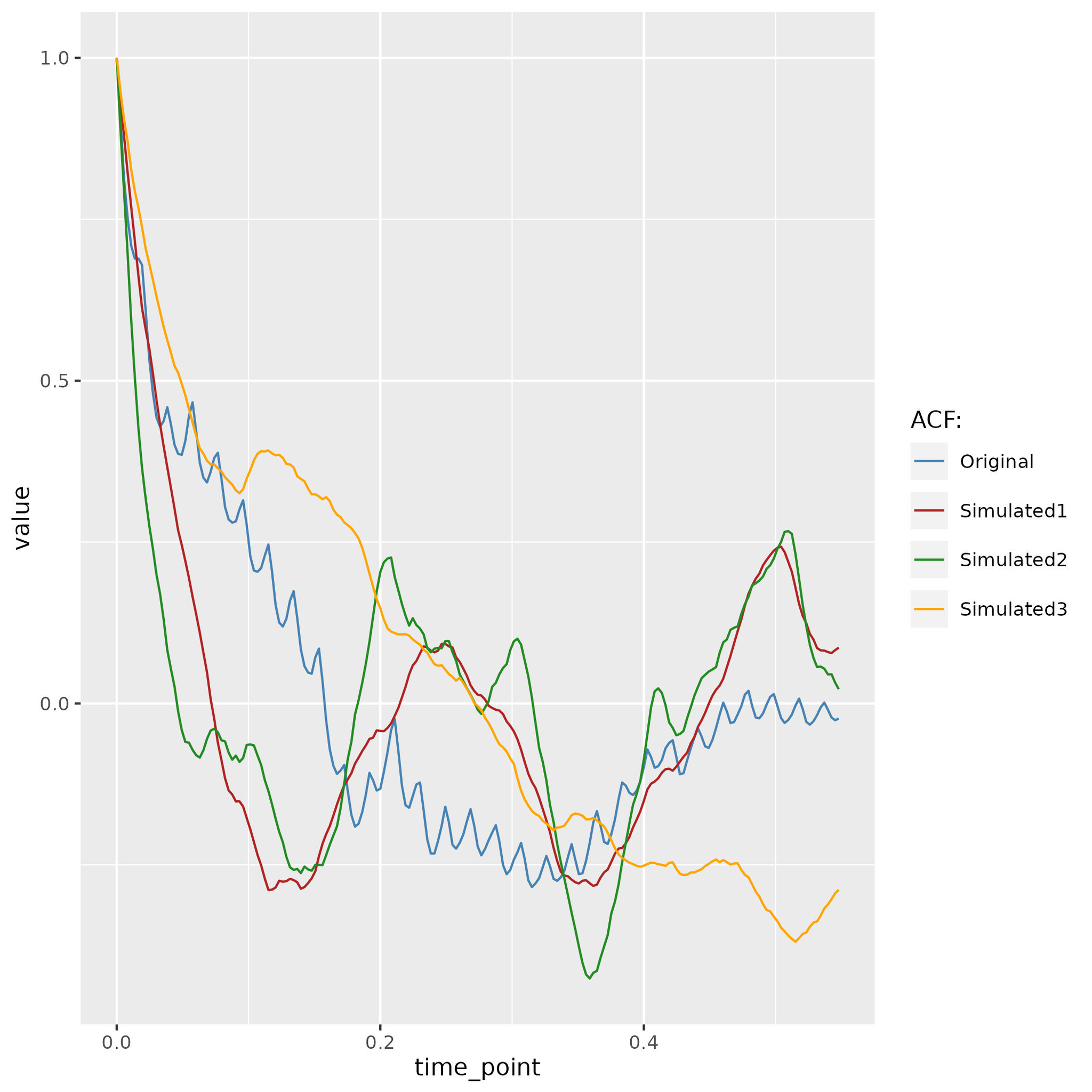}
    \caption{ACF for data and simulations}
    \end{subfigure}
  
\end{figure}    
 
\begin{figure}[H]\ContinuedFloat
     \begin{subfigure}{0.39\linewidth}
        \includegraphics[width=\linewidth]{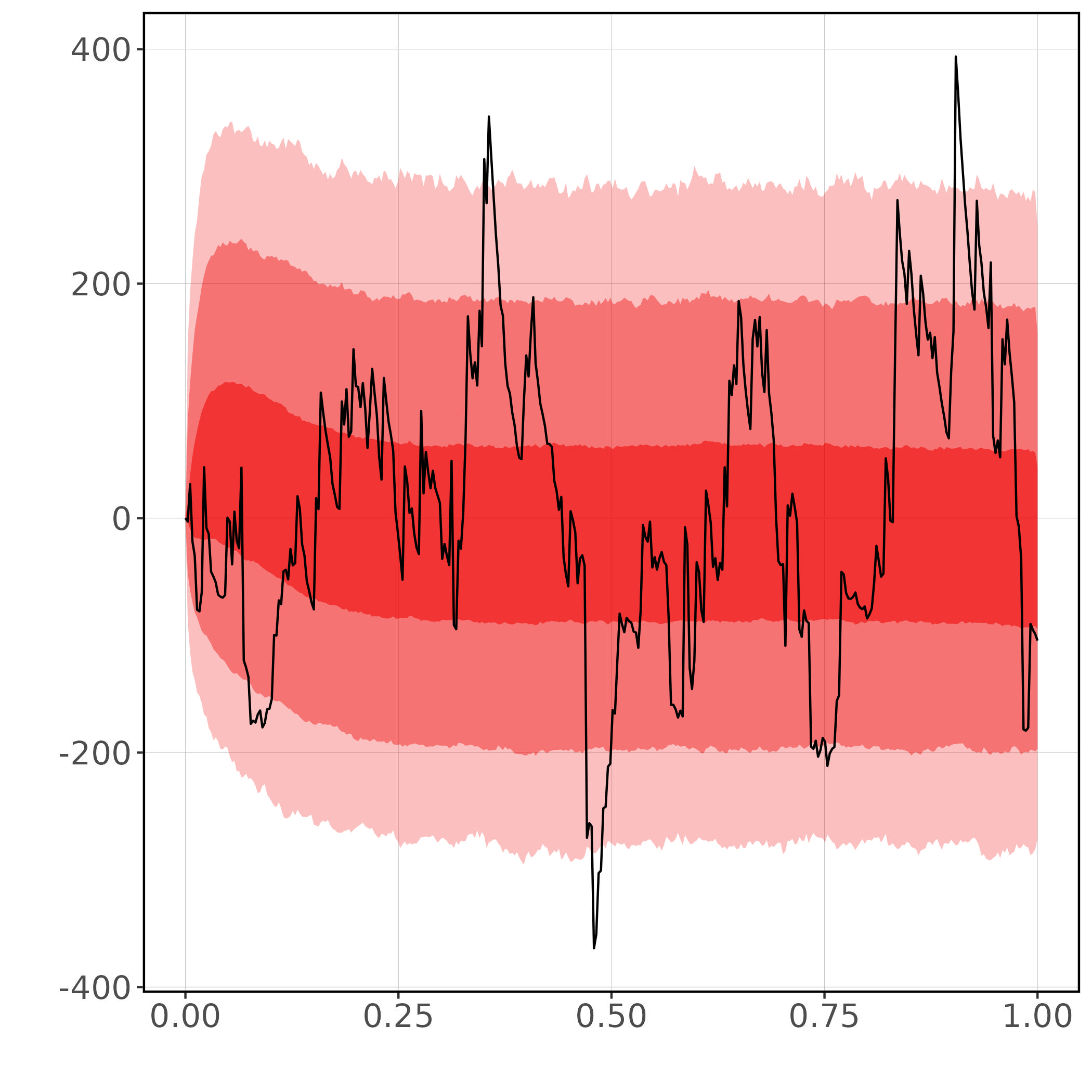}
    \caption{Simulated path and quantiles }
    \end{subfigure}   
\hfil
    \begin{subfigure}{0.39\linewidth}
        \includegraphics[width=\linewidth]{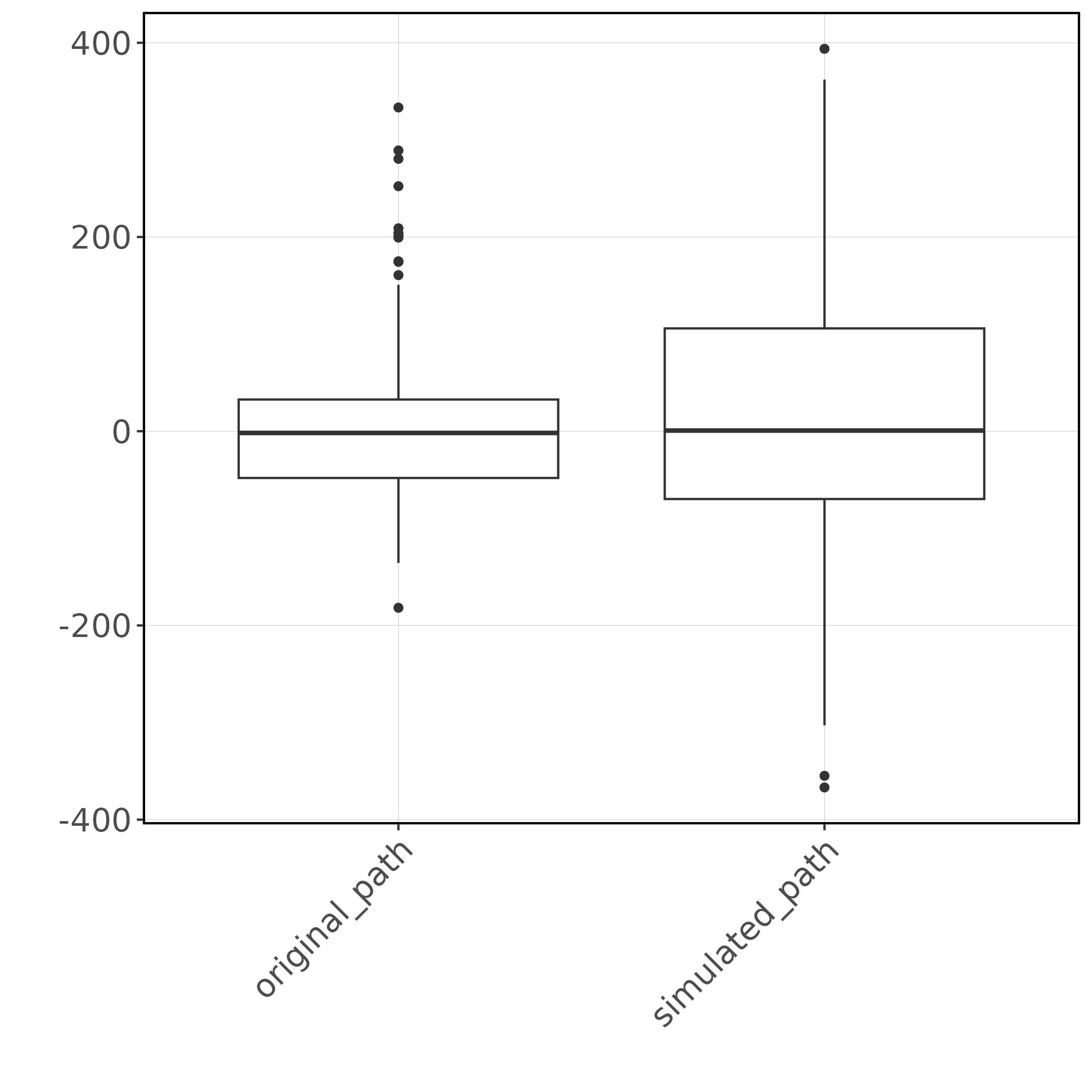}
    \caption{Boxplots for the ACF}
    \end{subfigure}

\caption{Statistical evaluation of the 3-factor model in the period 2021-23.}
\label{Fig3OU_21-23}
    \end{figure}    





When calibrating the $4$-factor model to the 2021-23 spot price data, it turns out, that the sample paths of the additional Gaussian OU-process fluctuate very closely around zero (cf. Fig. \ref{Fig21-23_4OU}(B)). This behavior is in line with the low volatility parameter in Table \ref{parameters_21_23}. Thus the second Gaussian component is completely insignificant from the modeling point of view. 

  \begin{figure}[H]
\centering
    \begin{subfigure}{0.45\linewidth}
        \includegraphics[width=\linewidth]{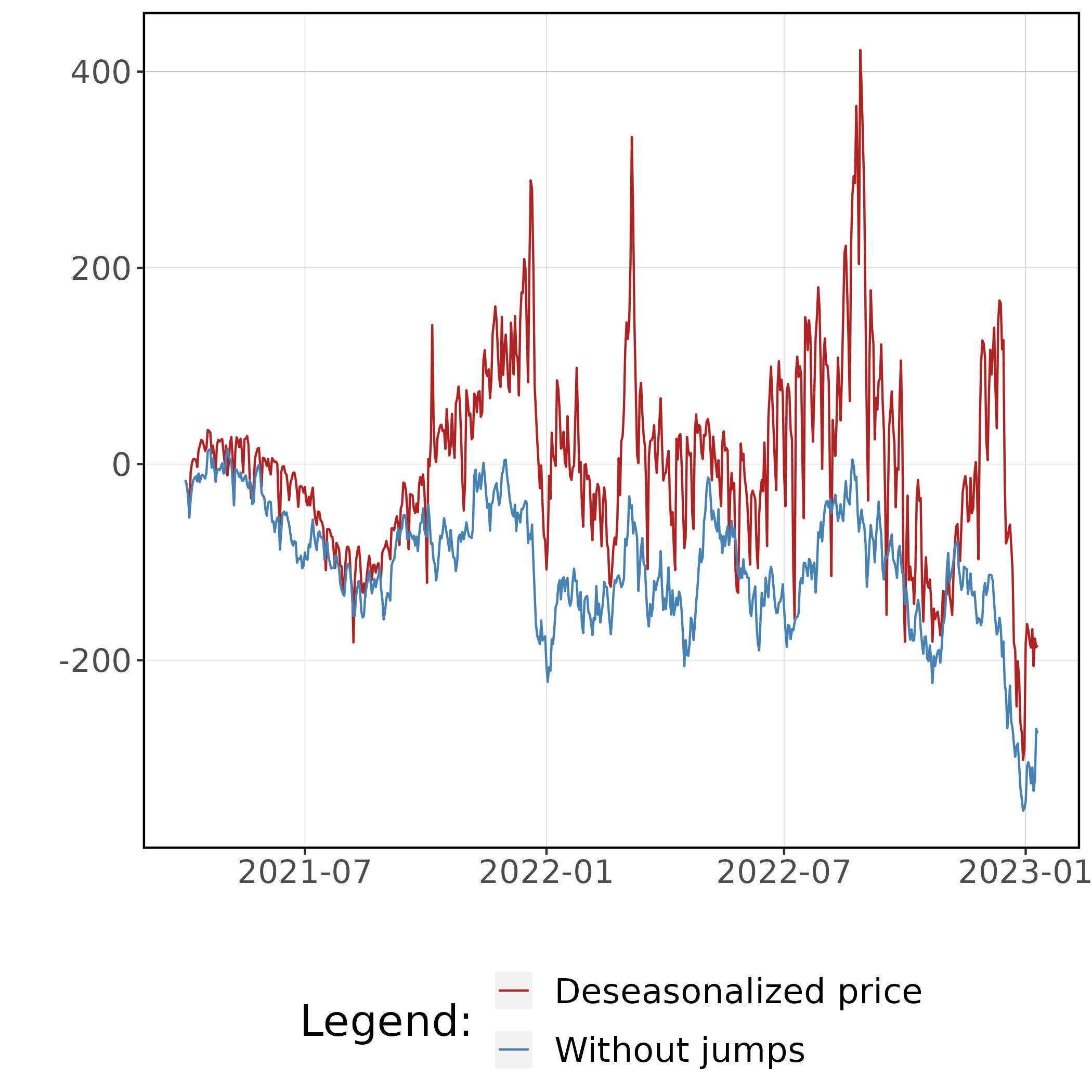}
    \caption{Data and Gaussian residuals}
    \end{subfigure}
\hfil
    \begin{subfigure}{0.45\linewidth}
        \includegraphics[width=\linewidth]{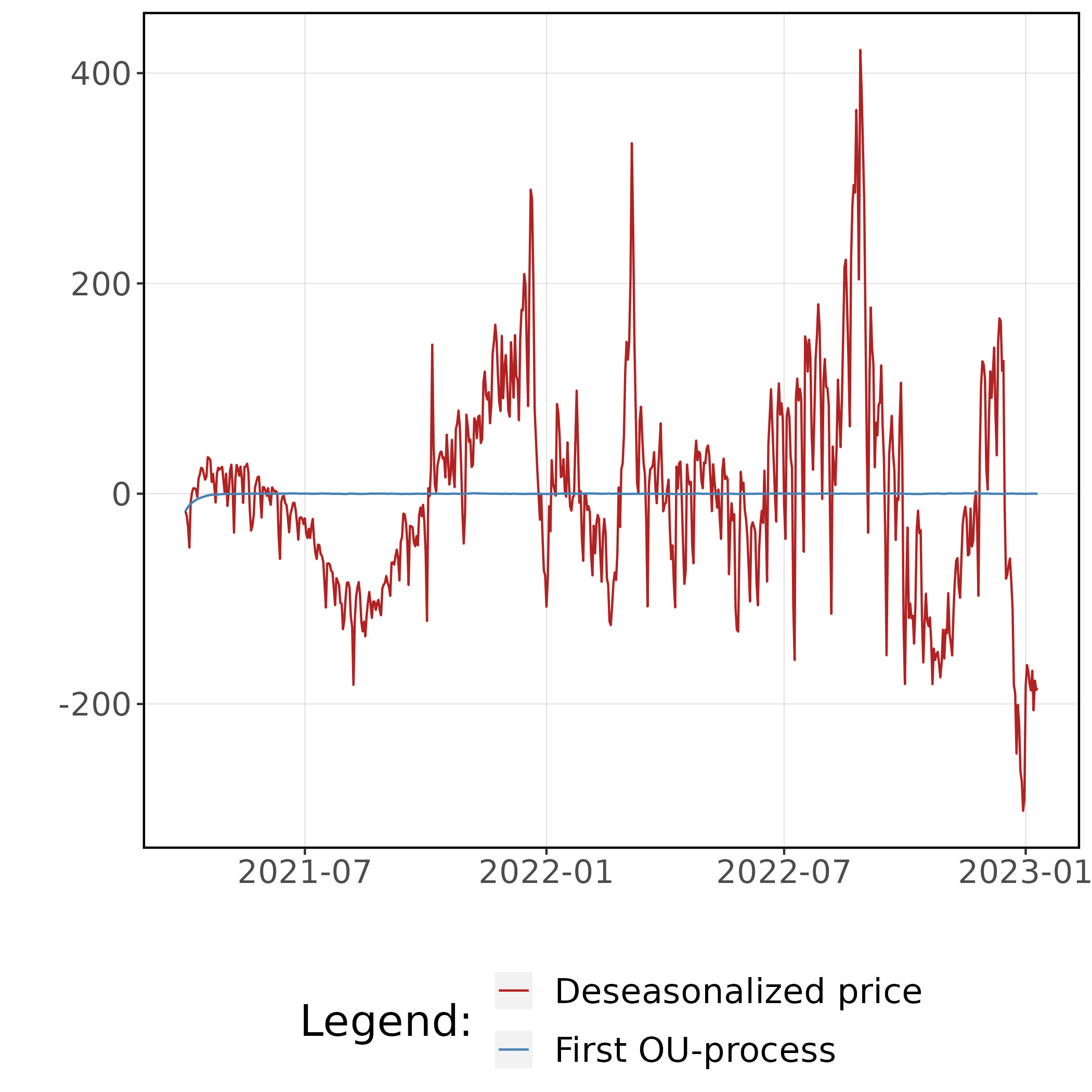}
    \caption{Sample path of the second OU-process}
    \end{subfigure}
\end{figure}
\begin{figure}[H]\ContinuedFloat
    \begin{subfigure}{0.45\linewidth}
        \includegraphics[width=\linewidth]{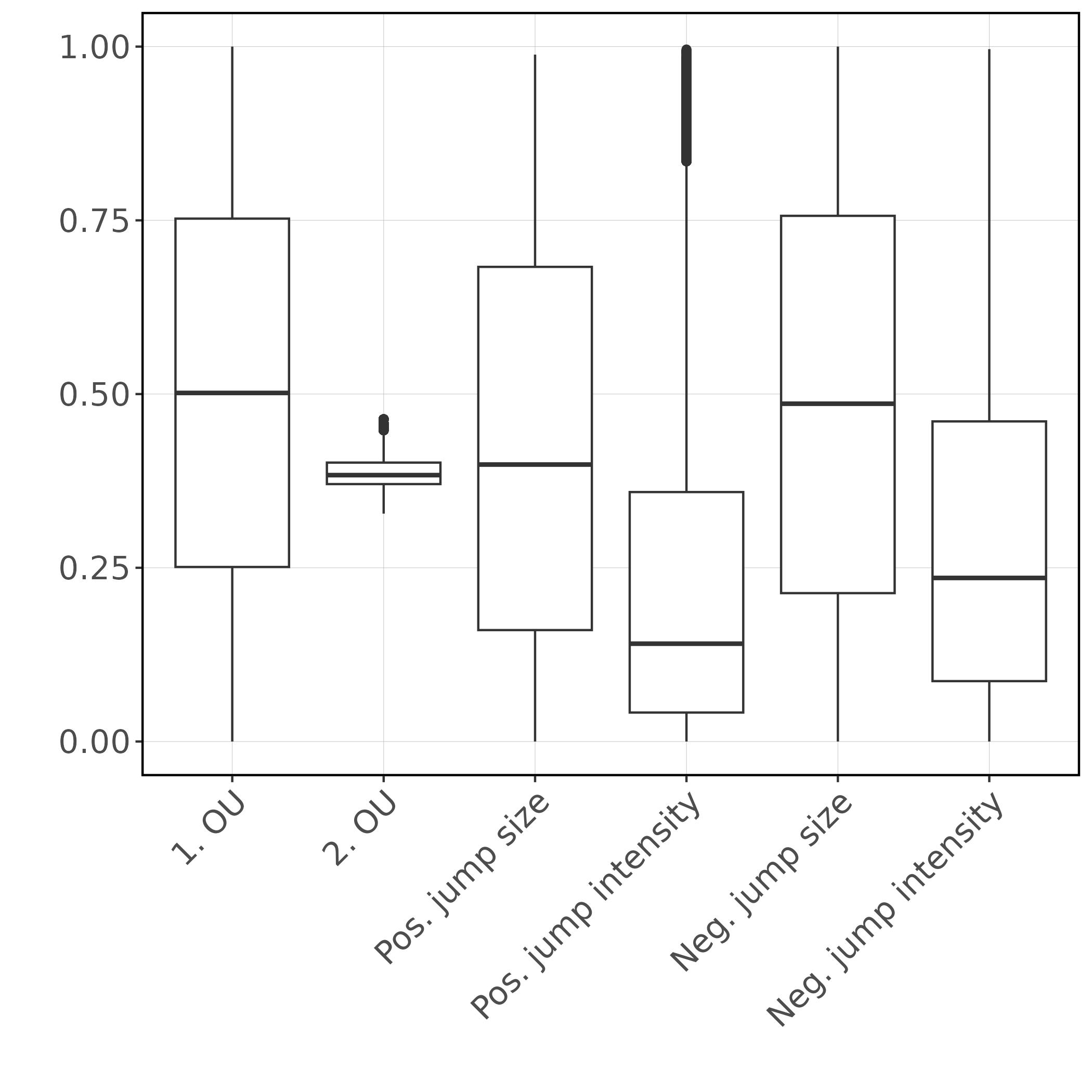}
    \caption{Boxplot of the $p$-values}
    \end{subfigure}
\hfil
    \begin{subfigure}{0.45\linewidth}
        \includegraphics[width=\linewidth]{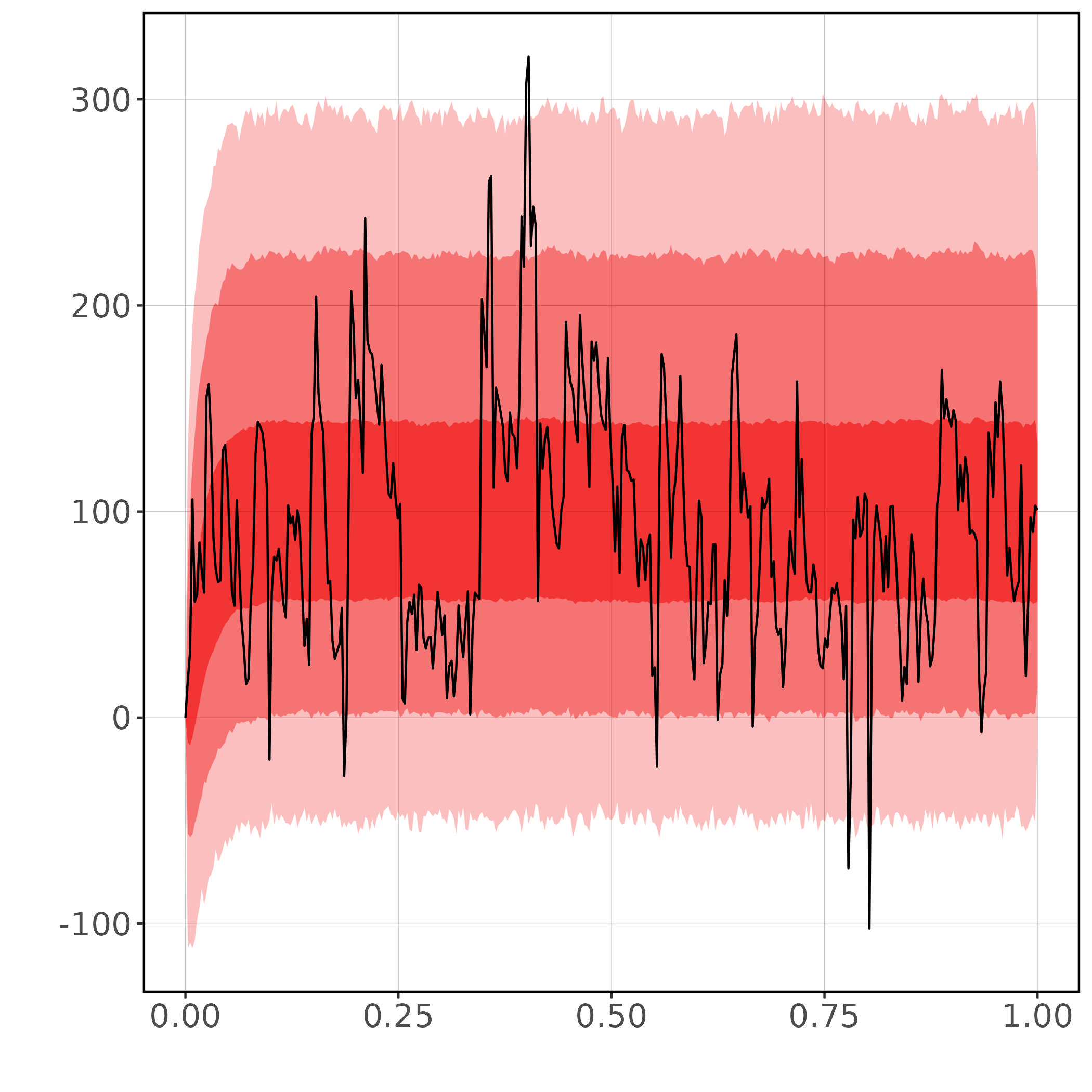}
    \caption{One simulated path and quantiles}
    \end{subfigure}
    
\caption{Statistical evaluation of the 4-factor model in the time period 2021-23.}
\label{Fig21-23_4OU}
    \end{figure}



\emph{Long-term Brownian motion:} Modeling the second Gaussian component as a Brownian motion without mean reversion again leads to larger long term deviations of the spot price process
from the mean level, as it can be observed in the simulations (cf. Figure \ref{Fig21-23_BB}(C)). As it was already the case for the 4-factor model in the previous subsection, also here the second Gaussian component is very close to zero which again indicates, that is sufficient to calibrate a model with only the arithmetic Brownian motion as Gaussian base signal.
  \begin{figure}[H]
\centering
    \begin{subfigure}{0.45\linewidth}
        \includegraphics[width=\linewidth]{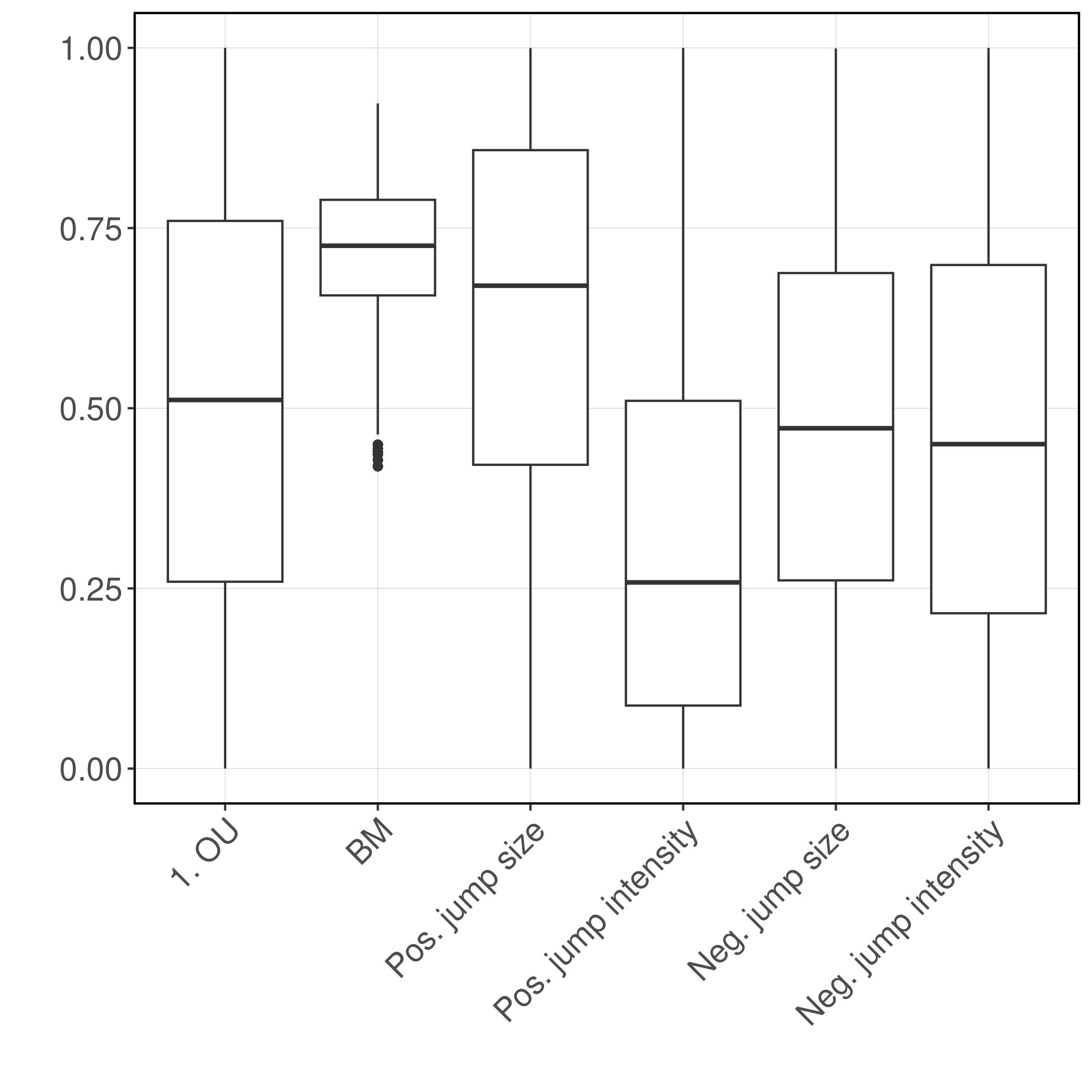}
    \caption{Boxplot of the $p$-values}
    \end{subfigure}
\hfil
    \begin{subfigure}{0.45\linewidth}
        \includegraphics[width=\linewidth]{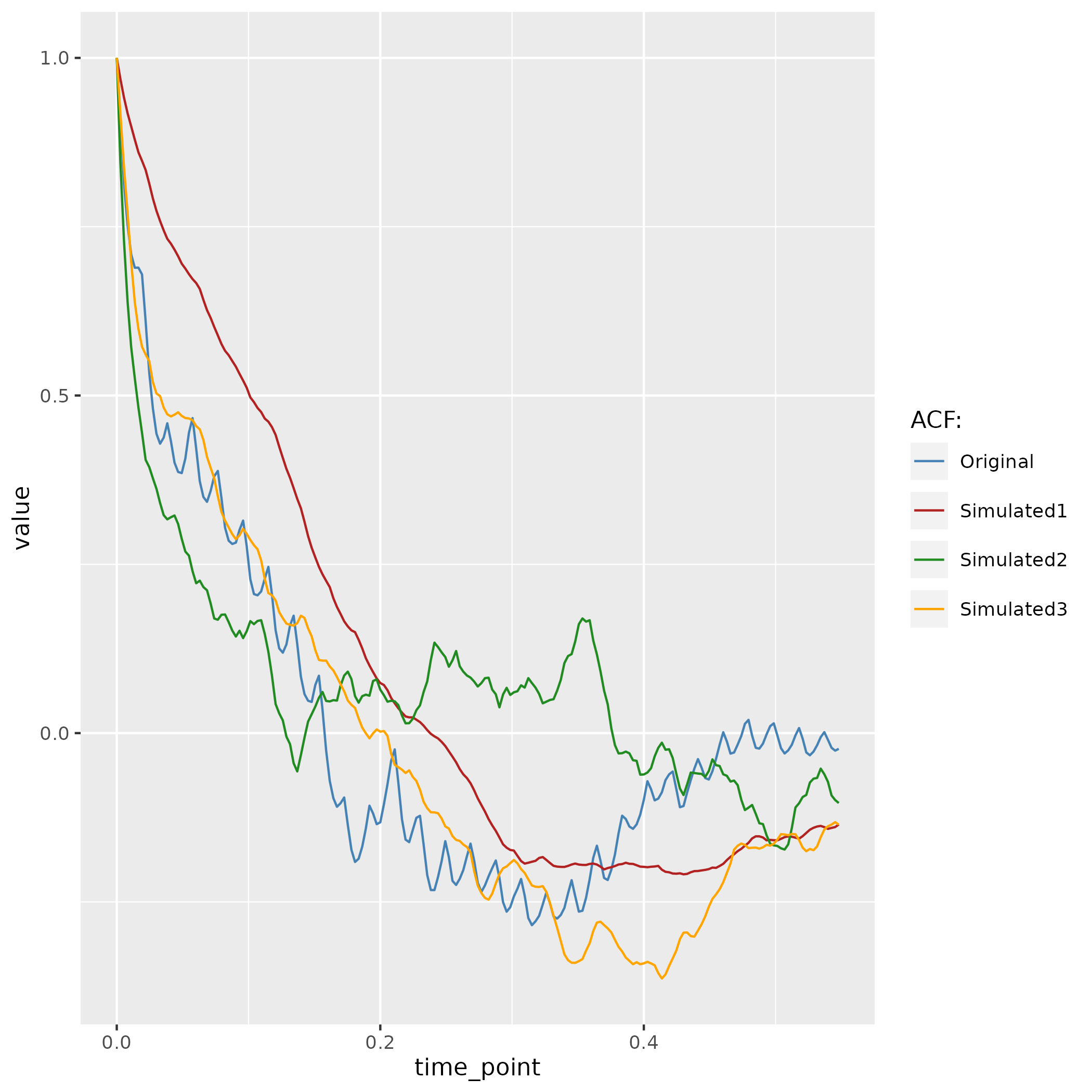}
    \caption{ACF for data and simulations}
    \end{subfigure}
\end{figure}
\begin{figure}[H]\ContinuedFloat
    \begin{subfigure}{0.45\linewidth}
        \includegraphics[width=\linewidth]{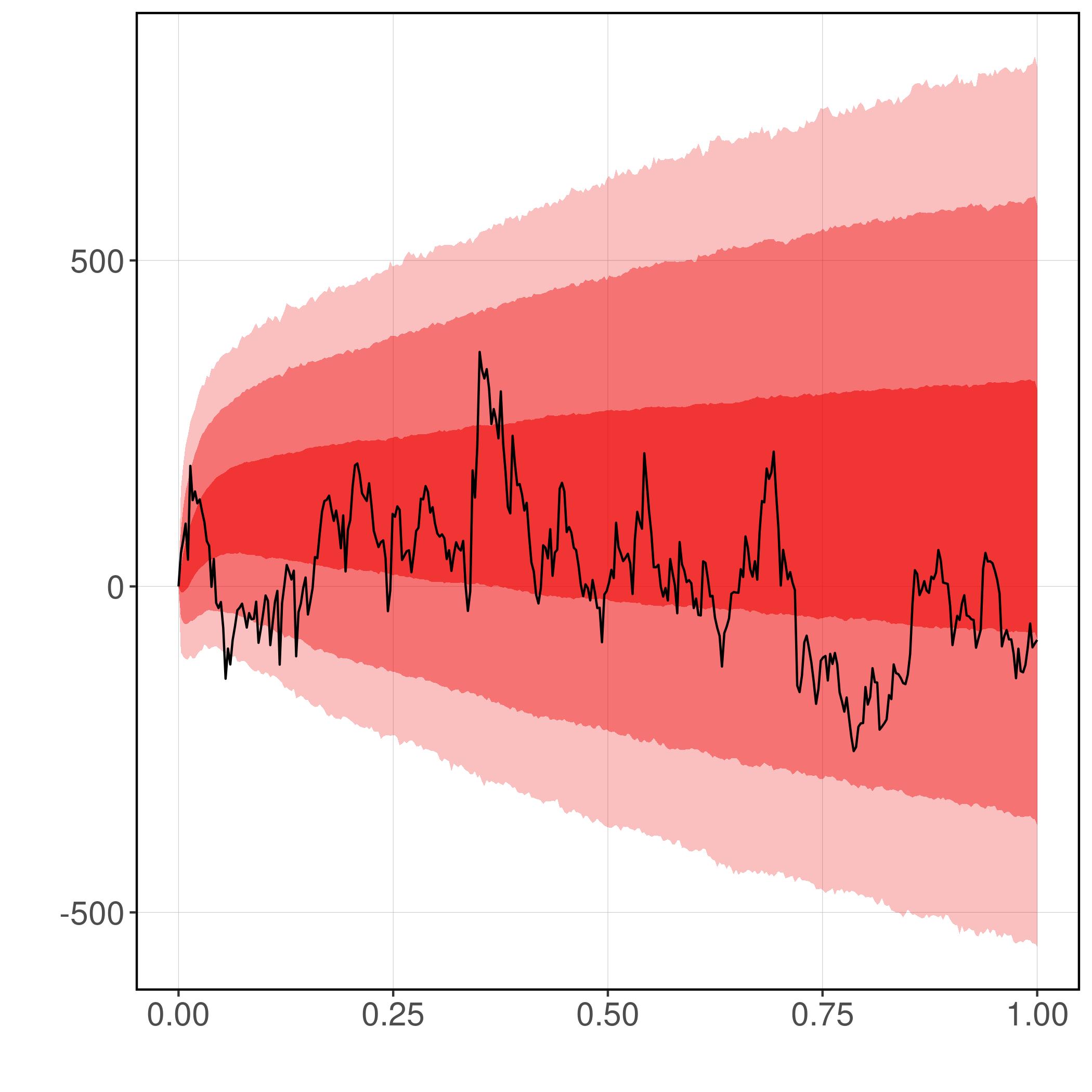}
    \caption{One simulated path and quantiles}
    \end{subfigure}
\hfil
    \begin{subfigure}{0.45\linewidth}
        \includegraphics[width=\linewidth]{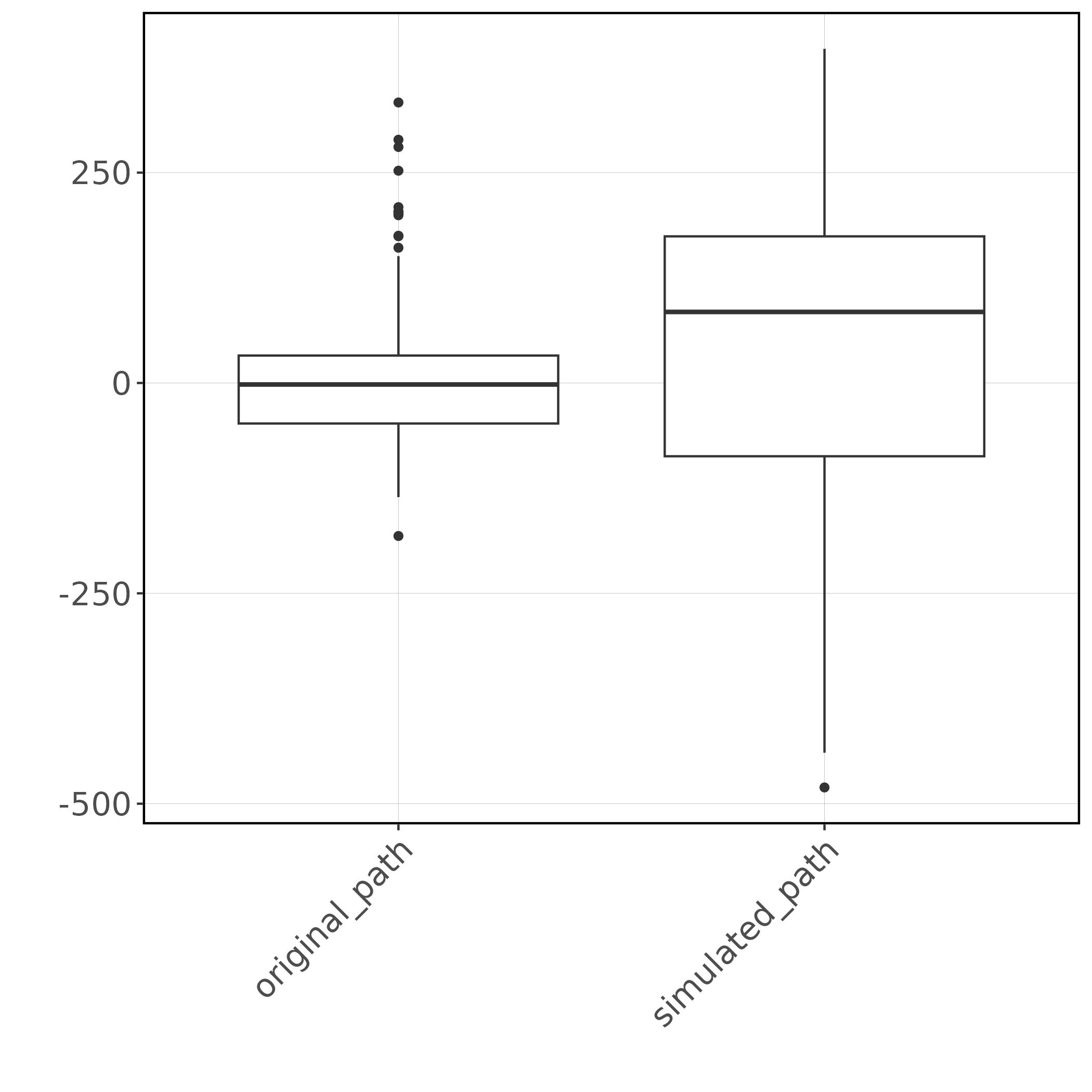}
    \caption{Boxplots for data and simulations}
    \end{subfigure}
    
\caption{Statistical evaluation of the 4-factor model with long-term Brownian motion in the period 2021-23}
\label{Fig21-23_BB}
    \end{figure}

\emph{Conclusion for the period 2021-23:} The 3-OU model performs well for the crisis data in the time period 2021-23 in our statistical evaluation (p-values, ACF) and the simulations fit the historical data. In contrast to the the pre-crisis period, long term fluctuations are now modeled as superpositions of exponentially distributed jumps of large size. This indicates, that the extreme deviations from the long term mean observed in the data with its very rapid up- and downward trends are no longer Gaussian. 
When calibrating the 4-OU model, it turns out that the additional Gaussian process is very close to zero and hence insignificant for the price modeling. Thus, the calibration of the 3-factor model is sufficient for this time period.

\subsubsection{2018-23 spot price data}
\begin{figure}[H]
\centering
    \includegraphics[width=0.7\linewidth]{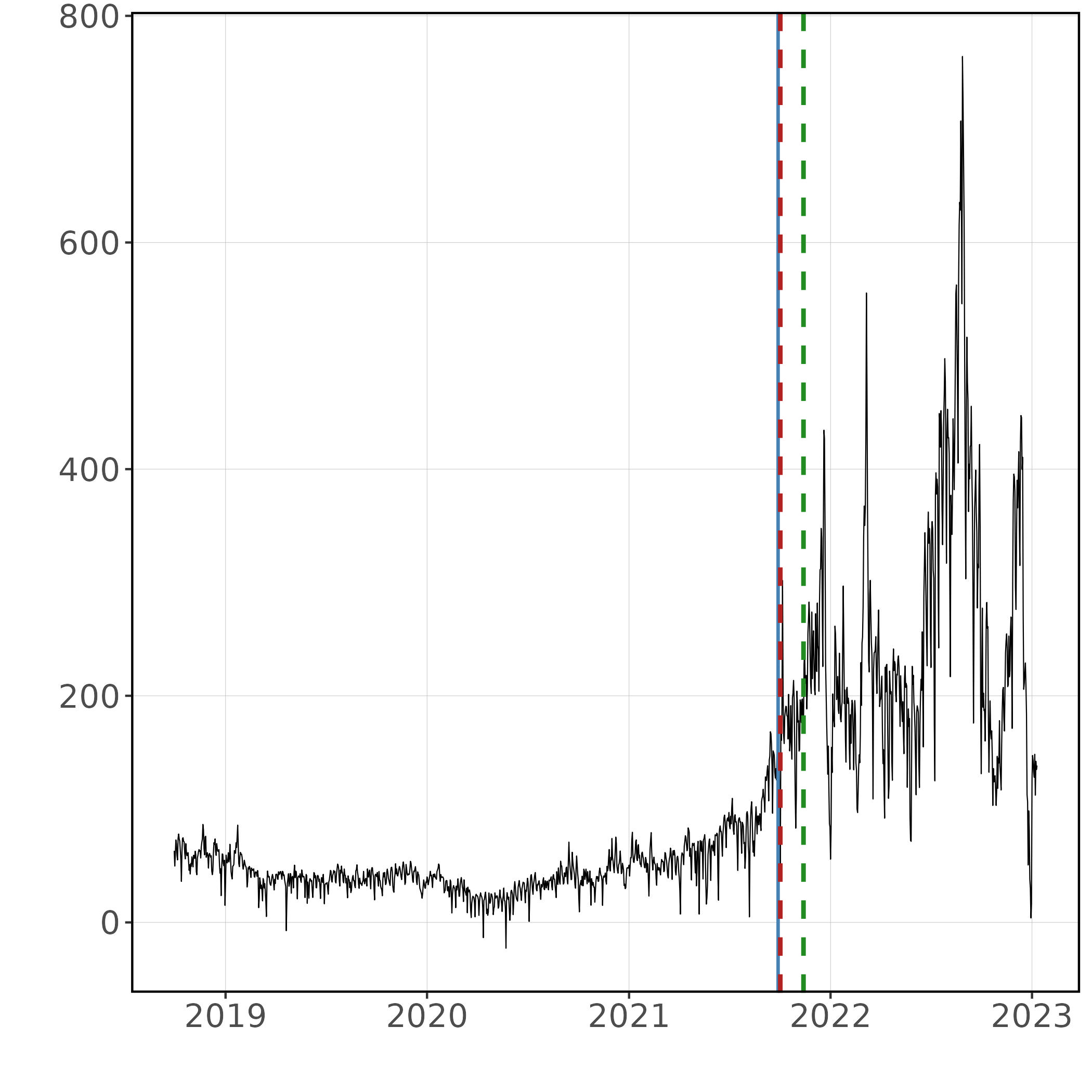}
    \caption{Time series with estimated change points via detecting changes in the mean (red dashed line at 2 October 2021) and changes in the variance (green dashed line at 13 November 2021). The blue vertical line is the change point that we used to run the MCMC algorithm. This change point is at 28 September 2021.}
    \label{change_points}
\end{figure}

\begin{figure}[H]
\centering
    \begin{subfigure}{0.45\linewidth}
        \includegraphics[width=\linewidth]{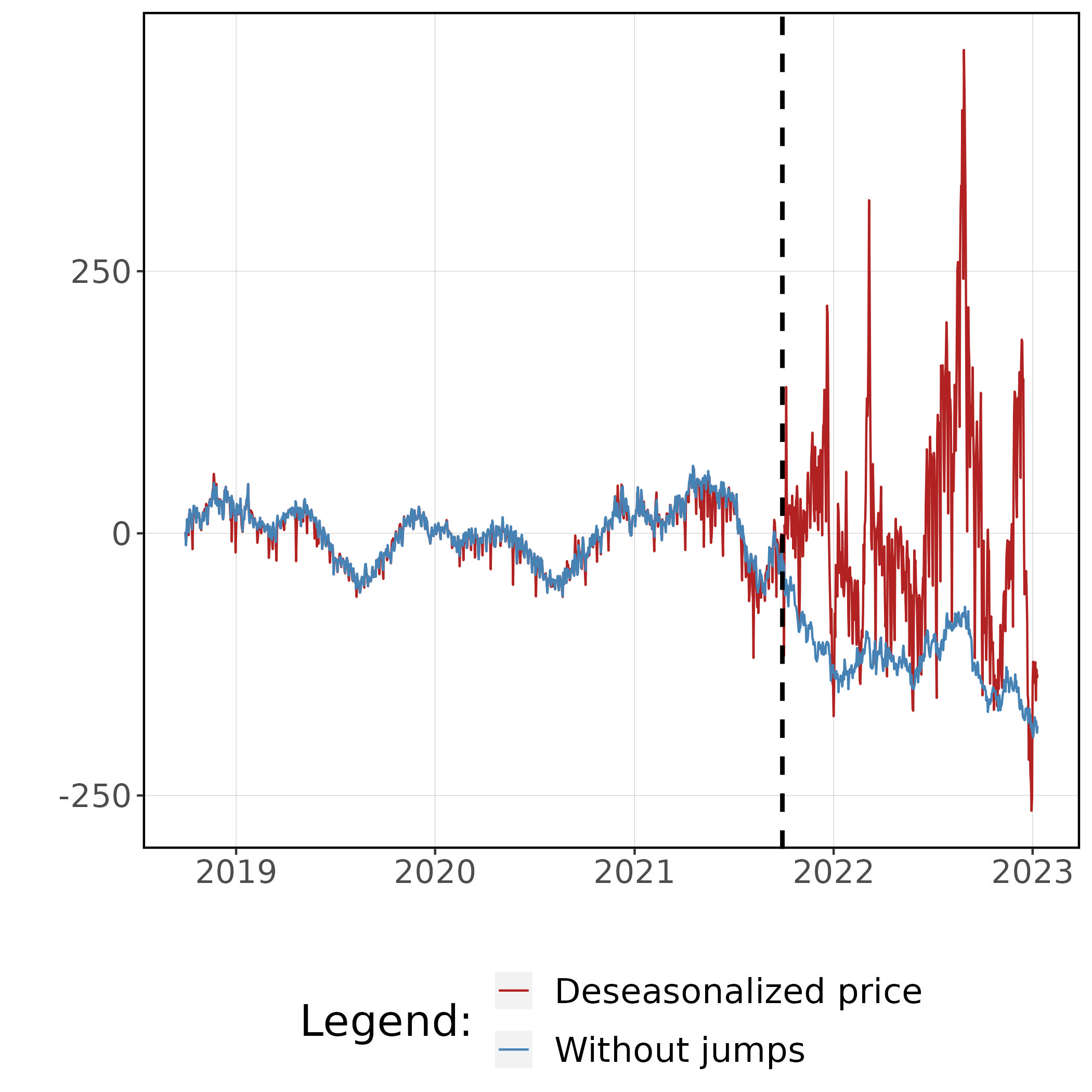}
    \caption{Gaussian residuals}
    \end{subfigure}
\hfil
    \begin{subfigure}{0.45\linewidth}
        \includegraphics[width=\linewidth]{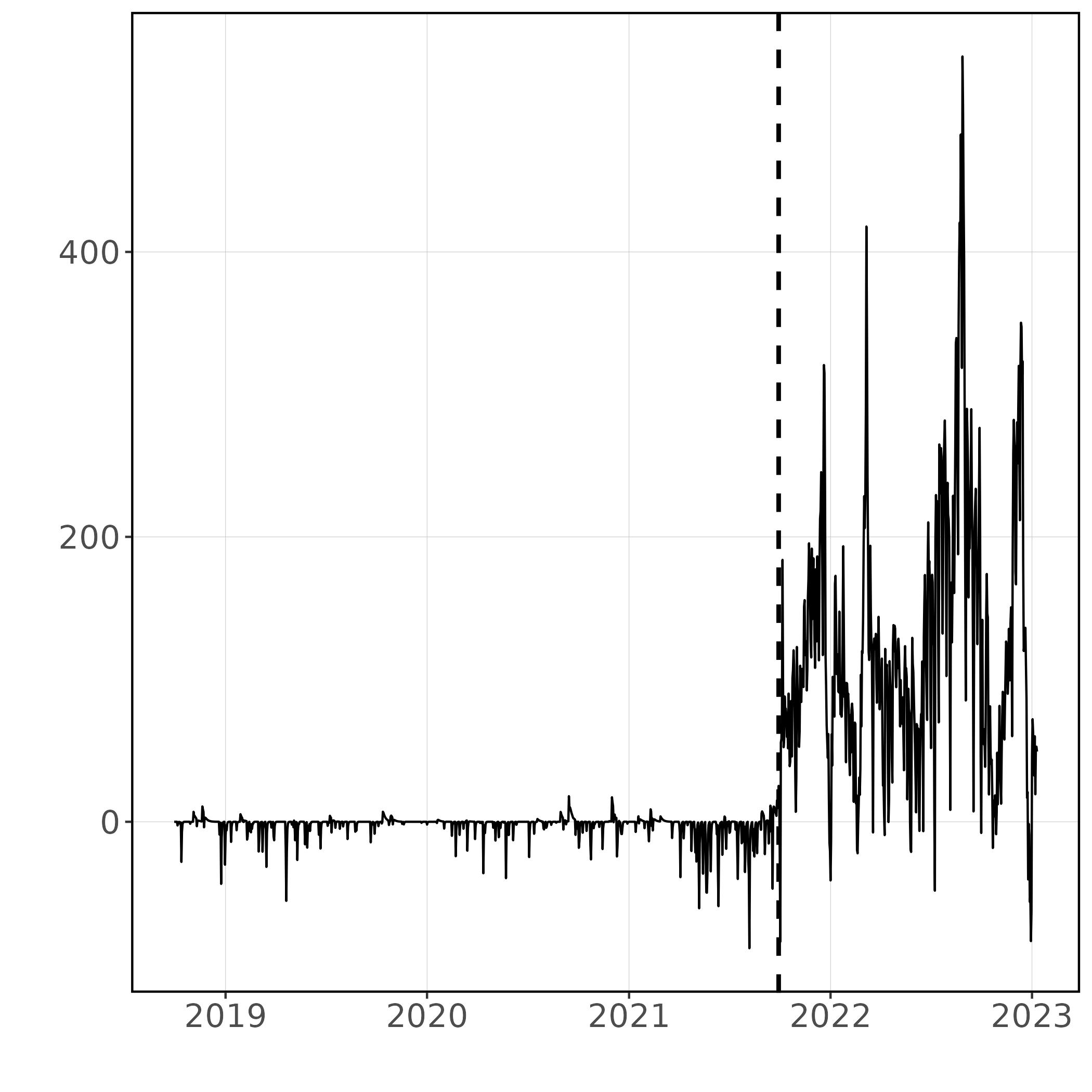}
    \caption{Jumps}
    \end{subfigure}

\caption{4-factor model with change point. The change point is given by the dashed vertical line.}
\label{Fig18-23_change}
    \end{figure}

\begin{figure}[H]
\centering
    \begin{subfigure}{0.45\linewidth}
        \includegraphics[width=\linewidth]{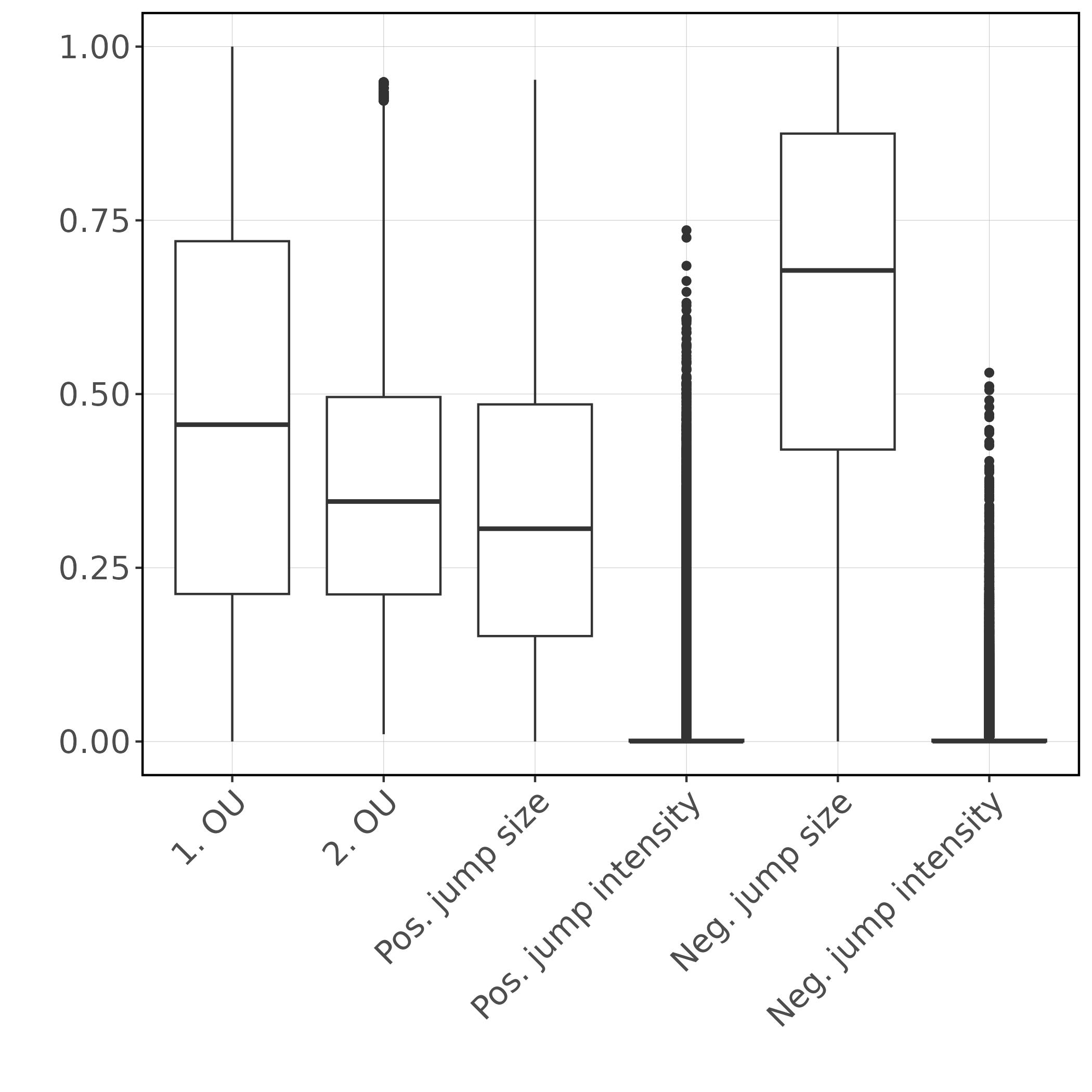}
    \caption{Boxplot of $p$-values fot the model without change point}
    \end{subfigure}
\hfil
    \begin{subfigure}{0.45\linewidth}
        \includegraphics[width=\linewidth]{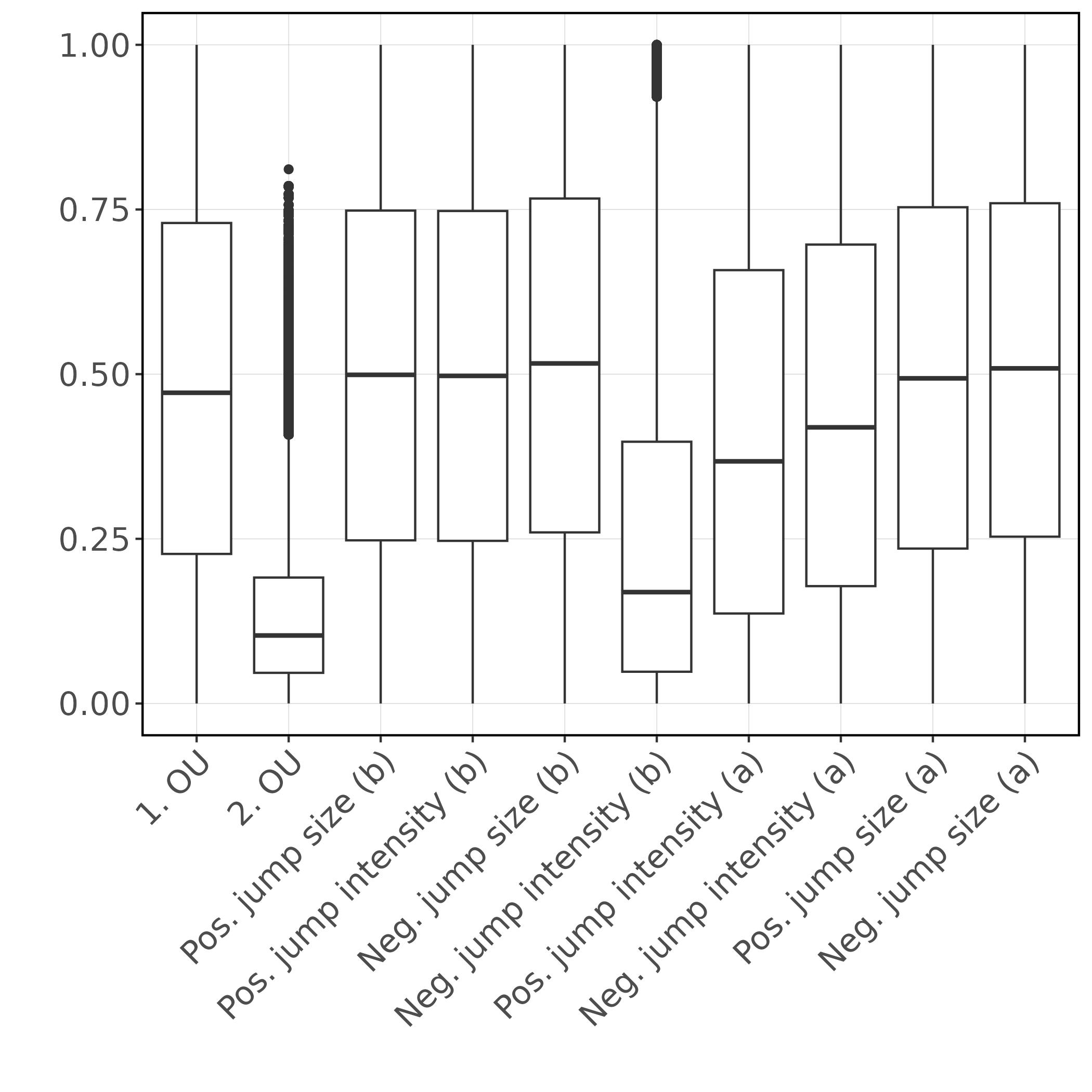}
    \caption{Boxplot of $p$-values fot the model with change point}
    \end{subfigure}

\caption{Fit of the 4-factor model in the time period 2018-23.}
\label{Fig18-23_p_val}
    \end{figure}

The calibration of the $3$-factor and $4$-factor model to the whole 2018-23 data yields very low $p$-values for the jump processes, due to the different jump patterns occurring before and after the start of the energy crisis (cf. Fig. \ref{Fig18-23_p_val}).
We therefore consider a change-point model, where we have different jump intensities and jump sizes in the respective time intervals (cf. Figure \ref{Fig18-23_change}). Figure~\ref{change_points} illustrates two possible change points that are detected with the default settings of the R-functions \texttt{cpt.mean} and \texttt{cpt.var} from the package \texttt{changepoint}. Both values lie in an area in which a significant increase in the spot price is apparent. For the MCMC-algorithm, we set the change point at 28 September 2021.

A detailed explanation of the jump update procedure in the change point model is presented in Appendix~B. This specification should allow a better fit of the jump processes, however, the convergence of the MCMC algorithm is quite slow due the additional parameters. 
We calibrate the 3-factor model and the 4-factor model with changepoint to the spot-price data in the whole time interval 2018-2023. We start with an overview of the posterior properties of the model parameters obtained from the MCMC procedure  described in Section~\ref{MCMC}. Later in this section, we present a more detailed analysis of our calibration results.

\begin{table}[H]

\centering
\begin{tabular}{c|cccc}
\toprule
\multicolumn{1}{c}{}  & \multicolumn{4}{c}{\textbf{2018-23}}\\
 \cmidrule(rl){2-5}
\multicolumn{1}{c}{}   & \multicolumn{2}{c}{\textbf{3-OU}} & \multicolumn{2}{c}{\textbf{4-OU}}\\

 \cmidrule(rl){2-5} 
\textbf{Parameter}   & {Mean} & {SD} & {Mean} & {SD}\\
\midrule
$\sigma_{\v}$ & 79.451 & 6.421 & 124.813 & 9.275\\
$\sigma_{\w}$  & - & -& 70.533 & 7.8973\\
$\lambda_{\v}$ & 27.472 & 771.510& 0.005 & 0.001\\
$\lambda_{\w}$  & - & -& 2.416 & 0.881\\
$\lambda_{J_1}$  & 0.016 & 0.002& 0.014 & 0.001\\
$\lambda_{J_2}$  & 0.003 & 0 & 0.002 & 0\\
$\theta_1^{(1)}$ & 205.563 & 37.208& 6.537 & 9.103\\
$\theta_2^{(1)}$ & 130.357 & 28.011& 92.919 & 21.170\\
$\beta_1^{(1)}$  &  5.240 & 0.587& 5.729 & 295.556\\
$\beta_2^{(1)}$ & 13.385 & 2.081& 17.586 & 2.927 \\
$\theta_1^{(2)}$ &  373.349 & 40.119& 368.858 & 47.868\\
$\theta_2^{(2)}$ &  614.171 & 59.463& 510.08 & 61.584\\
$\beta_1^{(2)}$  &  29.330 & 2.443& 31.157 & 2.928\\
$\beta_2^{(2)}$ &  26.523 & 2.805& 30.997 & 3.521\\
\bottomrule
\end{tabular}

\caption{Posterior properties of the change point model parameters in the whole 2018-23 time period. We present the mean and the standard deviation (SD) for all model parameters.}
\label{parameters_18_23}
\end{table}

In fact, the extremely high standard deviations for some parameters in Table \ref{parameters_18_23} indicate, that the algorithm has still not converged. The overall very bad calibration results for the whole time period 2018-23 confirm our approach to destinguish between pre-crisis and crisis data and separately calibrate the model to the 2018-21 and the 2021-23 data as it was done in the previous subsections.

\emph{Conclusion for the period 2018-23:}
The different structure of the data before and after the start of the crisis makes the calibration of the 3- and 4-factor model to the whole 2018-23 data difficult. The very different types of jump patters occurring in the respective time intervals lead to extremely bad $p$-values especially for the jump intensity rates. Also a change point model with separate jump intensities and jump sizes in the respective time intervals does not lead to satisfying results. Therefore, a separate calibration of the models to the data in two distinct time intervals as it was done in the previous subsections is strongly indicated. Of course, one could think of more advanced models for the spot price on the whole interval 2018-23, for example by introducing stochastic jump intensity rates. However, this would make the MCMC algorithm even more complicated and thus more time-consuming.    

\subsection{Outlook}
Since this is the first study that tests factor models for electricity spot prices in recent times of crisis, there is still space for improvements. But if  practitioners would like to develop a model by themselves, then our results suggest to use the 3-factor model as starting point. As already mentioned, a possible extension could be the use of a stochastic jump intensity rate. Alternatively, also regime-switching models could help to include larger price jumps. Indeed, if we have a look at the time series data again, then we see that in times of crises, at some points the price level jumps drastically followed by weaker fluctuations around the new price level. This could be an indicator of changing regimes.

\section*{Acknowledgements}

This research paper was inspired by a real-world industrial application. We therefore wish to express our gratitude to LINZ STROM GAS W{\"A}RME GmbH – Gesch{\"a}ftsbereich Energiemanagement.

F. Aichinger is supported by the Austrian Science Fund (FWF) Project P34808. For the purpose of open
access, the authors have applied a CC BY public copyright licence to any author accepted
manuscript version arising from this submission.

\section*{Appendix A: Detailed updated steps of the MCMC algorithm}\label{Appendix_details}

\subsubsection{Detailed updated steps} Here we give a detailed instruction for each updated step. As described in Algorithm \ref{GS}, if a parameter has already been updated within the cycle (1)-(13), then in the following steps, we condition on its updated value. Whenever it is possible, we directly sample from the conditional distribution, otherwise, we apply a random walk Metropolis step using the factorization \eqref{factorization} within the Gibbs procedure. The tuning procedure for most of the parameters follows \cite{GMP17}, explanations for the updates of our additional parameters are given in the respective steps. 
\vspace{3mm}\\
\textbf{Updated $\bf{\sigma_{\v}}$}: 

With the choice of an $\operatorname{IG}(a_{\sigma_{\v}},b_{\sigma_{\v}})$ distributed prior for $\sigma_{\v}$, for the posterior we get

\begin{equation*}
\resizebox{.9\hsize}{!}{$\sigma_{\v}^2|\lambda_{\v},\lambda_{\w},\sigma_{\w},\lambda_{J_1},\lambda_{J_2},\mathcal{X},\mathcal{E},\Phi_1,\Phi_2 \sim \operatorname{IG}\left(a_{\sigma_{\v}}+\frac{N}{2},b_{\sigma_{\v}}+\frac{1}{\lambda_{\v}} \sum\limits_{i=1}^N \frac{\left(y_{1,i}-y_{1,{i-1}} e^{-\lambda_{\v}^{-1}\Delta_i}\right)^2}{1-e^{-2\lambda_{\v}^{-1}\Delta_i}}\right).$}
\end{equation*}
\vspace{3mm}
\textbf{Updated $\bf{\sigma_{\w}}$}: 

Generate candidate $\sigma_{\w}'$ from $\sigma_{\w}'|\sigma_{\w}\sim\mathcal{N}(\sigma_{\w},\sigma,0,\infty)$. 
Here $\mathcal{N}(\mu,\sigma,a,b)$ denotes the truncated normal distribution with density
\[
f(x;\mu,\sigma,a,b)=\frac{1}{\sigma}\frac{\phi(\frac{x-\mu}{\sigma})}{\Phi(\frac{b-\mu}{\sigma})-\Phi(\frac{a-\mu}{\sigma})}, 
\]
where $\phi$ is the density function and $\Phi$ is the cumulative distribution function of the standard normal distribution. The truncated normal distribution was chosen since we want to get new samples around the previous parameter value, but greater than zero.
Perform Metropolis-step with acceptance ratio

\[
\alpha({\sigma_{\w}}',{\sigma_{\w}})=\operatorname{min}\left\{1,\frac{l(\mathcal{X}|\lambda_{\v},\sigma_{\v},\lambda_{\w},\sigma_{\w}',\lambda_{J_1},\lambda_{J_2},\Phi_1,\Phi_2,\mathcal{E})\pi(\sigma_{\w}')}{l(\mathcal{X}|\lambda_{\v},\sigma_{\v},\lambda_{\w},\sigma_{\w},\lambda_{J_1},\lambda_{J_2},\Phi_1,\Phi_2,\mathcal{E})\pi(\sigma_{\w})}\cdot\frac{1-\Phi(-\frac{\sigma_{\w}}{\sigma})}{1-\Phi(-\frac{\sigma_{\w}'}{\sigma})}\right\}.
\]

Since we do not know the magnitude of $\sigma_{\w}$ in relation to $\sigma_{\v}$, we decided to approximate the quotient of the apriori distributions $\frac{\pi(\sigma_{\w}')}{\pi(\sigma_{\w})}$  with $1$. This can, e.g.,~be justified by the choice of an inverse gamma distribution $\text{IG}(1,\beta)$ for the apriori distribution $\pi$ together with the limiting result $\lim\limits_{\beta\rightarrow\infty}\frac{\pi(\sigma_{\w}')}{\pi(\sigma_{\w})}=1$. The same argument holds by using an uninformative prior and considering the limiting case in which the interval length goes to infinity.

\vspace{3mm}
\noindent\textbf{Updated $\bf{\rho_{\v}=e^{-\lambda_{\v}^{-1}}}$}:

Generate candidate $\rho_{\v}'$ from $\rho_{\v}'|\rho_{\v}\sim\mathcal{N}(\rho_{\v},\sigma,0,1)$. Perform Metropolis-step with acceptance ratio

\begin{equation*}
\resizebox{.9\hsize}{!}{
$\alpha({\rho_{\v}}',{\rho_{\v}})=\operatorname{min}\left\{1,\frac{l(\mathcal{X}|\rho_{\v}',\sigma_{\v},\lambda_{\w},\sigma_{\w},\lambda_{J_1},\lambda_{J_2},\Phi_1,\Phi_2,\mathcal{E}) f_{\rho_{\v}}(\rho_{\v}')}{l(\mathcal{X}|\rho_{\v},\sigma_{\v},\lambda_{\w},\sigma_{\w},\lambda_{J_1},\lambda_{J_2},\Phi_1,\Phi_2,\mathcal{E})f_{\rho_{\v}}(\rho_{\v})}\cdot\frac{\Phi(\frac{1-\rho_{\v}}{\sigma})-\Phi(-\frac{\rho_{\v}}{\sigma})}{\Phi(\frac{1-\rho_{\v}'}{\sigma})-\Phi(-\frac{\rho_{\v}'}{\sigma})}\right\}.$}
\end{equation*}

Since $\lambda_{\v}$ has prior distribution $\operatorname{IG}(a_{\lambda_{\v}},b_{\lambda_{\v}})$,  $\rho_{\v}=e^{-\lambda_{\v}^{-1}}$ has apriori density

\[
f_{\rho_{\v}}(y)=\frac{1}{y}\frac{b_{\lambda_{\v}}}{\Gamma(a_{\lambda_{\v}})}(-\operatorname{ln}(y))^{a_{\lambda_{\v}}-1}e^{b_{\lambda_{\v}}\operatorname{ln}(y)}.
\]
\vspace{3mm}
\textbf{Updated $\bf{\rho_{\w}=e^{-\lambda_{\w}^{-1}}}$}:

Generate candidate $\rho_{\w}'$ from $\rho_{\w}'|\rho_{\w}\sim\mathcal{N}(\rho_{\w},\sigma,0,1)$. Perform Metropolis-step with acceptance ratio

\begin{equation*}
\resizebox{.9\hsize}{!}{
$\alpha({\rho_{\w}}',{\rho_{\w}})=\operatorname{min}\left\{1,\frac{l(\mathcal{X}|\lambda_{\v},\sigma_{\v},\rho_{\w}',\sigma_{\w},\lambda_{J_1},\lambda_{J_2},\Phi_1,\Phi_2,\mathcal{E})f_{\rho_{\w}}(\rho_{\w}')}{l(\mathcal{X}|\lambda_{\v},\sigma_{\v},\lambda_{\w},\rho_{\w},\lambda_{J_1},\lambda_{J_2},\Phi_1,\Phi_2,\mathcal{E}) f_{\rho_{\w}}(\rho_{\w})}\cdot\frac{\Phi(\frac{1-\rho_{\w}}{\sigma})-\Phi(-\frac{\rho_{\w}}{\sigma})}{\Phi(\frac{1-\rho_{\w}'}{\sigma})-\Phi(-\frac{\rho_{\w}'}{\sigma})}\right\}.$}
\end{equation*}
Since $\lambda_{\w}$ is $\operatorname{IG}(a_{\lambda_{\w}},b_{\lambda_{\w}})$ distributed,  $Y=e^{-\lambda_{\w}^{-1}}$ has density

\[
f_{\rho_{\w}}(y)=\frac{1}{y}\frac{b_{\lambda_{\w}}}{\Gamma(a_{\lambda_{\w}})}(-\operatorname{ln}(y))^{a_{\lambda_{\w}}-1}e^{b_{\lambda_{\w}}\operatorname{ln}(y)}.
\]
\vspace{3mm}
\textbf{Updated $\bf{\rho_{J_1}=e^{-\lambda_{J_1}^{-1}}}$}:

Generate candidate $\rho_{J_1}'$ from $\rho_{J_1}'|\rho_{J_1}\sim\mathcal{N}(\rho_{J_1},\sigma,0,1)$. Perform Metropolis-step with acceptance ratio

\begin{equation*}
\resizebox{.9\hsize}{!}{
$\alpha({\rho_{J_1}}',{\rho_{J_1}})=\operatorname{min}\left\{1,\frac{l(\mathcal{X}|\lambda_{\v},\sigma_{\v},\lambda_{\w},\sigma_{\w},\rho_{J_1}',\lambda_{J_2},\Phi_1,\Phi_2,\mathcal{E})f_{\rho_{J_1}}(\rho_{J_1}')}{l(\mathcal{X}|\lambda_{\v},\sigma_{\v},\lambda_{\w},\sigma_{\w},\rho_{J_1},\lambda_{J_2},\Phi_1,\Phi_2,\mathcal{E}) f_{\rho_{J_1}}(\rho_{J_1})}\cdot\frac{\Phi(\frac{1-\rho_{J_1}}{\sigma})-\Phi(-\frac{\rho_{J_1}}{\sigma})}{\Phi(\frac{1-\rho_{J_1}'}{\sigma})-\Phi(-\frac{\rho_{J_1}'}{\sigma})}\right\}.$}
\end{equation*}
Since $\lambda_{J_1}$ is $\operatorname{IG}(a_{\lambda_{J_1}},b_{\lambda_{J_1}})$ distributed,  $Y=e^{-\lambda_{J_1}^{-1}}$ has density
\[
f_{\rho_{J_1}}(y)=\frac{1}{y}\frac{b_{\lambda_{J_1}}}{\Gamma(a_{\lambda_{J_1}})}(-\operatorname{ln}(y))^{a_{\lambda_{J_1}}-1}e^{b_{\lambda_{J_1}}\operatorname{ln}(y)}.
\]
\vspace{3mm}
\textbf{Updated $\bf{\rho_{J_2}=e^{-\lambda_{J_2}^{-1}}}$}:

Generate candidate $\rho_{J_2}'$ from $\rho_{J_2}'|\rho_{J_2}\sim\mathcal{N}(\rho_{J_2},\sigma,0,1)$. Perform Metropolis-step with acceptance ratio

\begin{equation*}
\resizebox{.9\hsize}{!}{
$\alpha({\rho_{J_2}}',{\rho_{J_2}})=\operatorname{min}\left\{1,\frac{l(\mathcal{X}|\lambda_{\v},\sigma_{\v},\lambda_{\w},\sigma_{\w},\lambda_{J_1},\rho_{J_2}',\Phi_1,\Phi_2,\mathcal{E}) f_{\rho_{J_2}}(\rho_{J_2}')}{l(\mathcal{X}|\lambda_{\v},\sigma_{\v},\lambda_{\w},\sigma_{\w},\lambda_{J_1},\rho_{J_2},\Phi_1,\Phi_2,\mathcal{E}) f_{\rho_{J_2}}(\rho_{J_2})}\cdot\frac{\Phi(\frac{1-\rho_{J_2}}{\sigma})-\Phi(-\frac{\rho_{J_2}}{\sigma})}{\Phi(\frac{1-\rho_{J_2}'}{\sigma})-\Phi(-\frac{\rho_{J_2}'}{\sigma})}\right\}.$}
\end{equation*}
Since $\lambda_{J_2}$ is $\operatorname{IG}(a_{\lambda_{J_2}},b_{\lambda_{J_2}})$ distributed,  $Y=e^{-\lambda_{J_2}^{-1}}$ has density

\[
f_{\rho_{J_2}}(y)=\frac{1}{y}\frac{b_{\lambda_{J_2}}}{\Gamma(a_{\lambda_{J_2}})}(-\operatorname{ln}(y))^{a_{\lambda_{J_2}}-1}e^{b_{\lambda_{J_2}}\operatorname{ln}(y)}
\]
\vspace{3mm}
\textbf{Updated $\bf{\theta_1}$}:

With the choice of an $\operatorname{Ga}(a_{{\theta1}},b_{\theta1})$ distributed prior for $\theta_1$, for the posterior we get
\[
\theta_1|\Phi_1 \sim \operatorname{Ga}\left(a_{\theta_1}+N_T^1,b_{\theta_1}+T\right).
\]
\vspace{3mm}
\textbf{Updated $\bf{\theta_2}$}:

With the choice of an $\operatorname{Ga}\left(a_{{\theta_2}},b_{\theta_2}\right)$ distributed prior for $\theta_2$, for the posterior we get
\[
\theta_2|\Phi_2 \sim \operatorname{Ga}\left(a_{\theta_2}+N_T^2,b_{\theta_2}+T\right).
\]
\vspace{3mm}
\textbf{Updated $\bf{\beta_1}$}

With the choice of an $\operatorname{IG}(a_{{\beta_1}},b_{\beta_1})$ distributed prior for $\beta_1$, for the posterior we get
\[
\beta_1|\Phi_1 \sim \operatorname{IG}\left(a_{\beta_1}+N_T^1,b_{\beta_1}+\sum\limits_{i=1}^{N_T^1} \xi_{1i}\right).
\]
\vspace{3mm}
\textbf{Updated $\bf{\beta_2}$}:

With the choice of an $\operatorname{IG}(a_{{\beta_2}},b_{\beta_2})$ distributed prior for $\beta_2$, for the posterior we get
\[
\beta_2|\Phi_2 \sim \operatorname{IG}\left(a_{\beta_2}+N_T^2,b_{\beta_2}+\sum\limits_{i=1}^{N_T^2} \xi_{2i}\right).\vspace{3mm}
\]

\textbf{Updated $\bf{\mathcal{E}}$}:
For the update of the Brownian increments, we choose randomly, with equal probability, one of the following two proposals. The replacement step is essential for introducing new increments and thus producing a variety of possible trajectories of the hidden Gaussian process. The permutation step was inspired by the local displacement step in the jump update and is intended to improve the convergence of the algorithm.\\
\textit{Replacement step:}
For the apriori distribution of the increments $\mathcal{E}$, we choose the $N$-variate standard normal distribution with density 
\begin{equation*}
\prod_{i=1}^N \frac{1}{\sqrt{2\pi}}\operatorname{exp}\Big(-\frac{\epsilon_i^2}{2}\Big).
\end{equation*}
For $n\in\{1,\dots,N\}$ randomly choose indices $i_1,\dots,i_n\in\{1,\dots,N\}$ (there are $N\choose n$ possibilities to choose $n$ indices from a set of $N$ elements without repetition; each outcome has the same probability) and define $I:=\{i_1,\dots,i_n\}$, $\hat{I}:=\{1,\dots,N\}\setminus I$. Generate candidate $\bf{\mathcal{E}}'$ by setting $\epsilon_i'=\epsilon_i$ for $i\in \hat{I}$ and drawing $\epsilon_i' \sim\mathcal{N}(0,1)$ for $i\in I$. Since the proposal density is given by
\[
q(\mathcal{E}'|\mathcal{E})=\frac{1}{\binom{N}{n}}\prod_{i\in I} \frac{1}{\sqrt{2\pi}}\operatorname{exp}\Big(-\frac{(\epsilon_i')^2}{2}\Big),
\]
we have
\[
\frac{\pi(\mathcal{E}')}{\pi(\mathcal{E})}\cdot\frac{q(\mathcal{E}|\mathcal{E}')}{q(\mathcal{E}'|\mathcal{E})}=1.
\]
Thus for the Metropolis-step, the acceptance ratio is given by
\[
\alpha(\mathcal{E}',\mathcal{E})=\operatorname{min}\left\{1,\frac{l(\mathcal{X}|\lambda_{\v},\sigma_{\v},\lambda_{\w},\sigma_{\w},\lambda_{J_1},\lambda_{J_2},\Phi_1,\Phi_2,\mathcal{E}')}{l(\mathcal{X}|\lambda_{\v},\sigma_{\v},\lambda_{\w},\sigma_{\w},\lambda_{J_1},\lambda_{J_2},\Phi_1,\Phi_2,\mathcal{E})}\right\}.\vspace{3mm}
\]

\textit{Permutation step:}
For $n\in\{1,\dots,N\}$ randomly choose indices $i_1,\dots,i_n\in\{1,\dots,N\}$ (there are $N\choose n$ possibilities to choose $n$ indices from a set of $N$ elements without repetition; each outcome has the same probability) and define $I:=\{i_1,\dots,i_n\}$, $\hat{I}:=\{1,\dots,N\}\setminus I$. Let $\pi: I\rightarrow I$ be a permutation. Generate candidate $\bf{\mathcal{E}}'$ by setting $\epsilon_{i}'=\epsilon_i$ for $i\in \hat{I}$ and $\epsilon_i'=\epsilon_{\pi(i)}$ for $i\in {I}$. Here the proposal density is given by $q(\mathcal{E}'|\mathcal{E})=\frac{1}{\binom{N}{n}}\frac{1}{n!}$ and again we have
\[
\frac{\pi(\mathcal{E}')}{\pi(\mathcal{E})}\cdot\frac{q(\mathcal{E}|\mathcal{E}')}{q(\mathcal{E}'|\mathcal{E})}=1.
\]
Thus for the Metropolis-step, the acceptance ratio is given by
\[
\alpha(\mathcal{E}',\mathcal{E})=\operatorname{min}\left\{1,\frac{l(\mathcal{X}|\lambda_{\v},\sigma_{\v},\lambda_{\w},\sigma_{\w},\lambda_{J_1},\lambda_{J_2},\Phi_1,\Phi_2,\mathcal{E}')}{l(\mathcal{X}|\lambda_{\v},\sigma_{\v},\lambda_{\w},\sigma_{\w},\lambda_{J_1},\lambda_{J_2},\Phi_1,\Phi_2,\mathcal{E})}\right\}.\vspace{3mm}
\]



\vspace{3mm}
\textbf{Updated $\bf{\Phi_1}$}:
Following \cite{GMP17}, for the update of the jump trajectories, we choose randomly, with equal probability, one of the following three proposals.\\
\textit{Birth-and-death step}:
Choose birth-move with probability $p\in(0,1)$. Generate $(\tau_1,\xi_1)$, where $\tau_1\sim \mathcal{U}([0,T])$ and $\xi_1\sim \operatorname{Ex}(\beta_1)$. Proposal transition kernel $q(\Phi_1,\Phi_1\cup \{(\tau_1,\xi_1)\})$ has density 

\[
q(\Phi_1,\Phi_1\cup \{(\tau_1,\xi_1)\})=\beta_1^{-1}\operatorname{exp}(-(\beta_1^{-1}-1)\xi_1)
\]
with respect to the product of Lebesgue-measure on $[0,T]$ and $\operatorname{Ex}(1)$ measure on $(0,\infty)$.\\

Choose death-move with probability $1-p$. Select a randomly selected point $(\tau_{1i},\xi_{1i})$ being removed from $\Phi_1$ (if $\Phi_1$ is not empty). The proposal transition kernel with respect to the counting measure is
\[
q(\Phi_1,\Phi_1\setminus \{(\tau_{1i},\xi_{1i})\})=\frac{1}{N_T^1},
\]
where $N_T^1$ is the number of points in $\Phi_1$ before the death-move.\\
The acceptance ratio for birth-move from $\Phi_1$ to $\Phi_1\cup \{(\tau_1,\xi_1)\}$ is then given by
\[
\alpha(\Phi_1,\Phi_1\cup \{(\tau_1,\xi_1)\})=\operatorname{min}\{1,r(\Phi_1,(\tau_1,\xi_1))\}
\]\\
and the acceptance ratio for the death-move from $\Phi_1$ to $\Phi_1\setminus \{(\tau_{1i},\xi_{1i})$ is 
\[
\alpha(\Phi_1, \Phi_1\setminus \{(\tau_{1i},\xi_{1i})\})=\operatorname{min}\left\{1,\frac{1}{r(\Phi_1\setminus \{(\tau_{1i},\xi_{1i})\},(\tau_{1i},\xi_{1i}))}\right\},
\]
where

\begin{equation*}
    \begin{aligned}
    r(\hat{\Phi}_1,(\theta_1,\beta_1)){}
    & = \frac{l(\mathcal{X}|\lambda_{\v},\sigma_{\v},\lambda_{\w},\sigma_{\w},\lambda_{J_1},\hat{\Phi}_1\cup \{(\tau_1,\xi_1)\},\lambda_{J_2},\Phi_2,\mathcal{E})}{l(\mathcal{X}|\lambda_{\v},\sigma_{\v},\lambda_{\w},\sigma_{\w},\lambda_{J_1},\hat{\Phi}_1,\lambda_{J_2},\Phi_2,\mathcal{E})}\frac{\pi(\hat{\Phi}_1\cup \{(\tau_1,\xi_1)|\theta_1,\beta_1\})}{\pi(\hat{\Phi}_1|\theta_1,\beta_1\})}\\
    & \qquad \times\frac{1-p}{p}\frac{1}{(N_T^1+1)q(\hat{\Phi}_1,\hat{\Phi}_1\cup \{(\theta_1,\xi_1)\})}\\
    & = \frac{l(\mathcal{X}|\lambda_{\v},\sigma_{\v},\lambda_{\w},\sigma_{\w},\lambda_{J_1},\hat{\Phi}_1\cup \{(\tau_1,\xi_1)\},\lambda_{J_2},\Phi_2,\mathcal{E})}{l(\mathcal{X}|\lambda_{\v},\sigma_{\v},\lambda_{\w},\sigma_{\w},\lambda_{J_1},\hat{\Phi}_1,\lambda_{J_2},\Phi_2,\mathcal{E})}\frac{1-p}{p}\frac{T}{\hat{N}_T^1+1}\theta_1.
    \end{aligned}\vspace{3mm}
\end{equation*}

\textit{Local displacement move}:
Assume that the jump times are ordered, i.e.,~$\tau_{1,1}<\dots < \tau_{1,N_T^1}$. Choose randomly one of the jump times, say $\tau_{1,j}$, and generate a new jump time uniformly on $[\tau_{1,j-1},\tau_{1,j+1}]$. Displace and resize the point $(\tau_{1,j},\xi_{1,j})$ to $(\tau_1,\xi_1)$, where $\xi_1=e^{-\lambda_{J1}^{-1}(\tau_1-\tau_{1,j})}\xi_{1,j}$. Perform Metropolis-step with acceptance ratio $\alpha(\Phi_1,\Phi_1')=\operatorname{min}\{1,r(\Phi_1,\Phi_1')\}$, where

\begin{equation*}
    \begin{aligned}
    r({\Phi}_1,\Phi_1'){}
    & = \frac{l(\mathcal{X}|\lambda_{\v},\sigma_{\v},\lambda_{\w},\sigma_{\w},\lambda_{J_1},\lambda_{J_2},{\Phi}_2,\Phi_1',\mathcal{E})}{l(\mathcal{X}|\lambda_{\v},\sigma_{\v},\lambda_{\w},\sigma_{\w},\lambda_{J_1},\lambda_{J_2},{\Phi}_2,\Phi_1,\mathcal{E})}\frac{e^{-\beta_1^{-1}\xi_1}}{e^{-\beta_1^{-1}\xi_{1,j}}}e^{-\lambda_{J1}(\tau_1-\tau_{1,j})}.
    \end{aligned}\vspace{3mm}
\end{equation*}

\textit{Multiplicative jump size update}
For each jump $(\tau_{1,j},\xi_{1,j})$ propose a new jump size $\xi_{1,j}'=\xi_{1,j}\phi_{1,j}$, where $\operatorname{log}(\phi_{1,j})\sim \mathcal{N}(0,c_1^2)$ are i.i.d.~random variables. The variance $c_1^2$ is chosen inversely proportional to the current number of jumps. Perform Metropolis-step with acceptance ratio \footnote{The acceptance ratio depends on the chosen reference measure, see Appendix B.}

\begin{equation*}
    \begin{aligned}
    \alpha({\Phi}_1,\Phi_1'){}
    & = \frac{l(\mathcal{X}|\lambda_{\v},\sigma_{\v},\lambda_{\w},\sigma_{\w},\lambda_{J_1},\lambda_{J_2},{\Phi}_2,\Phi_1',\mathcal{E})}{l(\mathcal{X}|\lambda_{\v},\sigma_{\v},\lambda_{\w},\sigma_{\w},\lambda_{J_1},\lambda_{J_2},{\Phi}_2,\Phi_1,\mathcal{E})}\operatorname{exp}\Big(\minus(\beta_1^{-1}\minus 1)\sum\limits_{i=1}^{N_T^1}(\xi_{1,i}'-\xi_{1,i})\Big)\prod\limits_{i=1}^{N_T^1}\frac{\xi_{1,i}'}{\xi_{1,i}}.
    \end{aligned}
\end{equation*}
\vspace{3mm}
\textbf{Updated $\bf{\Phi_2}$}:

\textit{Birth-and-death step}:
Choose birth-move with probability $p\in(0,1)$. Generate $(\tau_2,\xi_2)$, where $\tau_2\sim \mathcal{U}([0,T])$ and $\xi_2\sim \operatorname{Ex}(\beta_2)$. Proposal transition kernel $q(\Phi_2,\Phi_2\cup \{(\tau_2,\xi_2)\})$ has density 

\[
q(\Phi_2,\Phi_2\cup \{(\tau_2,\xi_2)\})=\beta_2^{-1}\operatorname{exp}(-(\beta_2^{-1}-1)\xi_2)
\]
with respect to the product of Lebesgue-measure on $[0,T]$ and $\operatorname{Ex}(1)$ measure on $(0,\infty)$.\\

Choose death-move with probability $1-p$. Select a randomly selected point $(\tau_{2i},\xi_{2i})$ being removed from $\Phi_2$ (if $\Phi_2$ is not empty). The proposal transition kernel with respect to the counting measure is
\[
q(\Phi_2,\Phi_2\setminus \{(\tau_{2i},\xi_{2i})\})=\frac{1}{N_T^2},
\]
where $N_T^2$ is the number of points in $\Phi_2$ before the death-move.\\
The acceptance ratio for birth-move from $\Phi_2$ to $\Phi_2\cup \{(\tau_2,\xi_2)\}$ is then given by
\[
\alpha(\Phi_2,\Phi_2\cup \{(\tau_2,\xi_2)\})=\operatorname{min}\{1,r(\Phi_2,(\tau_2,\xi_2))\}
\]\\
and the acceptance ratio for the death-move from $\Phi_2$ to $\Phi_2\setminus \{(\tau_{2i},\xi_{2i})$ is 
\[
\alpha(\Phi_2, \Phi_2\setminus \{(\tau_{2i},\xi_{2i}))=\operatorname{min}\left\{1,\frac{1}{r(\Phi_2\setminus \{(\tau_{2i},\xi_{2i})\},(\tau_{2i},\xi_{2i}))}\right\},
\]
where

\begin{equation*}
    \begin{aligned}
    r(\hat{\Phi}_2,(\theta_2,\beta_2)){}
    & = \frac{l(\mathcal{X}|\lambda_{\v},\sigma_{\v},\lambda_{\w},\sigma_{\w},\lambda_{J_1},\hat{\Phi}_2\cup \{(\tau_2,\xi_2)\},\lambda_{J_2},\Phi_1,\mathcal{E})}{l(\mathcal{X}|\lambda_{\v},\sigma_{\v},\lambda_{\w},\sigma_{\w},\lambda_{J_1},\hat{\Phi}_2,\lambda_{J_2},\Phi_1,\mathcal{E})}\frac{\pi(\hat{\Phi}_2\cup \{(\tau_2,\xi_2)|\theta_2,\beta_2\})}{\pi(\hat{\Phi}_2|\theta_2,\beta_2\})}\\
    & \qquad \times\frac{1-p}{p}\frac{1}{(N_T^2+1)q(\hat{\Phi}_2,\hat{\Phi}_2\cup \{(\theta_2,\xi_2)\})}\\
    & = \frac{l(\mathcal{X}|\lambda_{\v},\sigma_{\v},\lambda_{\w},\sigma_{\w},\lambda_{J_1},\hat{\Phi}_2\cup \{(\tau_2,\xi_2)\},\lambda_{J_2},\Phi_1,\mathcal{E})}{l(\mathcal{X}|\lambda_{\v},\sigma_{\v},\lambda_{\w},\sigma_{\w},\lambda_{J_1},\hat{\Phi}_1,\lambda_{J_2},\Phi_2,\mathcal{E})}\frac{1-p}{p}\frac{T}{\hat{N}_T^2+1}\theta_2.
    \end{aligned}\vspace{3mm}
\end{equation*}

\textit{Local displacement move}:
Assume that the jump times are ordered, i.e. $\tau_{2,1}<\dots < \tau_{2,N_T^2}$. Choose randomly one of the jump times, say $\tau_{2,j}$, and generate a new jump time uniformly on $[\tau_{2,j-1},\tau_{2,j+1}]$. Displace and resize the point $(\tau_{2,j},\xi_{2,j})$ to $(\tau_2,\xi_2)$, where $\xi_2=e^{-\lambda_{J2}^{-1}(\tau_2-\tau_{2,j})}\xi_{2,j}$. Perform Metropolis-step with acceptance ratio $\alpha(\Phi_2,\Phi_2')=\operatorname{min}\{1,r(\Phi_2,\Phi_2')\}$, where

\begin{equation*}
    \begin{aligned}
    r({\Phi}_2,\Phi_2'){}
    & = \frac{l(\mathcal{X}|\lambda_{\v},\sigma_{\v},\lambda_{\w},\sigma_{\w},\lambda_{J_1},\lambda_{J_2},{\Phi}_2',\Phi_1,\mathcal{E})}{l(\mathcal{X}|\lambda_{\v},\sigma_{\v},\lambda_{\w},\sigma_{\w},\lambda_{J_1},\lambda_{J_2},{\Phi}_2,\Phi_1,\mathcal{E})}\frac{e^{-\beta_2^{-1}\xi_2}}{e^{-\beta_2^{-1}\xi_{2,j}}}e^{-\lambda_{J2}(\tau_2-\tau_{2,j})}.
    \end{aligned}\vspace{3mm}
\end{equation*}

\textit{Multiplicative jump size update}
For each jump $(\tau_{2,j},\xi_{2,j})$ propose a new jump size $\xi_{2,j}'=\xi_{2,j}\phi_{2,j}$, where $\operatorname{log}(\phi_{2,j})\sim \mathcal{N}(0,c_2^2)$ are i.i.d. random variables. The variance $c_2^2$ is chosen inversely proportional to the current number of jumps. Perform Metropolis-step with acceptance ratio

\begin{equation*}
    \begin{aligned}
    \alpha({\Phi}_2,\Phi_2'){}
    & = \frac{l(\mathcal{X}|\lambda_{\v},\sigma_{\v},\lambda_{\w},\sigma_{\w},\lambda_{J_1},\lambda_{J_2},{\Phi}_2',\Phi_1,\mathcal{E})}{l(\mathcal{X}|\lambda_{\v},\sigma_{\v},\lambda_{\w},\sigma_{\w},\lambda_{J_1},\lambda_{J_2},{\Phi}_2,\Phi_1,\mathcal{E})}\operatorname{exp}\Big(\minus(\beta_2^{-1}\minus 1)\sum\limits_{i=1}^{N_T^2}(\xi_{2,i}'-\xi_{2,i})\Big)\prod\limits_{i=1}^{N_T^2}\frac{\xi_{2,i}'}{\xi_{2,i}}.
    \end{aligned}
\end{equation*}
\vspace{5mm}

\vspace{5mm}
\section*{Appendix B: Jump update with change point}\label{AppendixA}

In this appendix, we give a detailed explanation for the update steps of the jump processes in case that jump sizes and intensity change over time.
We assume that there are two different jump regimes on the interval $[0,T]$, the change point is denoted by $T_C$. In the interval $[0,T_C]$, the jump arrival times have intensity $\theta_1^{(1)}$ and the jump size is $\beta_1^{(1)}$, whereas in the interval $(T_C,T]$, the jump intensity is $\theta_1^{(2)}$ and the size is $\beta_1^{(2)}$.\\

\textbf{Update $\bf{\theta_1}^{(1)}$}:

With the choice of an $\operatorname{Ga}(a_{{\theta_1^{(1)}}},b_{\theta_1^{(1)}})$ distributed prior for $\theta_1^{(1)}$, for the posterior we get
\[
\theta_1^{(1)}|\Phi_1 \sim \operatorname{Ga}\left(a_{\theta_1^{(1)}}+N_{T_C}^1,b_{\theta_1^{(1)}}+T_C^1\right).
\]
\vspace{3mm}

\textbf{Update $\bf{\theta_1}^{(2)}$}:

With the choice of an $\operatorname{Ga}(a_{{\theta_1^{(2)}}},b_{\theta_1^{(2)}})$ distributed prior for $\theta_1^{(2)}$, for the posterior we get
\[
\theta_1^{(2)}|\Phi_1 \sim \operatorname{Ga}\left(a_{\theta_1^{(2)}}+(N_T^1-N_{T_C}^1),b_{\theta_1^{(2)}}+(T-T_C^1)\right).
\]
\vspace{3mm}

\textbf{Update $\bf{\beta_1}^{(1)}$}

With the choice of an $\operatorname{IG}(a_{{\beta_1^{(1)}}},b_{\beta_1^{(1)}})$ distributed prior for $\beta_1^{(1)}$, for the posterior we get
\[
\beta_1^{(1)}|\Phi_1 \sim \operatorname{IG}\left(a_{\beta_1^{(1)}}+N_{T_C}^1,b_{\beta_1^{(1)}}+\sum\limits_{i=1}^{N_{T_C}^1} \xi_{1i}\right).
\]
\vspace{3mm}

\textbf{Update $\bf{\beta_1}^{(2)}$}

With the choice of an $\operatorname{IG}(a_{{\beta_1^{(2)}}},b_{\beta_1^{(2)}})$ distributed prior for $\beta_1^{(2)}$, for the posterior we get
\[
\beta_1^{(2)}|\Phi_1 \sim \operatorname{IG}\left(a_{\beta_1^{(2)}}+(N_T^1-N_{T_C}^1),b_{\beta_1^{(2)}}+\sum\limits_{N_{T_C}^1+1}^{N_T^1} \xi_{1i}\right).
\]
\vspace{3mm}

\textbf{Update $\Phi_1$}

\textit{Birth-and-death step}:

Choose birth-move with probability $p\in(0,1)$. Generate $(\tau_1,\xi_1)$ by the following procedure. With probability
\[
\frac{\theta_1^{(1)}T_C}{\theta_1^{(1)}T_C+\theta_1^{(2)}(T-T_C)}\text{ draw } \tau_1\sim\mathcal{U}([0,T_C]) \text{ and } \xi_1\sim\operatorname{Ex}(\beta_1^{1}),
\]
\[
\frac{\theta_1^{(2)}(T-T_C)}{\theta_1^{(1)}T_C+\theta_1^{(2)}(T-T_C)}\text{ draw } \tau_1\sim\mathcal{U}((T_C,T]) \text{ and } \xi_1\sim\operatorname{Ex}(\beta_1^{2}).
\]

The proposal transition kernel $q(\Phi_1,\Phi_1\cup \{(\tau_1,\xi_1)\})$ thus has density 
\[
q(\Phi_1,\Phi_1\cup \{(\tau_1,\xi_1)\})=
\begin{cases}
(\beta_1^{(1)})^{-1}\operatorname{exp}(-((\beta_1^{(1)})^{-1}-1)\xi_1) \frac{\theta_1^{(1)}}{\theta_1^{(1)}T_C+\theta_1^{(2)}(T-T_C)},\quad \tau_1\in[0,T_C]\\
(\beta_1^{(2)})^{-1}\operatorname{exp}(-((\beta_1^{(2)})^{-1}-1)\xi_1) \frac{\theta_1^{(2)}}{\theta_1^{(1)}T_C+\theta_1^{(2)}(T-T_C)},\quad \tau_1\in (T_C,T]
\end{cases}
\]
with respect to the product of Lebesgue-measure on $[0,T]$ and $\operatorname{Ex}(1)$ measure on $[0,\infty)$.\\

Choose death-move with probability $1-p$. Select a randomly selected point $(\tau_{1i},\xi_{1i})$ being removed from $\Phi_1$ (if $\Phi_1$ is not empty). The proposal transition kernel with respect to the counting measure is
\[
q(\Phi_1,\Phi_1\setminus \{(\tau_{1i},\xi_{1i})\})=\frac{1}{N_T^1},
\]
where $N_T^1$ is the number of points in $\Phi_1$ before the death-move.\\

Acceptance ratio for birth-move from $\Phi_1$ to $\Phi_1\cup \{(\tau_1,\xi_1)\}$ is 
\[
\alpha(\Phi_1,\Phi_1\cup \{(\tau_1,\xi_1)\})=\operatorname{min}\{1,r(\Phi_1,(\tau_1,\xi_1))\}.
\]\\

Acceptance ratio for the death-move from $\Phi_1$ to $\Phi_1\setminus \{(\tau_{1i},\xi_{1i})$ is 
\[
\alpha(\Phi_1, \Phi_1\setminus \{(\tau_{1i},\xi_{1i}))=\operatorname{min}\{1,\frac{1}{r(\Phi_1\setminus \{(\tau_{1i},\xi_{1i})\},(\tau_{1i},\xi_{1i}))}\}
\]

where

\textbf{Case $\tau_1\in [0,T_C]$:}

\begin{equation*}
    \begin{aligned}
    &r(\hat{\Phi}_1,(\tau_1,\xi_1)){}\\
    & = \frac{l(\mathcal{X}|\lambda_{\v},\sigma_{\v},\lambda_{\w},\sigma_{\w},\lambda_{J_1},\hat{\Phi}_1\cup \{(\tau_1,\xi_1)\},\lambda_{J_2},\Phi_2,\mathcal{B})}{l(\mathcal{X}|\lambda_{\v},\sigma_{\v},\lambda_{\w},\sigma_{\w},\lambda_{J_1},\hat{\Phi}_1,\lambda_{J_2},\Phi_2,\mathcal{B})}\frac{\pi(\hat{\Phi}_1\cup \{(\tau_1,\xi_1)|\theta_1^{(1)},\theta_1^{(2)},\beta_1^{(1)},\beta_1^{(2)}\})} 
    {\pi(\hat{\Phi}_1|\theta_1^{(1)},\theta_1^{(2)},\beta_1^{(1)},\beta_1^{(2)}\})}\\
    & \qquad \times  \frac{1-p}{p}\frac{1}{(N_T^1+1)q(\hat{\Phi}_1,\hat{\Phi}_1\cup \{(\theta_1,\xi_1)\})}\\
    & = \frac{l(\mathcal{X}|\lambda_{\v},\sigma_{\v},\lambda_{\w},\sigma_{\w},\lambda_{J_1},\hat{\Phi}_1\cup \{(\tau_1,\xi_1)\},\lambda_{J_2},\Phi_2,\mathcal{B})}{l(\mathcal{X}|\lambda_{\v},\sigma_{\v},\lambda_{\w},\sigma_{\w},\lambda_{J_1},\hat{\Phi}_1,\lambda_{J_2},\Phi_2,\mathcal{B})}\frac{1-p}{p}\frac{\theta_1^{(1)}T_C+\theta_1^{(2)}(T-T_C)}{\hat{N}_T^1+1}.
    \end{aligned}
\end{equation*}

\textbf{Case $\tau_1\in (T_C,T]$:}

\begin{equation*}
    \begin{aligned}
    &r(\hat{\Phi}_1,(\tau_1,\xi_1)){}\\
    & = \frac{l(\mathcal{X}|\lambda_{\v},\sigma_{\v},\lambda_{\w},\sigma_{\w},\lambda_{J_1},\hat{\Phi}_1\cup \{(\tau_1,\xi_1)\},\lambda_{J_2},\Phi_2,\mathcal{B})}{l(\mathcal{X}|\lambda_{\v},\sigma_{\v},\lambda_{\w},\sigma_{\w},\lambda_{J_1},\hat{\Phi}_1,\lambda_{J_2},\Phi_2,\mathcal{B})}\frac{\pi(\hat{\Phi}_1\cup \{(\tau_1,\xi_1)|\theta_1^{(1)},\theta_1^{(2)},\beta_1^{(1)},\beta_1^{(2)}\})}{\pi(\hat{\Phi}_1|\theta_1^{(1)},\theta_1^{(2)},\beta_1^{(1)},\beta_1^{(2)}\})}\\
    & \qquad \times\frac{1-p}{p}\frac{1}{(N_T^1+1)q(\hat{\Phi}_1,\hat{\Phi}_1\cup \{(\theta_1,\xi_1)\})}\\
    & = \frac{l(\mathcal{X}|\lambda_{\v},\sigma_{\v},\lambda_{\w},\sigma_{\w},\lambda_{J_1},\hat{\Phi}_1\cup \{(\tau_1,\xi_1)\},\lambda_{J_2},\Phi_2,\mathcal{B})}{l(\mathcal{X}|\lambda_{\v},\sigma_{\v},\lambda_{\w},\sigma_{\w},\lambda_{J_1},\hat{\Phi}_1,\lambda_{J_2},\Phi_2,\mathcal{B})}\frac{1-p}{p}\frac{\theta_1^{(1)}T_C+\theta_1^{(2)}(T-T_C)}{\hat{N}_T^1+1}.
    \end{aligned}
\end{equation*}

\textit{Multiplicative jump size update}

For each jump $(\tau_{1,j},\xi_{1,j})$ propose a new jump size $\xi_{1,j}'=\xi_{1,j}\phi_{1,j}$, where $\operatorname{log}(\phi_{1,j})\sim \mathcal{N}(0,c_1^2)$ are i.i.d. random variables. The variance $c_1^2$ is chosen inversely proportional to the current number of jumps. Perform Metropolis-step with acceptance ratio

\begin{equation*}
    \begin{aligned}
    \alpha({\Phi}_1,\Phi_1'){}
    & = \operatorname{min}\Big\{1,\frac{l(\mathcal{X}|\lambda_{\v},\sigma_{\v},\lambda_{\w},\sigma_{\w},\lambda_{J_1},\lambda_{J_2},{\Phi}_2,\Phi_1',\mathcal{B})}{l(\mathcal{X}|\lambda_{\v},\sigma_{\v},\lambda_{\w},\sigma_{\w},\lambda_{J_1},\lambda_{J_2},{\Phi}_2,\Phi_1,\mathcal{B})}\prod\limits_{i=1}^{N_T^1}\frac{\xi_{1,i}'}{\xi_{1,i}}\\
    &\, \times \operatorname{exp}\Big(-((\beta_1^{(1)})^{-1}-1)\sum\limits_{i=1}^{N_C^1}(\xi_{1,i}'-\xi_{1,i})\Big) \operatorname{exp}\Big(-((\beta_1^{(2)})^{-1}-1)\sum\limits_{i=N_C^1+1}^{N_T^1}(\xi_{1,i}'-\xi_{1,i})\Big)\Big\}.
    \end{aligned}
\end{equation*}

\textit{Local displacement move}

Assume that the jump times are ordered, i.e. $\tau_{1,1}<\dots < \tau_{1,N_T^1}$. Choose randomly one of the jump times, say $\tau_{1,j}$, and generate a new jump time $\tau_1$ uniformly on $[\tau_{1,j-1},\tau_{1,j+1}]$. Displace and resize the point $(\tau_{1,j},\xi_{1,j})$ to $(\tau_1,\xi_1)$, where $\xi_1=e^{-\lambda_{J1}^{-1}(\tau_1-\tau_{1,j})}\xi_{1,j}$. Perform Metropolis-step with acceptance ratio $\alpha(\Phi_1,\Phi_1')=\operatorname{min}\{1,r(\Phi_1,\Phi_1')\}$, where\\

Case $\tau_1 ,\tau_{1,j}\in [0,T_C]$:

\begin{equation*}
    \begin{aligned}
    r({\Phi}_1,\Phi_1'){}
    & = \frac{l(\mathcal{X}|\lambda_{\v},\sigma_{\v},\lambda_{\w},\sigma_{\w},\lambda_{J_1},\lambda_{J_2},{\Phi}_2,\Phi_1',\mathcal{B})}{l(\mathcal{X}|\lambda_{\v},\sigma_{\v},\lambda_{\w},\sigma_{\w},\lambda_{J_1},\lambda_{J_2},{\Phi}_2,\Phi_1,\mathcal{B})}\frac{e^{-(\beta_1^{(1)})^{-1}\xi_1}}{e^{-(\beta_1^{(1)})^{-1}\xi_{1,j}}}e^{-\lambda_{J1}(\tau_1-\tau_{1,j})},
    \end{aligned}
\end{equation*}

Case $\tau_1\in [0,T_C] ,\tau_{1,j}\in (T_C,T]$:

\begin{equation*}
    \begin{aligned}
    r({\Phi}_1,\Phi_1'){}
    & = \frac{l(\mathcal{X}|\lambda_{\v},\sigma_{\v},\lambda_{\w},\sigma_{\w},\lambda_{J_1},\lambda_{J_2},{\Phi}_2,\Phi_1',\mathcal{B})}{l(\mathcal{X}|\lambda_{\v},\sigma_{\v},\lambda_{\w},\sigma_{\w},\lambda_{J_1},\lambda_{J_2},{\Phi}_2,\Phi_1,\mathcal{B})}\frac{\theta_1^{(1)}}{\theta_1^{(2)}}\frac{e^{-(\beta_1^{(1)})^{-1}\xi_1}}{e^{-(\beta_1^{(2)})^{-1}\xi_{1,j}}}e^{-\lambda_{J1}(\tau_1-\tau_{1,j})},
    \end{aligned}
\end{equation*}

Case $\tau_1\in (T_C,T] ,\tau_{1,j}\in [0,T_C]$:

\begin{equation*}
    \begin{aligned}
    r({\Phi}_1,\Phi_1'){}
    & = \frac{l(\mathcal{X}|\lambda_{\v},\sigma_{\v},\lambda_{\w},\sigma_{\w},\lambda_{J_1},\lambda_{J_2},{\Phi}_2,\Phi_1',\mathcal{B})}{l(\mathcal{X}|\lambda_{\v},\sigma_{\v},\lambda_{\w},\sigma_{\w},\lambda_{J_1},\lambda_{J_2},{\Phi}_2,\Phi_1,\mathcal{B})}\frac{\theta_1^{(2)}}{\theta_1^{(1)}}\frac{e^{-(\beta_1^{(2)})^{-1}\xi_1}}{e^{-(\beta_1^{(1)})^{-1}\xi_{1,j}}}e^{-\lambda_{J1}(\tau_1-\tau_{1,j})},
    \end{aligned}
\end{equation*}

Case $\tau_1 ,\tau_{1,j}\in (T_C,T]$:

\begin{equation*}
    \begin{aligned}
    r({\Phi}_1,\Phi_1'){}
    & = \frac{l(\mathcal{X}|\lambda_{\v},\sigma_{\v},\lambda_{\w},\sigma_{\w},\lambda_{J_1},\lambda_{J_2},{\Phi}_2,\Phi_1',\mathcal{B})}{l(\mathcal{X}|\lambda_{\v},\sigma_{\v},\lambda_{\w},\sigma_{\w},\lambda_{J_1},\lambda_{J_2},{\Phi}_2,\Phi_1,\mathcal{B})}\frac{e^{-(\beta_1^{(2)})^{-1}\xi_1}}{e^{-(\beta_1^{(2)})^{-1}\xi_{1,j}}}e^{-\lambda_{J1}(\tau_1-\tau_{1,j})}.
    \end{aligned}
\end{equation*}
\vspace{5mm}
\section*{Appendix C: Poisson Processes}\label{Poisson}

Let $\Pi$ be a Poisson process on $[0,T]$ with constant intensity $\theta$. For each point $\tau$ of the random set $\Pi$, we associate an $\operatorname{Ex}(\beta)$ distributed random variable $\xi_{\tau}$ (the mark of $\tau$), taking values in $\R^+$. The pair $(\tau,\xi_{\tau})$ can then be regarded as a random point in the product space $[0,T]\times \R^+$. Then by the Marking Theorem \cite{K92}, the random set 
\[
\Pi^*=\{(\tau,\xi_{\tau})| \tau\in\Pi\}
\]
is a Poisson process on $[0,T]\times \R^+$ and its intensity measure is given by
\[
\Lambda(C)=\int\int_{(t,x)\in C} \theta(dt)p_{\xi}(dx) = \int\int_{(t,x)\in C} \theta \frac{1}{\beta} e^{-\frac{1}{\beta}x}dtdx.
\]
Thus the sets $\Phi_1=\{(\tau_{1,j},\xi_{1,j})\}_{1\leq j\leq N_{T_1}}$ and $\Phi_2=\{(\tau_{2,j},\xi_{2,j})\}_{1\leq j\leq N_{T_2}}$ of jump times and corresponding jump sizes can be interpreted as realisations of marked Poisson processes taking values in $[0,T]\times \R^+$. 
We are now interested in the likelihood $l(\Phi_i|\theta_i,\beta_i)$ with respect to a so called dominating measure. 
Let therefore $\mathcal{X}=[0,T]\times\mathbb{R^+}$ and
denote by $\mathcal{M}$ the set of $\sigma$-finite measures defined on $(\mathcal{X},\mathcal{B}(\mathcal{X})$ and by $\mathcal{M}_0$ the subspace of integer-valued point measures $X=\sum_i \epsilon_{x_i}$, where $x_i\in \mathcal{X}$ and $\epsilon_x$ is the point mass located in $x$. By $\mathcal{B}(\mathcal{M}_0)$ we denote the smallest $\sigma$-algebra of subsets of $\mathcal{M}$ with respect to which all mappings 
\[
\Pi_B:\mathcal{M}_0\rightarrow \{0,1,2,\dots,\infty\}\text{ with } \Pi_B(X)=X(B),\, B\in\mathcal{B}(\mathcal{X}),
\]
are measurable.
Let now $\Pi_0$ be a Poisson process taking values in $\mathcal{X}$ with constant intensity $\theta_0$ on $[0,T]$ and $\operatorname{Ex}(\beta_0)$ distributed jump sizes. 
Then $\Pi_0$ induces a measure $P_0$ on $(\mathcal{M}_0,\mathcal{B}(\mathcal{M}_0))$ via 
\[
P_0(X\in\mathcal{M}_0:X(B)=x)=\mathbb{P}(\omega\in\Omega:\Pi_0(\omega)(B)=x).
\]
The measure $P_0$ will now be our reference measure on $\mathcal{B}(\mathcal{M}_0)$ and will be referred to as the dominating measure. For convenience, the dominating measure is often chosen as the one induced by a Poisson process with unit intensity on $[0,T]$ and exponential jump sizes with parameter $1$, however, the choices of $\theta_0$ and $\beta_0$ can also be different \cite{RPD04}.
For another Poisson process $\Pi$ taking values in $\mathcal{X}$ with constant intensity $\theta$ on $[0,T]$ and $\operatorname{Ex}(\beta)$ distributed jump sizes, the density of $P$ with reference to the dominating measure $P_0$ can now be calculated by \cite[Theorem 1.3]{K12} as 

\begin{equation*}
    \begin{aligned}
    \frac{dP}{dP_0}(X){}
    & = \operatorname{exp}\left(\int_{\mathcal{X}} \operatorname{ln}(S(x))X(dx)-\int_{\mathcal{X}} (S(x)-1)\Lambda_0(dx)\right),\quad X\in \mathcal{M}_0\\
    \end{aligned}
\end{equation*}
where 
\[
S(x):=\frac{d\Lambda}{d\Lambda_0}((x_1,x_2))=\frac{\theta}{\theta_0} \frac{\beta_0}{\beta} \operatorname{exp}\left(-\left(\frac{1}{\beta}-\frac{1}{\beta_0}\right)x_2\right).
\]
Thus for a set $\Phi=\{(\tau_1,\xi_1),\dots, (\tau_{N_T},\xi_{N_T})\}$, $\tau_i\in [0,T]$, $\xi_i\in \R^+$ we can now calculate the likelihood of $\Phi$ with respect to the dominating measure as
\begin{equation*}
    \begin{aligned}
    l(\Phi|\theta,\beta)=\frac{dP}{dP_0}(\Phi){}
    & = \operatorname{exp}\left(\int_{\mathcal{X}} \operatorname{ln}(S(x))\Phi(dx)-\int_{\mathcal{X}} (S(x)-1)\Lambda_0(dx))\right)\\
    & = \operatorname{exp}\left(\sum\limits_{i=1}^{N_T}\operatorname{ln}(S(\tau_i,\xi_i))-\left(\int_{\mathcal{X}} \Lambda(dx)-\int_{\mathcal{X}} \Lambda_0(dx)\right)\right)\\
    & = \prod\limits_{i=1}^{N_T} S(\tau_i,\xi_i)\operatorname{exp}\left(\Lambda_0(\mathcal{X})-\Lambda(\mathcal{X})\right)\\
    & = \prod_{i=1}^{N_T} \frac{\theta}{\theta_0} \frac{\beta_0}{\beta} \operatorname{exp}\left(-\left(\frac{1}{\beta}-\frac{1}{\beta_0}\right)\xi_i\right)\cdot \operatorname{exp}\left(\theta_0 T-\theta T\right)\\
    & = \left(\frac{\theta}{\theta_0}\right)^{N_T} \left(\frac{\beta}{\beta_0}\right)^{-N_T}  \operatorname{exp}\left(-\left(\frac{1}{\beta}-\frac{1}{\beta_0}\right)\sum\limits_{i=1}^{N_T}\xi_i\right)\cdot \operatorname{exp}\left(-(\theta -\theta_0) T\right).
    \end{aligned}
\end{equation*}

If the dominating measure is chosen as a Poisson process with unit intensity on $[0,T]$ and exponential jump sizes with parameter $1$ (cf. \cite[Section 3.1]{GMP17}), $\Phi$ has likelihood

\begin{equation*}
    \begin{aligned}
    l(\Phi|\theta,\beta)=
    & = \theta^{N_T} \beta^{-N_T}  \operatorname{exp}\left(-\left(\frac{1}{\beta}-1\right)\sum\limits_{i=1}^{N_T}\xi_i\right)\cdot \operatorname{exp}\left(-(\theta -1) T\right).
    \end{aligned}\vspace{3mm}
\end{equation*}

\emph{Note:} The acceptance ratio in the multiplicative update step of the jump process
\begin{equation*}
    \begin{aligned}
    \alpha({\Phi}_1,\Phi_1'){}
    & = \frac{l(\mathcal{X}|\lambda_{\v},\sigma_{\v},\lambda_{\w},\sigma_{\w},\lambda_{J_1},\lambda_{J_2},{\Phi}_2,\Phi_1',\mathcal{E})}{l(\mathcal{X}|\lambda_{\v},\sigma_{\v},\lambda_{\w},\sigma_{\w},\lambda_{J_1},\lambda_{J_2},{\Phi}_2,\Phi_1,\mathcal{E})}\frac{l(\Phi_1'|\beta_1,\theta_1)}{l(\Phi_1|\beta_1,\theta_1)}\frac{q(\Phi_1,\Phi_1')}{q(\Phi_1',\Phi_1)}.
    \end{aligned}
\end{equation*}
in Section \ref{MCMC} depends on the chosen reference measure. 
If the jump size parameter of the Poisson process inducing the dominating measure is $1$, the acceptance ratio is given as 

\begin{equation*}
    \begin{aligned}
    \alpha({\Phi}_1,\Phi_1'){}
    & = \frac{l(\mathcal{X}|\lambda_{\v},\sigma_{\v},\lambda_{\w},\sigma_{\w},\lambda_{J_1},\lambda_{J_2},{\Phi}_2,\Phi_1',\mathcal{E})}{l(\mathcal{X}|\lambda_{\v},\sigma_{\v},\lambda_{\w},\sigma_{\w},\lambda_{J_1},\lambda_{J_2},{\Phi}_2,\Phi_1,\mathcal{E})}\operatorname{exp}\Big\{\minus(\beta_1^{-1}\minus 1)\sum\limits_{i=1}^{N_T^1}(\xi_{1,i}'-\xi_{1,i})\Big\}\prod\limits_{i=1}^{N_T^1}\frac{\xi_{1,i}'}{\xi_{1,i}},
    \end{aligned}
\end{equation*}
whereas, if the jump size parameter is changed to (arbitrarily large) $c$, the term $\frac{l(\Phi_1'|\beta_1,\theta_1)}{l(\Phi_1|\beta_1,\theta_1)}$ evaluates to $\operatorname{exp}\Big(\minus(\beta_1^{-1}\minus c^{-1})\sum\limits_{i=1}^{N_T^1}(\xi_{1,i}'-\xi_{1,i})\Big)$, where $c^{-1}$ can be arbitrarily close to $0$.

\section*{Appendix D: Proof of Theorem \ref{ThmFutures}}
By the linearity of the conditional expectation, we obtain
\[
\mathbb{E}^{\mathbb{Q}}[P_T|\mathcal{F}_t]=f(T)+\mathbb{E}^{\mathbb{Q}}[\v(T)|\mathcal{F}_t]+\mathbb{E}^{\mathbb{Q}}[\w(T)|\mathcal{F}_t]+\mathbb{E}^{\mathbb{Q}}[J_{1}(T)|\mathcal{F}_t]-\mathbb{E}^{\mathbb{Q}}[J_{2}(T)|\mathcal{F}_t].
\]
Under the risk neutral measure $\mathbb{Q}$, equation \eqref{L-Q} has the solution
\[
\v(T)=\v(t) e^{-\lambda_{\v}^{-1}(T-t)}-\frac{\phi_{\v}}{\lambda_{\v}^{-1}}(1-e^{-\lambda_{\v}^{-1}(T-t)})+\sigma_{\v}\int_t^T e^{-\lambda_{\v}^{-1}(T-s)}dW_{\v}^{\mathbb{Q}}(s).
\]
Thus we get
\[
\mathbb{E}_t^{\mathbb{Q}}[\v(T)]:=\mathbb{E}^{\mathbb{Q}}[\v(T)|\mathcal{F}_t]=\v(t) e^{-\lambda_{\v}^{-1}(T-t)}-\frac{\phi_{\v}}{\lambda_{\v}^{-1}}(1-e^{-\lambda_{\v}^{-1}(T-t)}).
\]
Analogously we have
\[
\mathbb{E}_t^{\mathbb{Q}}[\w(T)]=\w(t) e^{-\lambda_{\w}^{-1}(T-t)}-\frac{\phi_{\w}}{\lambda_{\w}^{-1}}(1-e^{-\lambda_{\w}^{-1}(T-t)}).
\]
Let $N(t,T)$ denote the number of jumps of the Poisson process in the interval $(t,T)$ and let $\tau_1^*<\dots < \tau_{N(t,T)}^*$ be the jump arrival times with corresponding jump sizes $\xi_1^*,\dots,\xi_{N(t,T)}^*$, then: 
\[
J_T=J_t e^{-\lambda_J^{-1}(T-t)}+\sum\limits_{i=1}^{N(t,T)}e^{-\lambda_J^{-1}(T-\tau_i^*)}\xi_i^*,
\]
where the jump sizes are exponentially distributed with parameter $\beta^*$ and the jump intensity $\theta^*=\theta$ does not change under the risk neutral measure. For the conditional expectation we get 
$\mathbbm{E}_t^{\mathbb{Q}}[J_T]=\mathbb{E}_t^{\mathbb{Q}}\Big[J_t e^{-\lambda_J^{-1}(T-t)}+\sum\limits_{i=1}^{N(t,T)}e^{-\lambda_J^{-1}(T-\tau_i^*)}\xi_i^*\Big]=J_t e^{-\lambda_J^{-1}(T-t)}+\mathbb{E}_t^{\mathbb{Q}}\Big[\sum\limits_{i=1}^{N(t,T)}e^{-\lambda_J^{-1}(T-\tau_i^*)}\xi_i^*\Big]$,
and since
\begin{equation*}
    \begin{aligned}
    {}
    & \mathbb{E}_t^{\mathbb{Q}}[J_T]-J_t e^{-\lambda_J^{-1}(T-t)}=\mathbb{E}_t^{\mathbb{Q}}\Big[\sum\limits_{i=1}^{N(t,T)}e^{-\lambda_J^{-1}(T-\tau_i^*)}\xi_i^*\Big]= e^{-\lambda_J^{-1}T}\mathbb{E}_t^{\mathbb{Q}}\Big[\sum\limits_{i=1}^{N(t,T)}e^{\lambda_J^{-1}\tau_i^*}\xi_i^*\Big]\\
    & = e^{-\lambda_J^{-1}T}\mathbb{E}_t^{\mathbb{Q}}\Big[\mathbb{E}^{\mathbb{Q}}\Big[\sum\limits_{i=1}^{N(t,T)}e^{\lambda_J^{-1}\tau_i^*}\xi_i^*|N(t,T)\Big]\Big]=e^{-\lambda_J^{-1}T}\mathbb{E}_t^{\mathbb{Q}}\Big[\sum\limits_{i=1}^{N(t,T)}\mathbb{E}^{\mathbb{Q}}[e^{\lambda_J^{-1}\tau_i^*}\xi_i^*]\Big]\\
    &= e^{-\lambda_J^{-1}T}\mathbb{E}_t^{\mathbb{Q}}\Big[\sum\limits_{i=1}^{N(t,T)}\mathbb{E}^{\mathbb{Q}}[e^{\lambda_J^{-1}\tau_i^*}]\mathbb{E}^{\mathbb{Q}}[\xi_i^*]\Big]=e^{-\lambda_J^{-1}T}\frac{1}{\beta^*}\mathbb{E}_t^{\mathbb{Q}}\Big[\sum\limits_{i=1}^{N(t,T)}\mathbb{E}^{\mathbb{Q}}[e^{\lambda_J^{-1}\tau_i^*}]\Big],
    \end{aligned}
\end{equation*}
and
\begin{align*} 
& \sum\limits_{i=1}^{N(t,T)}\mathbb{E}^{\mathbb{Q}}\Big[e^{\lambda_J^{-1}\tau_i^*}\Big]=\mathbb{E}^{\mathbb{Q}}\Big[\sum\limits_{i=1}^{N(t,T)}e^{\lambda_J^{-1}\tau_i^*}|N(t,T)\Big]=N(t,T)\mathbb{E}^{\mathbb{Q}}\Big[e^{\lambda_J^{-1}\tau}\Big]
\end{align*}
where $\tau\sim\mathcal{U}(t,T)$ and the last equality follows by the fact that $\tau_1^{*},\dots,\tau_{N(t,T)}^{*}$ are the jump times of a Poisson process with $N(t, T)$ jumps in the interval $(t, T)$, i.e., they are distributed like the order statistics of $N(t, T)$ independent random variables that are uniformly distributed on the interval $(t, T)$. Hence we end up with
\begin{equation*}
    \begin{aligned}
    \mathbb{E}_t^{\mathbb{Q}}[J_T]-J_t e^{-\lambda_J^{-1}(T-t)} {}
    & =e^{-\lambda_J^{-1}T}\frac{1}{\beta^*}\mathbb{E}_t^{\mathbb{Q}}\Big[N(t,T)\mathbb{E}^{\mathbb{Q}}[e^{\lambda_J^{-1}\tau}]\Big]= e^{-\lambda_J^{-1}T}\frac{1}{\beta^*}\mathbb{E}_t^{\mathbb{Q}}[N(t,T)]\mathbb{E}^{\mathbb{Q}}[e^{\lambda_J^{-1}\tau}]\\\
    & = e^{-\lambda_J^{-1}T}\frac{1}{\beta^*}\theta (T-t)\frac{1}{T-t}\int_t^T e^{\lambda_J^{-1}s}ds=\frac{\theta}{\beta^*}\lambda_J(1-e^{-\lambda_J^{-1}(T-t)}).
    \end{aligned}
\end{equation*}
Thus 
\[
\mathbb{E}_t^{\mathbb{Q}}[J_T]=J_t e^{-\lambda_J^{-1}(T-t)}+\frac{\theta}{\beta^*}\lambda_J(1-e^{-\lambda_J^{-1}(T-t)}).
\]\qed


\begin{thebibliography}{}
\bibliographystyle{alpha}

       

\bibitem{BGJM11}
     {\sc S. Brooks, A. Gelman, G. Jones and X. Meng}, {\em Handbook of Markov Chain Monte Carlo}, (2011)

\bibitem{BNS01}
     {\sc O.E. Barndorff-Nielsen and N. Shephard}, {\em Non-{G}aussian {O}rnstein-{U}hlenbeck-based models and some of their uses in financial economics}, Journal of the Royal Statistical Society. Series B. Statistical Methodology, 63 (2001), pp.~167--241.    

\bibitem{BKMB07}
     {\sc F.E. Benth, J. Kallsen and T. Meyer-Brandis}, {\em A non-{G}aussian {O}rnstein-{U}hlenbeck process for electricity spot price modeling and derivatives pricing}, Applied Mathematical Finance, 14 (2007), pp.~153--169. 

\bibitem{BKN12}
     {\sc F.E. Benth, R. Kiesel and A. Nazarova}, {\em A critical empirical study of three electricity spot price models}, Energy Economics, 34 (2012), pp.~1589--1616.           

\bibitem{CF05}
     {\sc A. Cartea and M. Figueroa}, {\em A mean reverting jump diffusion model with seasonality}, Applied Mathematical Finance, 12 (2005), pp.~313--335.      

\bibitem{FS09}
     {\sc S. Fr{\"u}hwirth-Schnatter and L. S{\"o}gner}, {\em Bayesian estimation of stochastic volatility models based on OU processes with marginal Gamma law}, Annals of the Institute of Statistical Mathematics, 61 (2009), pp.~159--179. 

\bibitem{GCDVR13}
     {\sc A. Gelman, J. Carlin, H. Stern, D. Dunson, A. Vehtari and D. Rubin}, {\em Bayesian Data Analysis}, (2013) 

\bibitem{GR06}
     {\sc H. Geman and A. Roncoroni}, {\em Understanding the fine structure of electricity prices}, The Journal of Business, 79 (2006), pp.~1225--1261.    
     
\bibitem{GMP17}
   {\sc J. Gonzalez, J. Moriarty and J. Palczewski}, {\em Bayesian calibration and number of jump components in electricity spot price models}, Energy Economics, 65 (2017), pp.~375--388. 

\bibitem{GN08}
     {\sc R. Green and M. Nossman}, {\em Markov chain Monte Carlo estimation of a multi-factor jump diffusion model for power prices}, The Journal of Energy Markets, 1 (2008), pp.~65--91.          

\bibitem{HW20}
   {\sc W.J. Hinderks and A. Wagner}, {\em Factor models in the German electricity market: Stylized facts, seasonality, and calibration}, Energy Economics, 85 (2020), pp.~1043--1051.   

\bibitem{K92}
     {\sc J.F. Kingman}, {\em Poisson processes}, 3 (1992)      

\bibitem{K12}
     {\sc Y. A. Kutoyants}, {\em Statistical inference for spatial Poisson processes}, 134 (2012)      

\bibitem{LS02}
   {\sc J. Lucia and E. Schwartz}, {\em Electricity prices and power derivatives: Evidence from the Nordic Power Exchange}, Review of derivatives research, 5 (2002), pp.~5--50.       

\bibitem{MBT08}
   {\sc T. Meyer-Brandis and P. Tankov}, {\em Multi-factor jump-diffusion models of electricity prices}, International Journal of Theoretical and Applied Finance, 11 (2008), pp.~503--528.      
   
\bibitem{RPD04}
     {\sc G. O. Roberts, O. Papaspiliopoulos and P. Dellaportas}, {\em Bayesian inference for non-{G}aussian {O}rnstein-{U}hlenbeck stochastic volatility processes}, Journal of the Royal Statistical Society. Series B.
              Statistical Methodology, 66 (2004), pp.~369--393.         

\bibitem{R84}
     {\sc D.B. Rubin}, {\em Bayesianly justifiable and relevant frequency calculations for the applied statistician}, The Annals of Statistics, 12 (1984), pp.~1151--1172.       

\bibitem{SS00}
     {\sc E. Schwartz and J. Smith}, {\em Short-term variations and long-term dynamics in commodity prices}, Management Science, 46 (2000), pp.~893--911.      

\bibitem{SUH07}
   {\sc J. Seifert and M. Uhrig-Homburg}, {\em Modelling jumps in electricity prices: theory and empirical evidence}, Review of Derivatives Research, 10 (2007), pp.~59--85.          

\bibitem{V03}
     {\sc P. Villaplana}, {\em Pricing power derivatives: A two-factor jump-diffusion approach}, Available at SSRN 493943 (2003)  

\end{thebibliography}
\end{document}